\documentclass[12pt,preprint]{aastex}
\usepackage{graphicx}
\shorttitle{Radial dust properties of SINGS galaxies}
\shortauthors{Mu\~{n}oz-Mateos et al.}

\begin{document}

\title{Radial distribution of stars, gas and dust in SINGS galaxies: II. Derived dust properties}

\author{J.C. Mu\~{n}oz-Mateos\altaffilmark{1},
A. Gil de Paz\altaffilmark{1},
S. Boissier\altaffilmark{2},
J. Zamorano\altaffilmark{1},
D.A. Dale\altaffilmark{3},
P.G. P\'{e}rez-Gonz\'{a}lez\altaffilmark{1},
J. Gallego\altaffilmark{1},
B.F. Madore\altaffilmark{4},
G. Bendo\altaffilmark{5},
M. D. Thornley\altaffilmark{6},
B. T. Draine\altaffilmark{7},
A. Boselli\altaffilmark{2},
V. Buat\altaffilmark{2},
D. Calzetti\altaffilmark{8},
J. Moustakas\altaffilmark{9},
R. C. Kennicutt, Jr.\altaffilmark{10,11}
}

\altaffiltext{1}{Departamento de Astrof\'{\i}sica y CC$.$ de la Atm\'osfera, Universidad Complutense de Madrid, Avda$.$ de la Complutense, s/n, E-28040 Madrid, Spain; jcmunoz, agpaz, jaz, pgperez, jgm@astrax.fis.ucm.es}
\altaffiltext{2}{Laboratoire d'Astrophysique de Marseille, OAMP, Universit\'e Aix-Marseille \& CNRS UMR 6110, 38 rue Fr\'ed\'eric Joliot-Curie, 13388 Marseille cedex 13, France; samuel.boissier, alessandro.boselli, veronique.buat@oamp.fr}
\altaffiltext{3}{Department of Physics and Astronomy, University of Wyoming, Laramie, WY; ddale@uwyo.edu}
\altaffiltext{4}{Observatories of the Carnegie Institution of Washington, 813 Santa Barbara Street, Pasadena, CA 91101; barry@ociw.edu}
\altaffiltext{5}{Astrophysics Group, Imperial College, Blackett Laboratory, Prince Consort Road, London SW7 2AZ; g.bendo@imperial.ac.uk}
\altaffiltext{6}{Department of Physics and Astronomy, Bucknell University, Lewisburg, PA 17837, m.thornley@bucknell.edu}
\altaffiltext{7}{Princeton University Observatory, Princeton, NJ 08544-1001; draine@astro.princeton.edu}
\altaffiltext{8}{Department of Astronomy, University of Massachusetts, Amherst, MA 01003; calzetti@astro.umass.edu}
\altaffiltext{9}{Department of Physics, New York University, 4 Washington Place, New York, NY 10003, USA}
\altaffiltext{10}{Institute of Astronomy, University of Cambridge, Madingley Road, Cambridge CB3 0HA, UK}
\altaffiltext{11}{Steward Observatory, University of Arizona, Tucson, AZ 85721}
\begin{abstract}
We present a detailed analysis of the radial distribution of dust
properties in the SINGS sample, performed on a set of UV, IR and HI
surface brightness profiles, combined with published molecular gas
profiles and metallicity gradients. The internal extinction, derived
from the TIR-to-FUV luminosity ratio, decreases with radius, and is
larger in Sb-Sbc galaxies. The TIR-to-FUV ratio correlates with the UV
spectral slope $\beta$, following a sequence shifted to redder UV
colors with respect to that of starbursts. The star formation history
(SFH) is identified as the main driver of this departure. Both
$L_{\mathrm{TIR}}/L_{\mathrm{FUV}}$ and $\beta$ correlate well with
metallicity, especially in moderately face-on galaxies. The relation
shifts to redder colors with increased scatter in more edge-on
objects. By applying physical dust models to our radial SEDs, we have
derived radial profiles of the total dust mass surface density, the
fraction of the total dust mass contributed by PAHs and the intensity
of the radiation field heating the grains. The dust profiles are
exponential, their radial scale-length being constant from Sb to Sd
galaxies (only $\sim 10$\% larger than the stellar scale-length). Many
S0/a-Sab galaxies have central depressions in their dust radial
distributions. The PAH abundance increases with metallicity for
$12+\log(\mathrm{O/H})<9$, and at larger metallicities the trend
flattens and even reverses, with the SFH being a plausible underlying
driver for this behavior. The dust-to-gas ratio is also well
correlated with metallicity and therefore decreases with
galactocentric radius. Although most of the total emitted IR power
(especially in the outer regions of disks) is contributed by dust
grains heated by diffuse starlight with a similar intensity as the
local Milky Way radiation field, a small amount of the dust mass
($\sim 1$\%) is required to be exposed to very intense starlight in
order to reproduce the observed fluxes at 24\,$\micron$, accounting
for $\sim 10$\% of the total integrated IR power.
\end{abstract}


\section{Introduction}
Understanding the spatial distribution of interstellar dust is of
particular importance for two reasons: it affects our view of galaxies
at different wavelengths, by absorbing UV and optical light and
reemitting it in the infrared, and it also constitutes an important
element in the chemical evolution of the interstellar medium (ISM). As
for the first issue, correcting for dust extinction is usually the
main source of uncertainty when deriving properties such as the star
formation rate (SFR), age or metallicity (Calzetti et al$.$ 1994; Buat
\& Xu 1996; Calzetti 2001; P\'erez-Gonz\'alez et al$.$ 2003). This
limitation not only applies to integrated data, but also to surface
brightness profiles and color gradients. For instance, the so-called
inside-out scenario for the formation of galactic disks predicts that
the timescale of gas infall and conversion into stars increases with
radius, leading to radial variations of the star formation history
(SFH), which are in turn observationally translated into color
gradients (de Jong 1996; Bell \& de Jong 2000; MacArthur et al$.$
2004; Taylor et al$.$ 2005). Given that the dust content also changes
with radius, it is key to properly quantify that variation with direct
tracers of dust extinction; otherwise, derived parameters such as the
radial growth rate of disks might be biased (see Mu\~{n}oz-Mateos et
al$.$ 2007 and references therein).

While being an inconvenience when studying star formation, dust itself
is also a key ingredient in the chemical enrichment of the ISM. Metals
resulting from stellar nucleosynthesis are returned to the ISM, where
they condense to form dust grains, some of which are later destroyed
and incorporated into new generations of stars. Elements that later
condense to form dust grains are injected into the ISM at different
rates (see e.g$.$ Dwek et al$.$ 2009 and references therein). For
instance, carbon now locked up into Polycyclic Aromatic Hydrocarbons
(PAHs) might have been originally produced in AGB stars, whose typical
lifetimes are a few Gyr. Other elements now constituting larger grains
may have been synthesized in more massive stars, which die as
supernovae in shorter timescales. Thus, the PAH abundance and the
dust-to-gas ratio are expected to vary with the age of the stellar
populations and correlate with the metal (oxygen) abundance of the gas
(see e.g$.$ Galliano et al$.$ 2008 and references therein). Note,
however, that dust formation results from a long chain of
poorly-understood processes, of which stellar lifetimes and yields are
only the first link. Grain growth and destruction in the ISM must be
also considered. Indeed, it is thought that only 10\% of interstellar
dust is directly formed in stellar sources, with the remaining 90\%
being later condensed in the ISM (see Draine 2009 for a recent review
on the subject).

Addressing these important issues requires a multi-wavelength,
multi-object spatially resolved analysis, which is now possible for
nearby galaxies thanks to the {\it Spitzer} Infrared Nearby Galaxies
Survey (SINGS, Kennicutt et al$.$ 2003). The SINGS project has made
use of the {\it Spitzer} Space Telescope (Werner et al$.$ 2004) to
collect IR data, as well as ancillary data from other facilities, for
a sample of 75 nearby galaxies, representative of the galaxy
population in the local universe. Together with UV images from the
Galaxy Evolution Explorer (GALEX, Martin et al$.$ 2005) and HI maps
from The HI Nearby Galaxies Survey (THINGS, Walter et al$.$ 2008),
this data-set provides the scientific community with unprecedented
spectral and spatial coverage of the most representative nearby
galaxies.

In this paper we exploit the unique data-set resulting from the
GALEX-SINGS-THINGS collaboration by obtaining multi-wavelength radial
profiles for the 57 galaxies in the SINGS sample which are detected
and resolved at the longest wavelengths. Our subsample still includes
galaxies representative of different morphological types (ellipticals,
lenticulars, spirals and irregulars). The profiles themselves, along
with other observational parameters such as asymptotic magnitudes,
concentration indexes and asymmetries, can be found in the
accompanying paper Mu\~{n}oz-Mateos et al$.$ (2009, Paper I
hereafter). The present paper focuses on the radial distribution of
dust properties. Finally, by comparing the profiles of the spirals in
the sample with models for the chemical and spectro-photometric
evolution of spiral galaxies (Boissier \& Prantzos 1999), we are
studying the radial variation of the star formation history in these
galaxies in a self-consistent frame, where radial changes in the
gas-infall and chemical enrichment are considered. The attenuation
profiles turn out to be essential to properly accommodate the model
predictions with the observed profiles. Ignoring extinction not only
biases the overall flux level in the UV and optical bands, but also
the different radial scale-lengths at these wavelengths, which are a
direct result of the inside-out assembly of disks. The results of this
comparison with theoretical models will be presented in a forthcoming
paper and, combined with the present study of the dust properties,
should help to better understand how the different components of
spiral galaxies get assembled together.

This paper is organized as follows. In Section~\ref{sample} we briefly
outline the main properties of the sample, describe the different data
used in this work, and explain how the radial profiles were
obtained. Section~\ref{extinction_profs} deals with the radial
distribution of internal extinction, and how it relates with other
properties such as the UV color, inclination or metallicity. In
Section~\ref{models} we compare our profiles with the dust models of
Draine \& Li (2007) (DL07 hereafter), and analyze the radial variation
of the PAH abundance, the dust mass and luminosity surface densities,
the intensity of the heating radiation field and the dust-to-gas
ratio. Finally, we summarize our main conclusions in
Section~\ref{summary}. In Appendix~\ref{app_empirical} we provide some
empirical relationships between the dust properties derived from the
model and the observed flux densities at the near- to far-infrared
bands. In Appendix~\ref{app_systematic} we discuss the possible
systematic factors that might affect our results in deriving the
parameters of the dust models of DL07.

\section{The sample, data and procedure}\label{sample}
The SINGS sample (Kennicutt et al$.$ 2003) consists of 75 nearby
galaxies which span the range in morphological type, luminosity and
FIR/optical luminosity observed in the local universe. The SINGS
galaxies were also selected to cover a reasonably wide range in other
additional properties, like nuclear activity, spiral and bar
structure, inclination, surface brightness and environment. Note,
however, that no significant luminous or ultra-luminous infrared
galaxy (that is, with $L_{\mathrm{IR}}>10^{11}\mathrm{\,L}_{\odot}$)
is included the sample. The median distance of the SINGS galaxies is
10\,Mpc, with all objects being closer than 30\,Mpc. Since the sample
is neither flux- nor volume-limited, its statistical power as a whole
is limited. Nevertheless, the wealth of panchromatic data available
for these galaxies, together with their proximity, makes it possible
to carry out detailed studies on the physics of star formation at kpc
scales, including the interplay between star formation and the ISM.

Eighteen SINGS galaxies were not suitable for our purposes and were
excluded from the present study:

\begin{enumerate}
\item Seven galaxies, DDO~154, DDO~165, Holmberg~IX, M81~Dw~a,
M81~Dw~b, NGC~0584 and NGC~4552, are not (or marginally) detected in
at least one of the MIPS bands.

\item Nine galaxies, Mrk~33, NGC~1266, NGC~1377, NGC~1482, NGC~2798,
NGC~3265, NGC~3773, NGC~5195 and NGC~7552, are unresolved in the MIPS
bands. Thus, radial profiles for these galaxies would reflect the
shape of the PSF rather than real structures. In NGC~7552 some
spatially extended emission can be discerned, but it is too faint
compared to the extremely bright nucleus.

\item The only evident IR source within the optical extent of NGC~1404
at 70 and 160\,$\micron$ is an off-center object at the northeast of
the frame. Since it might not be related to NGC~1404, we also excluded
this galaxy.

\item The bright center of NGC~3034 saturates the MIPS 24\,$\micron$
detector, thus precluding reliable photometry.

Therefore, the final subsample consists of 57 galaxies, whose main
properties are summarized in Table~\ref{sample_tab}.

\end{enumerate}

\subsection{GALEX data}
Nearly all SINGS galaxies have been observed in the FUV
($\lambda_{eff}=151.6$\,nm) and NUV ($\lambda_{eff}=226.7$\,nm) by
GALEX (Martin et al$.$ 2005) (see Table~\ref{sample_tab}). The
observations are performed simultaneously at both bands thanks to a
beam splitter, but some galaxies lack FUV images since the
corresponding detector had to be turned off due to intense solar
activity or overcurrent events. The delivered images have a final
pixel-scale of 1.5$\arcsec$. Although the size of the PSF depends
slightly on the position on the detector and the brightness of the
source, the typical FWHM is 5-6$\arcsec$. This resolution is similar
to that of the MIPS 24\,$\micron$ images, and corresponds to a
physical scale of $\sim$300\,pc at 10\,Mpc, the median distance of the
SINGS galaxies. The flux calibration is based on white dwarf standard
stars. For the pipeline version used here (the same as in Gil de Paz
et al$.$ 2007), the estimated zero-point uncertainty is 0.15\,mag.

\subsection{{\it Spitzer} data}
The reader is referred to Paper~I for a more detailed description of
the {\it Spitzer} data used here. Mid- and far-infrared observations
of the SINGS sample were carried out using the {\it Spitzer} Space
Telescope (Werner et al$.$ 2004).  The Infrared Array Camera (IRAC,
Fazio et al$.$ 2004) was used to image the SINGS galaxies at 3.6, 4.5,
5.8 and 8.0\,$\micron$. We made use of the images provided in the
SINGS Fourth Data Delivery, which are based on the Version 13 Basic
Calibrated Data produced by the {\it Spitzer} Science Center. The
delivered images have a pixel scale of 0.75$\arcsec$, and the FWHM of
the PSF at each band are 1.7$\arcsec$, 1.7$\arcsec$, 1.9$\arcsec$ and
2.0$\arcsec$ respectively. These correspond to spatial scales of
80-100\,pc at the median distance of the sample. The estimated
photometric uncertainty is $\sim 2$\% (Reach et al$.$ 2005). However,
the photometry needs to be corrected for the diffuse scattering of
incoming photons throughout the IRAC array\footnotemark[1], and these
corrections have an associated uncertainty of $\sim 10$\%.

The Multi-band Imaging Photometer (MIPS, Rieke et al$.$ 2004) was used
in scan-mapping mode to observe the SINGS galaxies at 24, 70 and
160$\micron$. The final frames are delivered with pixel scales of
1.5$\arcsec$, 4.5$\arcsec$ and 9.0$\arcsec$, respectively, thus being
integer multiples of the pixel-scale of the IRAC frames while still
properly mapping the MIPS PSF. The corresponding FHWM are
5.7$\arcsec$, 16$\arcsec$ and 38$\arcsec$ at each band, probing
physical scales of 0.28, 0.78 and 1.84\,kpc at 10\,Mpc. The estimated
zero-point errors are 4\%, 5\% and 12\% at 24, 70 and 160$\micron$,
respectively (Engelbracht et al$.$ 2007; Gordon et al$.$ 2007;
Stansberry et al$.$ 2007).

\footnotetext[1]{\tt http://ssc.spitzer.caltech.edu/irac/calib/extcal/}

\subsection{HI data}
The HI Nearby Galaxy Survey (THINGS, Walter et al$.$ 2008) used the
Very Large Array (VLA) to map HI 21-cm line emission from 34 nearby
($D < 15$\,Mpc) galaxies, most of which were also targets of SINGS and
the GALEX Nearby Galaxies Survey. The observations were done using the
B-, C- and D-array configurations. For details of data reduction and
processing, see Walter et al$.$ (2008). Here, we use profiles derived
from natural-weighted moment-zero (integrated intensity) maps. These
have a typical angular resolution of $11\arcsec$ and sensitivity to
surface densities as low as $4 \times 10^{19}$~cm$^{-2}$ once
convolved to our $38\arcsec$ working resolutions (see below). These
radial profiles were kindly provided by A$.$ Leroy and F$.$
Walter. Because THINGS includes data from the compact B array
configuration, the maps comfortably recover extended structure in our
sources. The column densities in the THINGS maps are estimated to be
correct to within $\pm 10\%$.

\subsection{Surface brightness profiles}\label{profiles}

The reader is referred to Paper~I for a more in-depth description of
the procedure followed to obtain the surface brightness radial
profiles; here we will just briefly describe the most important steps
relevant to this paper. In our study of the spatial distribution of
the dust properties, we are limited by the resolution of the MIPS
160\,$\micron$ images (FWHM of 38$\arcsec$). We convolved the GALEX,
IRAC, MIPS and HI images with different kernels (see Gordon et al$.$
2008) in order to match the shapes and resolution of their PSFs to the
MIPS 160\,$\micron$ PSF. Prior to convolving the images, foreground
stars, background galaxies and artifacts were masked and interpolated
over to avoid contamination when degrading the images.

The radial profiles were obtained using the IRAF\footnotemark[2] task
{\sc ellipse}, measuring the mean intensity along elliptical isophotes
with fixed ellipticity and position angle, equal to those of the
$\mu_{B}$ = 25 mag arcsec$^{-2}$ isophote from the RC3
catalog\footnotemark[3] (de Vaucouleurs et al$.$ 1991). For those
objects for which these parameters were not included in the RC3
catalog, we used the major and minor axis diameters and position
angles available in NED. The centers of these elliptical isophotes
were set at the coordinates shown in Table~\ref{sample_tab}. The
semi-major axis of these ellipses were successively incremented by
48$\arcsec$ (a step larger than the PSF FWHM), to a final radius at
least 1.5 times the R25 radius (depending on the extension of each
particular galaxy). While using radially-varying ellipticities and
position angles is useful in detailed studies of galactic structure at
a specific wavelength, a panchromatic analysis requires using the same
set of fixed elliptical isophotes in all bands to measure the
different fluxes in the same regions of each galaxy, and for that
matter a fixed PA and ellipticity was found to be as appropriate as
any other set of values obtained from a given band.

\footnotetext[2]{IRAF is distributed by the National Optical Astronomy
Observatories, which are operated by the Association of Universities
for Research in Astronomy, Inc., under cooperative agreement with the
National Science Foundation.}

\footnotetext[3]{Except for NGC~5194, whose original values were
highly affected by its companion galaxy, NGC~5195.}

Uncertainties in the surface photometry include the error of the mean
intensity within each isophote, computed assuming Poisson statistics,
and the uncertainty in the sky level. The latter comes from high
spatial frequency errors (Poisson noise, pixel-to-pixel variations)
and low spatial frequency ones (flat-fielding errors) (Gil de Paz \&
Madore 2005).

The radial profiles were corrected for Galactic extinction as in Dale
et al$.$ (2007), using the color excesses from the maps of Schlegel et
al$.$ (1998) and the extinction curve of Li \& Draine (2001), assuming
$R_{V}=3.1$. The final profiles are shown in
Table~\ref{phot_profiles_tab}.

Note that our radial profiles should be taken with caution in very
inclined galaxies. First, although the radial step along the major
axis is larger than the FWHM of the PSF, this is not the case along
the minor axis in edge-on or close to edge-on galaxies. Moreover, the
outer regions might be contaminated by emission from the central ones
when performing the azimuthal average. Finally, the observed UV and IR
radiation in these systems will likely probe different spatial regions
within each galaxy, due to the large amount of dust along the line of
sight.

\section{Radial distribution of dust attenuation}\label{extinction_profs}
\subsection{Radial extinction profiles}\label{extinction}
As part of our analysis of the radial distribution of dust properties,
we will first determine the radial variation of internal extinction,
which is also necessary to recover the intrinsic radial profiles from
FUV to NIR. These kinds of studies have been addressed by several
authors following different methodologies. Boissier et al$.$ (2004,
2005, 2007) used the radial change in the total-infrared (TIR) to
far-ultraviolet (FUV) ratio to derive radial extinction profiles for
nearby galaxies. A similar procedure was followed by Popescu et al$.$
(2005) on a pixel-to-pixel basis for M~101. Prescott et al$.$ (2007)
obtained extinction profiles in H$\alpha$ for the SINGS galaxies by
comparing the H$\alpha$ and 24$\micron$ fluxes of individual
star-forming regions. The number of distant galaxies seen through a
spiral disk can also provide an independent estimation of the
extinction (e.g$.$ Holwerda et al$.$ 2005 and references therein).

Here we follow the first method to infer the radial distribution of
dust attenuation for the SINGS galaxies. Several studies (Buat \& Xu
1996; Meurer et al$.$ 1999; Gordon et al$.$ 2000; Witt \& Gordon 2000;
Buat et al$.$ 2005) have shown that the TIR-to-ultraviolet ratio is a
robust tracer of the internal extinction in star forming galaxies, in
the sense that it depends weakly on details such as the relative
geometry of stars and dust, the shape of the extinction curve, or the
star formation history (SFH). Regarding this latter issue, Buat et
al$.$ (2005) only found significant deviations in systems with very
quiescent SFH at present-day (that is, havig decayed with exponential
time-scales $\leq 2$\,Gyr). In these galaxies, the general starlight
radiation field would become an even more important source of
dust-heating than it already is in normal star-forming galaxies. This
precludes computing the internal extinction in ellipticals,
lenticulars and also the bulges of spirals with the same
calibration. Using synthetic SEDs of galaxies with different SFHs and
attenuations, Cortese et al$.$ (2008) further investigated the
dependence of the TIR-to-FUV ratio on the mean age of the stellar
populations, and confirmed that quiescent systems exhibit larger
TIR-to-FUV ratios than more actively star-forming ones with the same
extinction. To account for the extra dust heating contributed by older
stars, they provide a SFH-dependent calibration to estimate the UV
attenuation. Note that, as pointed out by these authors, this recipe
should not be blindly applied to ellipticals, since their FUV emission
does not seem to be linked with recent star formation activity. We
include these galaxies here just for completeness, but any further
interpretation of their observed TIR-to-FUV in terms of attenuation
should be done with caution.

Using the low ($48\arcsec$-step) resolution profiles at 8, 24, 70 and
160\,$\micron$, we built TIR (3-1100\,$\micron$) profiles using the
weighted sum proposed by DL07, after having subtracted the stellar
emission at 8 and 24\,$\micron$ (see Section~\ref{fitting}). This
estimator of the total-infrared luminosity constitutes a slight
improvement over that of Dale \& Helou (2002) using the MIPS bands
alone, to the degree that including the 8\,$\micron$ flux seems to
reduce the scatter associated with differences in PAH
abundances\footnotemark[4].

\footnotetext[4]{We have checked that the particular choice of this
recipe is not critical, since both estimators yield almost equal TIR
luminosities (on average, the TIR values derived from the DL07
calibration are 5\% larger, with a rms of 5\%). The DL07 recipe is a
proxy for the actual TIR luminosity derived from the DL07 models,
designed to be valid over a wide range of starlight intensities and
PAH abundances. In Section~\ref{models} we will derive
$L_{\mathrm{TIR}}$ from the model fitting, along with the remaining
model parameters. On average, for the particular range of starlight
intensities and PAH abundances in our sample, the TIR luminosities
yielded by the estimator are only 6\% larger than those from the
model, with a rms of 3\%.}

We then measured low resolution profiles on the convolved GALEX FUV
and NUV images and estimated $A_{\mathrm{FUV}}$ and $A_{\mathrm{NUV}}$
from the TIR-to-FUV and TIR-to-NUV ratios, respectively.  The
attenuation at both wavelengths was computed using both the fits of
Buat et al$.$ (2005) and those of Cortese et al$.$ (2008). Note that
at small subgalactic scales, radiative transfer could limit the
usefulness of the TIR-to-UV ratio as an extinction tracer, in the
sense that UV photons emerging from a given region could heat dust in
another region. Considering that our radial sampling is relatively
coarse and that we average our data within elliptical annuli, this
effect $-$if present$-$ should be small (except maybe in extreme
objects with very different UV and IR distributions, which are not
common in the SINGS sample)

\begin{figure}
\resizebox{1\hsize}{!}{\includegraphics{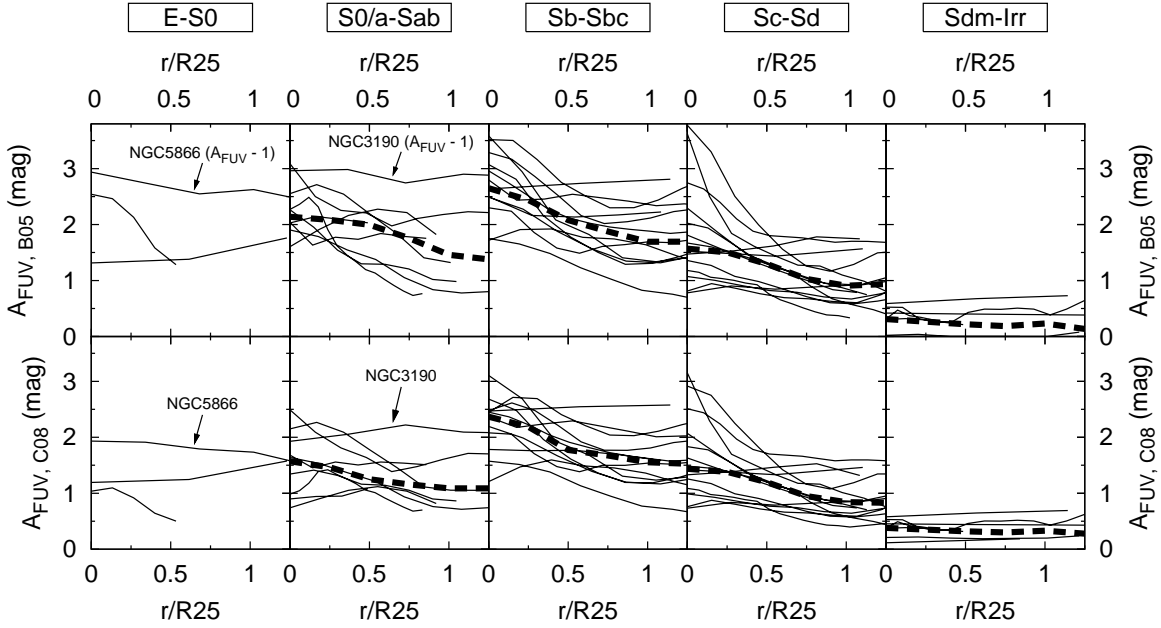}}
\caption{(a): Radial extinction profiles for the SINGS galaxies,
computed from the TIR-to-FUV ratio. The horizontal axis is normalized
to the optical size of each galaxy. The dashed curves show the median
profiles in each bin of morphological types, when enough galaxies are
available. Top row: Extinction computed using the fit of Buat et al$.$
(2005), which is independent of the SFH. Bottom row: Extinction
computed from the SFH-dependent calibration of Cortese et al$.$
(2008). The conversion law is explicitly derived for each annulus
depending on its observed $(\mathrm{FUV}-3.6\,\micron)$ color, in
order to take into account the dust-heating contributed by old
stars. For clarity, the profiles of the edge on galaxies NGC~5866 and
NGC~3190 have been offsetted by $-$1\,mag in the top panels (but not
in the bottom ones).\label{all_extprofs}}
\end{figure}

The extinction profiles we derive are shown in Table~\ref{extprof_tab}. In
Fig.~\ref{all_extprofs} we show all the FUV extinction profiles for
the galaxies in the sample, normalized to the optical radius R25. A
general trend of $A_{\mathrm{FUV}}$ decreasing with radius is clearly
seen. The NUV profiles are not shown, since they exhibit similar
behavior as the FUV ones.

The extinction in the top panels has been computed with the
calibration of Buat et al$.$ (2005), whereas in the bottom ones we
have used the SFH-dependent recipe of Cortese et al$.$ (2008). While
the equation relating $L_{\mathrm{TIR}}/L_{\mathrm{FUV}}$ and
$A_{\mathrm{FUV}}$ is unique in the case of Buat et al$.$ (2005),
Cortese et al$.$ (2008) parameterized the coefficients of such a
conversion as a function of observable colors that depend on the
overall SFH. We followed the prescriptions given by these authors and
determined the particular conversion between
$L_{\mathrm{TIR}}/L_{\mathrm{FUV}}$ and $A_{\mathrm{FUV}}$ for each
annulus depending on its observed $(\mathrm{FUV}-H)$ color. The latter
was estimated assuming that $(H-3.6\,\micron)\sim-1.03$ which,
according to the stellar populations models of Bruzual \& Charlot
(2003), is the typical color exhibited by star-forming galaxies, and
it is quite independent of their SFH.

In S0/a-Sab galaxies and the bulges of later-type galaxies, the
attenuation derived from the age-dependent calibration is
$\sim0.5$\,mag lower than the one yielded by the age-independent
recipe. The difference is negligible in the disk-dominated regions.
Given that the SINGS galaxies span a wide range in star formation
histories, both among them and within different regions of the same
object, the results discussed in the rest of the paper refer to the
extinction derived following the method of Cortese et al$.$ (2008),
except when mentioned otherwise.

The overall level of extinction varies along the Hubble sequence,
reaching a maximum in Sb-Sbc galaxies. On average, in these spirals
the attenuation in the FUV ranges from $\sim2.5$\,mag in the central
regions to $\sim1.5$\,mag in the outer ones, although with large
scatter. The extinction is $\sim1$\,mag lower in earlier spirals, and
it goes below 0.5\,mag in Sdm spirals and irregulars. Note again that
the attenuation in ellipticals and lenticulars is highly uncertain,
since even the age-dependent recipes of Cortese et al$.$ (2008) might
fail in these galaxies.

\begin{figure}
\resizebox{1\hsize}{!}{\includegraphics{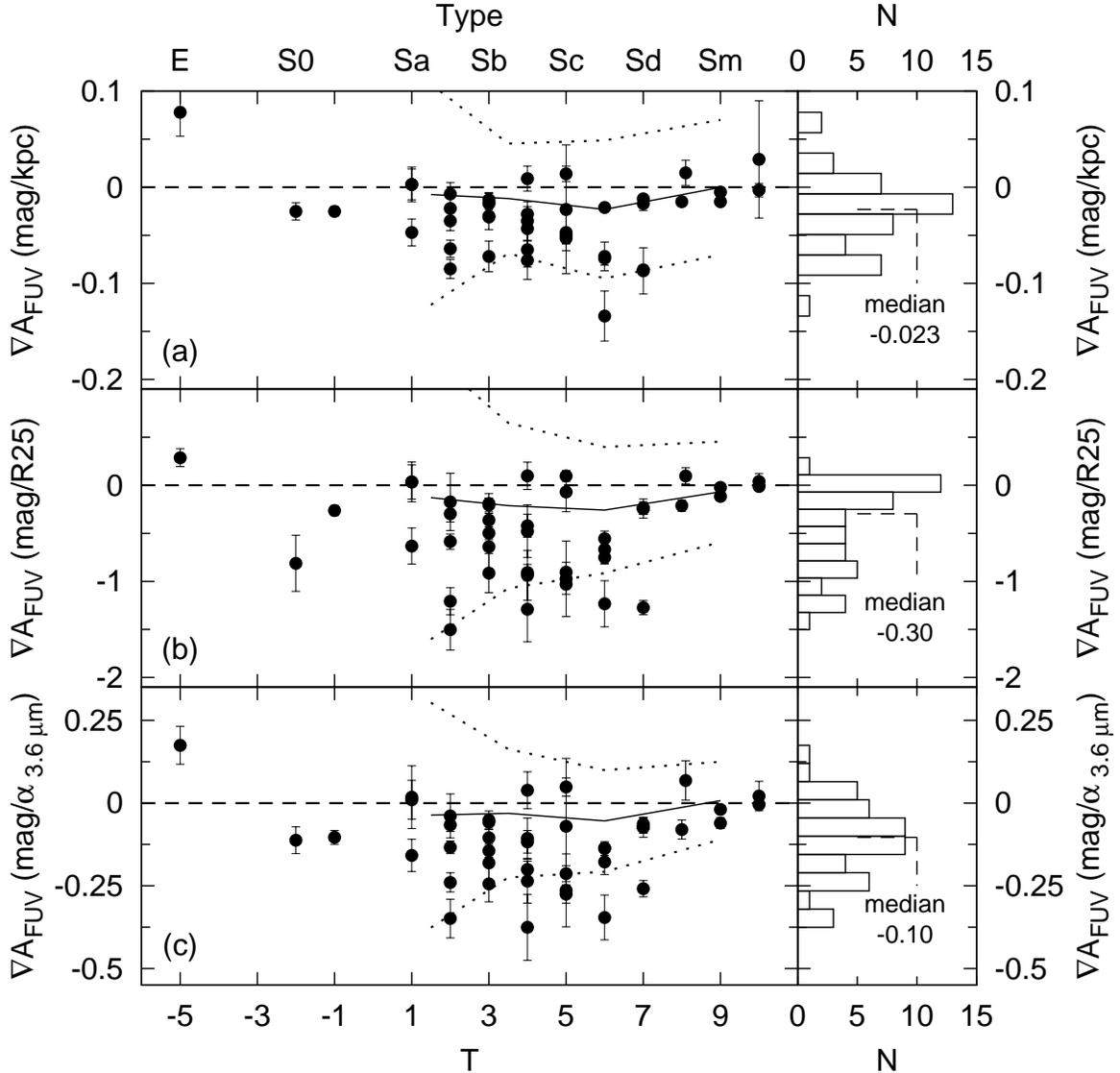}}
\caption{Radial gradients of the attenuation in the FUV, measured by
applying a linear fit to the $A_{\mathrm{FUV}}$ profiles, after
excluding the bulges. The gradients are expressed in mag/kpc (a), in
units of the R25 radius (b), and in units of the radial exponential
scale-length of the stellar mass distribution (c). The latter was
derived from the profiles measured on the degraded 3.6\,$\micron$
images (bulges excluded). The overall distribution of gradients for
all Hubble types is shown in the histograms at the right, along with
the corresponding median values. For comparison, the solid and dotted
lines in each panel show the mean and 1-$\sigma$ limits of the
distribution of attenuation gradients derived by Mu\~{n}oz-Mateos et
al$.$ (2007) for a sample of 161 nearby, face on spirals. In that
paper we relied on the $(\mathrm{FUV}-\mathrm{NUV})$ profiles as
indirect tracers of the extinction, and 2MASS $K$-band ones as a proxy
for the stellar mass.\label{AFUV_grad}}
\end{figure}

In order to quantify the attenuation radial gradients, we performed a
linear fit to the $A_{\mathrm{FUV}}(r)$ profiles, without including
the bulges of spirals for the reasons mentioned above. This exclusion
was done visually: bulges produce a steep central rise of the
3.6\,$\micron$ luminosity above the exponential disk, and the FUV
emission is significantly reduced in the central regions. This yields
a sharp change in the (FUV$-$3.6\,$\micron$) that, together with a
visual inspection of the image, can be used to roughly delimit the
bulge- and disk-dominated regions of the profiles (see also
Paper~I). The results are shown in Fig.~\ref{AFUV_grad}, where the
gradients are also expressed in terms of the R25 radius and the radial
exponential scale-length of the 3.6\,$\micron$ profiles, defined so
that $I_{3.6\micron} \propto e^{-r/\alpha_{3.6\micron}}$. The
scale-length $\alpha_{3.6\micron}$ was computed by fitting the
profiles measured on the convolved 3.6\,$\micron$ images (again, after
excluding the bulges). Since at this wavelength the luminosity traces
the stellar mass, this can be considered as the stellar mass
scalelength. Most galaxies exhibit negative attenuation gradients, and
the dispersion is larger in spirals of intermediate types.

\subsection{The IRX-$\beta$ relation in normal disks}
\subsubsection{Estimating the extinction from UV data alone}
In spite of the importance of correcting UV data for internal
extinction, when FIR data are not available it is not possible to
apply such corrections following the methods described in
Section~\ref{extinction}. In this regard, the slope of the UV spectrum
$\beta$ $-$or, equivalently, the (FUV$-$NUV) color$-$ has been
proposed as an indirect tracer of dust attenuation in starburst
galaxies (Calzetti et al$.$ 1994; Heckman et al$.$ 1995; Meurer et
al$.$1995, 1999). While the infrared excess (IRX) relative to the UV
seems to be tightly correlated with $\beta$ in starbursts, later
studies (Bell 2002; Buat et al$.$ 2005; Seibert et al$.$ 2005; Cortese
et al$.$ 2006; Gil de Paz et al$.$ 2007; Dale et al$.$ 2007) have
shown that the so-called IRX-$\beta$ relation shows rather large
scatter in normal star-forming galaxies. A similar behavior has also
been observed for star-forming regions within galaxies (Calzetti et
al$.$ 2005). The correlation shows a wider spread, and is also
systematically shifted to redder (FUV$-$NUV) colors. Using radial
profiles from GALEX and IRAS, Boissier et al$.$ (2007) confirmed that
the IRX-$\beta$ relation differs between starburst and normal
star-forming galaxies. The relation is clearly shifted with respect to
that for starburst galaxies, but the dispersion in their case seems to
be reduced when using radial profiles instead of integrated
photometry, possibly due to not mixing different stellar populations
from the bulge and the disk.

Here we extend the work carried out in that study by deriving a new
radially resolved IRX-$\beta$ diagram using data from {\it Spitzer},
which has better angular resolution than IRAS, and also probes colder
dust by reaching a bit further into the FIR. An analysis of the
IRX-$\beta$ plot for the SINGS galaxies, derived from integrated
photometry, can be found in Dale et al$.$ (2007).

Throughout this paper we assume the definition of the UV spectral
slope $\beta_{\mathrm{GLX}}$ given by Kong et al$.$ (2004):
\begin{equation}
\beta_{\mathrm{GLX}}=\frac{\log(f_{\lambda,\mathrm{FUV}})-\log(f_{\lambda,\mathrm{NUV}})}{\log(\lambda_{\mathrm{FUV}})-\log(\lambda_{\mathrm{NUV}})}=2.201(\mathrm{FUV}-\mathrm{NUV})-2
\end{equation}

In Fig.~\ref{irxbeta}a we show the IRX-$\beta$ plot for each of the
low resolution radial profiles for the SINGS galaxies, classified
according to their morphological type. Each data-point represents a
given radial bin, and those belonging to spiral galaxies are also
color-coded according to the (FUV$-$3.6\,$\micron$) color of that bin
(corrected for internal extinction), which can be used as a measure of
the specific star formation rate (SFR per unit of stellar mass, sSFR
hereafter) or, equivalently, the present to past-averaged star
formation rate, usually referred to as the birthrate parameter $b$
(Scalo 1986). This color scheme cannot be applied to ellipticals,
since their FUV flux is not necessarily linked to star formation
(Burstein et al$.$ 1988; O'Connell 1999; Boselli et al$.$ 2005). As
for the irregulars, the relative distribution of dust and stars is
usually patchy (and not necessarily axisymmetric), so their
$L_{\mathrm{TIR}}/L_{\mathrm{FUV}}$ profiles must be considered with
caution. As a reference, the right vertical axis shows the extinction
in the FUV derived from $L_{\mathrm{TIR}}/L_{\mathrm{FUV}}$ with the
fit of Buat et al$.$ (2005), although the color-coding is based on the
(FUV$-$3.6\,$\micron$) color corrected for attenuation using the
SFH-dependent calibration of Cortese et al$.$ (2008). We also show the
mean IRX-$\beta$ for the starburst galaxies studied by Meurer et al$.$
(1999), as given by Kong et al$.$ (2004), and the fit provided by
Boissier et al$.$ (2007) for the radial profiles of normal spirals.

Although differences in SFH can introduce systematic deviations in
this plot (see next section), it is desirable to get at least a rough
estimate of the internal extinction for normal spirals when FIR data
are not available, as has been traditionally done for
starbursts. Since we are interested in obtaining a fit valid for normal
star-forming spirals, we have excluded three galaxies with intense
starburst activity (NGC~4536, NGC~4631 and NGC~5713). Moreover, we
have only considered regions with $(\mathrm{FUV}-\mathrm{NUV})<0.9$,
thus avoiding the large scatter in the TIR-to-FUV ratio at redder UV
colors. We applied a non-linear least-squares algorithm to the
remaining data-points, giving:
\begin{equation}
L_{\mathrm{TIR}}/L_{\mathrm{FUV}}=10^{0.30+1.15(\mathrm{FUV}-\mathrm{NUV})}-1.64\label{eq_irxbeta}
\end{equation}

The resulting fit is shown in Fig.~\ref{irxbeta}a, and allows for the
recovery of $\log(L_{\mathrm{TIR}}/L_{\mathrm{FUV}})$ for normal
star-forming galaxies with a residual rms uncertainty of $\pm
0.27$\,dex. Instead of imposing a UV-color condition, we could have
used the extinction-corrected
$(\mathrm{FUV}-3.6\,\micron)_{\mathrm{corr}}$ color to exclude the
most quiescent systems. However, such a criterion cannot be applied by
an observer lacking FIR data, which are necessary to correct that
color for internal extinction. While our UV condition includes a few
`red' systems [in terms of their
$(\mathrm{FUV}-3.6\,\micron)_{\mathrm{corr}}$ color], they lie well
within the main relation delineated by the rest of points. The
resulting fit is not significantly sensitive to these few points, and
this selection on the observed UV color is more justified from a
purely empirical point of view. Beyond
$(\mathrm{FUV}-\mathrm{NUV})=0.9$\,mag, the observed dispersion is too
large to reliably estimate $A_{\mathrm{FUV}}$ using this fit.

Note that our fit differs from the one derived by Boissier et al$.$
(2007) from GALEX and IRAS data, the latter having a somewhat flatter
slope at the reddest UV colors, possibly due to data-points embedded
within the bulges (no conditions on the UV color were imposed when
computing that fit). Nevertheless, both fits are in very good
agreement in the regions least affected by contamination from older
stellar populations. By comparing the TIR-to-FUV profiles of the
galaxies we have in common with the sample of Boissier et al$.$
(2007), we have checked that the agreement between the TIR-to-FUV
ratios is excellent. On average, our values are 0.04\,dex higher, with
a scatter of 0.13\,dex.

In brief, in order to determine the attenuation in the FUV one must
first estimate the $\log(L_{\mathrm{TIR}}/L_{\mathrm{FUV}})$ ratio. If
FIR data are available, they should be used to directly compute
$\log(L_{\mathrm{TIR}}$ using, for instance, the calibrations of Dale
\& Helou (2002) or DL07. Once
$\log(L_{\mathrm{TIR}}/L_{\mathrm{FUV}})$ is known, it can be
translated into $A_{\mathrm{FUV}}$ using any of the recipes available
in the literature (e.g$.$ Buat et al$.$ 2005). In order to prevent the
attenuation in early-type spirals from being overestimated, one can
rely on age-dependent calibrations such as that of Cortese et al$.$
(2008), which requires additional constraints such as optical and/or
near-IR measurements. In the absence of FIR data, $A_{\mathrm{FUV}}$
can be estimated from the (FUV$-$NUV) colour via Eq.~\ref{eq_irxbeta},
but such a relation should be only employed in a statistical sense for
relatively large samples of galaxies. Given the large scatter, the use
of Eq.~\ref{eq_irxbeta} is discouraged for single objects.

\subsubsection{Dependence on the star formation history}\label{irxbeta_sfh}
\begin{figure}
\resizebox{1\hsize}{!}{\includegraphics{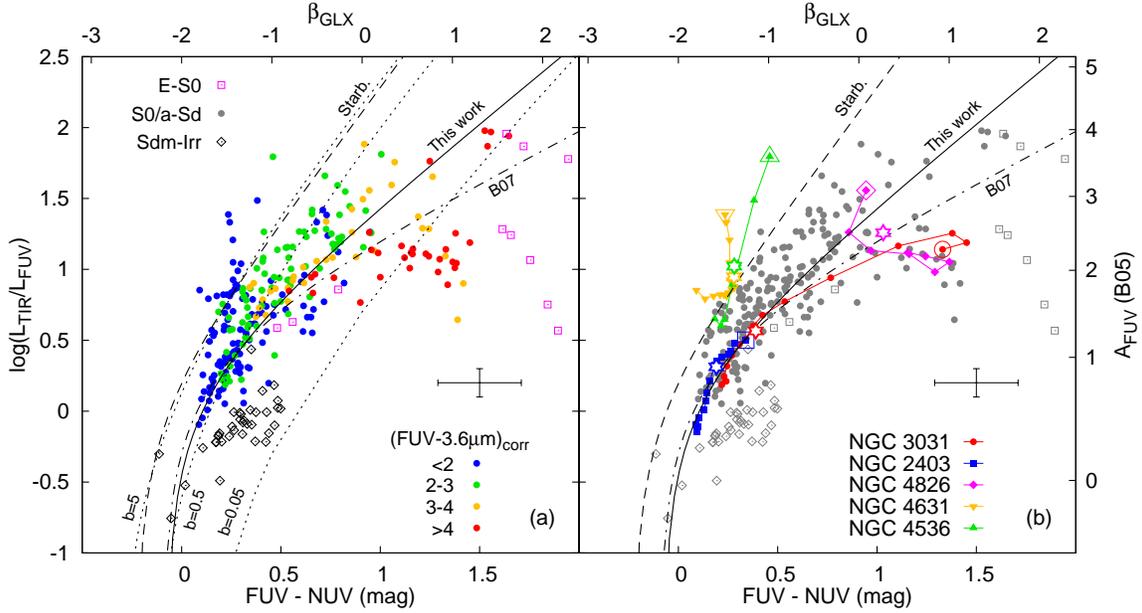}}
\caption{Ratio of the total infrared to FUV luminosity as a function
of the UV spectral slope ($\beta_{\mathrm{GLX}}$) and the UV color,
for all the profiles in the sample. The rightmost y-axis shows the
internal extinction in the FUV for a given TIR-to-FUV ratio, according
to the fits of Buat et al$.$ (2005), only valid for late-type systems
(see text). (a): Galaxies are sorted out into different morphological
types. For the disk-like galaxies, a color scheme is used to map the
intrinsic $(\mathrm{FUV}-3.6\micron)$ color, corrected for internal
extinction using the age-dependent calibration of Cortese et al$.$
(2008). The solid line corresponds to a fit to the data-points bluer
than (FUV$-$NUV)=0.9, excluding three starburst galaxies (see
text). The dot-dashed line shows the relation found by Boissier et
al$.$ (2007) for spiral galaxies using GALEX and IRAS data. The dashed
line is the mean relation for the starburst galaxies of Meurer et
al$.$ (1999). The three dotted lines are model predictions by Kong et
al$.$ (2004) for different values of the birthrate parameter $b$. The
mean error bars (including zero-point errors in all bands) are also
shown. (b): Individual radial tracks followed by some galaxies in the
diagram: an early-type spiral (NGC~3031), a late-type one (NGC~2403),
an anemic spiral (NGC~4826; van den Bergh 1976), and two starbursts
(NGC~4536 and NGC~4631, the latter being edge-on). The innermost point
of each profile is marked for reference. The integrated colors of each
galaxy measured by Dale et al$.$ (2007) are shown with
stars.\label{irxbeta}}
\end{figure}

In Fig.~\ref{irxbeta}a, only a few data-points follow the starbursts
relation, and they indeed correspond to galaxies with intense star
formation activity (e.g$.$ NGC~4536, NGC~4631). Most galaxies,
however, lie to the right in the diagram: that is, for the same amount
of attenuation, they have redder UV colors than
starbursts. Differences in the star formation history are the most
likely explanation for this broadening (Kong et al$.$ 2004, Calzetti
et al$.$ 2005). Besides the effects of dust, regions with more
quiescent SFHs will be intrinsically redder in the UV due to their
more evolved stellar populations. Indeed, there is a clear trend with
the $(\mathrm{FUV}-3.6\,\micron)_{\mathrm{corr}}$ color, which, as
noted above, is a proxy for the present to past-averaged star
formation rate. Note that the `present' SFR, as derived from the FUV,
is not instantaneous, but represents an average over the last $\sim
100$\,Myr. Data-points with
$(\mathrm{FUV}-3.6\,\micron)_{\mathrm{corr}}<3$\,mag follow a
reasonably well-defined sequence, parallel to the starburst one; but
redder regions (usually embedded within bulges) depart towards the
zone populated by ellipticals.

The three dotted curves in Fig.~\ref{irxbeta}a show the IRX-$\beta$
relation predicted by Kong et al$.$ (2004) for galaxies with different
values of the birthrate parameter $b$. The comparison with
$(\mathrm{FUV}-3.6\,\micron)_{\mathrm{corr}}$ should be done with
caution, since the birthrate parameter of the models is the
instantaneous one. The empirical relation for starbursts closely
follows the model predictions for $b=5$, which falls in the range
where starbursts are commonly found ($b>2$-3; see Brinchmann et al$.$
2004 and references therein). The bulk of our data-points are
consistent with lower values of $b$. These model predictions for
$A_{\mathrm{FUV}}$, however, should be considered just as average
approximations, given that their uncertainties range from $\pm
0.3$\,mag for $b\gtrsim0.3$ to $\pm 1$\,mag for lower values, owing to
differences in the particular details of the SFH and the dust content
(Kong et al$.$ 2004).

\begin{figure}
\resizebox{1\hsize}{!}{\includegraphics{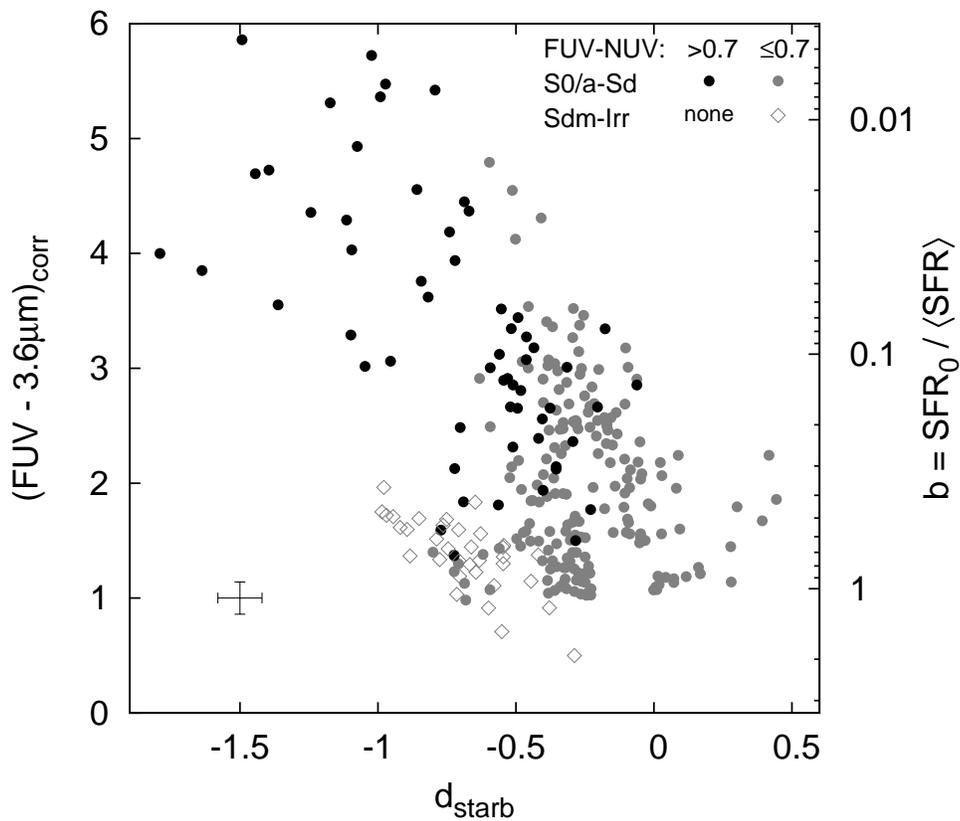}}
\caption{Relation between the intrinsic $(\mathrm{FUV}-3.6\,\micron)$
color (i.e$.$ corrected for internal extinction) and the perpendicular
distance to the mean relation for starbursts in the IRX-$\beta$
diagram (see Fig.~\ref{irxbeta}). Negative distances correspond to
data-points below the starbursts curve. Different symbols and colors
are used depending on the Hubble type and the observed UV
color.\label{dstarb}}
\end{figure}

We further explore this trend in Fig.~\ref{dstarb}, where we have
plotted the $(\mathrm{FUV}-3.6\,\micron)_{\mathrm{corr}}$ color as a
function of the perpendicular (i.e$.$ shortest) distance from each
data-point to the starburst relation, $d_{\mathrm{starb}}$ (see Kong
et al$.$ 2004). Although with considerable dispersion, the trend is
rather evident: regions with intrinsically redder
$(\mathrm{FUV}-3.6\,\micron)_{\mathrm{corr}}$ colors $-$having then
lower current-to-past star formation activity$-$ are clearly located
further away from the relation for starbursts. Interestingly,
selection effects can blur or even make this trend vanish. For
instance, if we only consider regions bluer than
$(\mathrm{FUV}-\mathrm{NUV})\sim 0.7$, no clear correlation between
SFH and $d_{\mathrm{starb}}$ can be inferred. Therefore, one needs to
explore a wide range in SFHs in order to see this trend. This could
imply that although SFH seems to be driving this departure from the
starburst relation, other factors such as dust geometry or the
extinction law might be also contributing in different ways. It should
be noted as well that if we choose another IRX-$\beta$ curve as a
reference instead of the starburst one (i.e$.$ MW-type dust with
different geometries, for instance), all perperincular distances will
change accordingly.

This could explain the different conclusions reached by authors
studying integrated properties of galaxies. Kong et al$.$ (2004) found
a correlation between $d_{\mathrm{starb}}$ and different indicators of
the SFH, like the $D_{n}(4000)$ break and the H$\alpha$ equivalent
width. Cortese et al$.$ (2006) estimated the birthrate parameter from
H$\alpha$ and $H$-band luminosities for an optically selected sample
of normal star-forming galaxies, and found a weak correlation between
$b$ and $d_{\mathrm{starb}}$, with considerable scatter. The fact that
their galaxies belong to nearby clusters might be in part responsible
for the observed scatter. Interactions with the the intra-cluster
medium and the cluster potential well may likely affect their SFHs by
removing gas from the disks and quenching their SF activity, thus
progressively turning them into anemic spirals (see e.g$.$ Boselli \&
Gavazzi 2006 and references therein). In their analysis of the
integrated properties of the SINGS galaxies, Dale et al$.$ (2007)
found that most of the scatter towards redder UV colors in the
IRX-$\beta$ diagram was due to ellipticals and early-type
spirals. Panuzzo et al$.$ (2007) studied a UV selected sample of
galaxies, and did not find any systematic deviation from the starburst
relation that depended on $b$, computed from NUV and $H$-band
luminosities. However, their UV selected sample did not contain
objects redder than $(\mathrm{FUV}-\mathrm{NUV})\sim 0.7$, thus making
it difficult to infer any correlation, as Fig.~\ref{dstarb}
demonstrates.

In order to further illustrate these issues, in Fig.~\ref{irxbeta}b we
have highlighted some tracks for particular galaxies, connecting the
data-points of annular regions at different galactocentric distances
within each particular galaxy. The innermost point of each profile is
marked for clarity, and the integrated colors of each galaxy are shown
with open stars. `Normal' spirals follow their own IRX-$\beta$
relation, which is quite similar in shape to the average one (modulo
global offsets in the overall UV color and/or extinction). In early
type spirals with well defined bulges, such as NGC~3031 (M~81), the
track clearly deviates towards redder colors as we move closer to the
center, due to the contribution of more evolved stars in the
bulge. This does not happen in late-type galaxies, like NGC~2403,
given their smaller bulge-to-disk ratios, especially at UV
wavelengths.

The SINGS sample also includes some peculiar objects that do not
follow these smooth trends. This is the case of NGC~4826, which is a
clear example of an anemic spiral (van den Bergh 1976). Although it
has a global radial extent of $\sim 13$\,kpc at 3.6\,$\micron$, the
bulk of the star formation activity $-$as traced by the FUV or
24\,$\micron$ images, for instance$-$ seems to be restricted to the
central 5\,kpc. Therefore, as we move away from the center towards the
outer and more quiescent regions, the track followed by this galaxy in
the IRX-$\beta$ diagram heads towards the region populated by
bulges. Something similar happens with NGC~4569, another anemic galaxy
in the Virgo cluster, where ram pressure stripping seems to have
quenched the star formation in the outer regions (Boselli et al$.$
2006).

Systems hosting starburst activity, like NGC~4536, lie close to the
empirical relation given by Kong et al$.$ (2004). In very edge-on
systems like NGC~4631 (which also happens to host starburst activity),
geometry might also play an important role. The observed TIR-to-FUV
ratio is probably larger than the one we would measure if the galaxy
was face-on (besides the fact that this galaxy is probably more
dust-rich than the average spiral galaxy). However, we do not find any
significant trend between the position in the IRX-$\beta$ diagram and
inclination, apart from this extreme case. At fixed metallicity,
however, inclination does seem to play a role (see
Section~\ref{S_ext_metal}).

Kong et al$.$ (2004) suggest that galaxies in their sample departing
too much from the starburst relation require very quiescent SFHs, with
very short time-scales of star-formation, not typical of spirals. They
invoke an additional mechanism that can contribute to broadening the
IRX-$\beta$ relation, consisting of an extra burst of star formation
at some point during the galaxy's lifetime, superimposed on top of an
otherwise smooth SFH. While this can certainly add more scatter, we
note that in our radial analysis most regions lying at large distances
from the starburst relation are either bulges or regions of anemic
spirals with clear signs of star-formation quenching. Indeed, there
are 18 galaxies in our sample having points with
$(\mathrm{FUV}-3.6\,\micron)_{\mathrm{corr}}>3$. Five of them are
ellipticals or lenticulars, and 11 are early-type spirals (S0/a-Sb),
including the already mentioned anemic ones. Therefore, at least for
these very quiescent systems, a rapidly declining SFH might be a more
reasonable possibility.

From the above analysis we can conclude that the star-formation
history is possibly driving the departure of star-forming regions from
the locus of starburst galaxies in the IRX-$\beta$ diagram. However,
the trends with different tracers of SFH (both here and in other
studies) seem to be quite noisy and not always evident $-$they can
actually disappear if the range of explored SFHs is not wide
enough. Several reasons might explain this. First of all, observational
errors might blur the offset between regions if their SFHs are not
different enough. Secondly, realistic SFHs cannot be parametrized with
a single quantity like $b$; indeed, the intermediate bursts proposed
by Kong et al$.$ (2004) were shown to contribute to the observed
scatter. Besides, a number of additional factors such as the relative
geometry of dust and stars, the shape of the internal extinction law
and the IMF can also have a great impact (Burgarella et al$.$ 2005,
Panuzzo et al$.$ 2007).

\subsection{Influence of metallicity and inclination}\label{S_ext_metal}
Attenuation and metallicity are known to be correlated in starburst
galaxies (Calzetti et al$.$ 1994; Heckman et al$.$ 1998), which can be
interpreted in terms of increasing extinction at larger dust-to-gas
ratios (see the discussion and references given in
Section~\ref{S_dust2gas}). Such a trend has been also observed in
normal star-forming galaxies, both with integrated data (see e.g$.$
Cortese et al$.$ 2006) and with radial profiles (Boissier et al$.$
2007). These studies find an offset with respect to the starbursts
relation, with normal galaxies being less attenuated at a given
metallicity.

\begin{figure}
\resizebox{1\hsize}{!}{\includegraphics{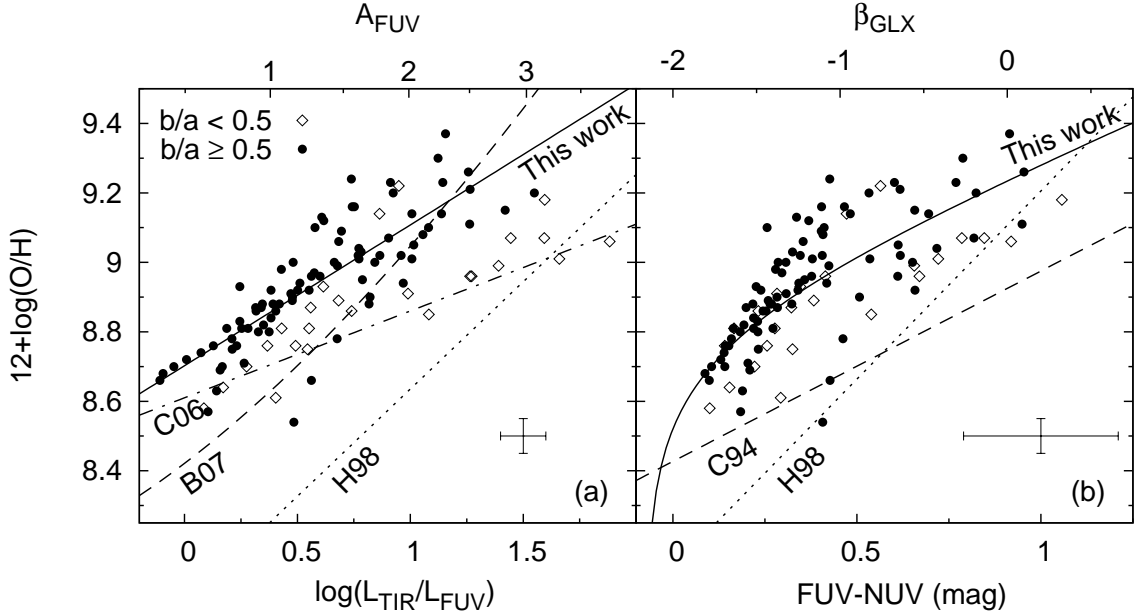}}
\caption{(a): Oxygen abundances as a function of the TIR-to-FUV ratio
for the radial profiles of 22 SINGS galaxies with metallicity
gradients in Moustakas et al$.$ (2009, in preparation). Data-points
lying outside the spatial range covered by the HII regions used to
compute O/H (typically beyond the optical size) have not been plotted,
as their extrapolated metallicities are not reliable. Filled circles
correspond to moderately face-on galaxies ($b/a \geq 0.5$), whereas
open diamonds indicate those being more edge-on ($b/a < 0.5$). The
solid line is a fit to the values of moderately face-on galaxies. The
empirical relations of Cortese et al$.$ (2006), Boissier et al$.$
(2007) and Heckman et al$.$ (1998) $-$ this latter one valid for
starbursts alone $-$ are also shown. (b): Relation between metallicity
and the observed UV color. The symbol code is the same as in panel
(a). The solid curve is not a fit to the data in this plot; it is the
result of the combination of the fit shown in panel (a) with the
IRX-$\beta$ relation in Eq.~\ref{eq_irxbeta}. Median errobars
(including zero-point errors) are shown in each panel. Fits to the
starburst galaxies of Calzetti et al$.$ (1994) and Heckman et al$.$
(1998) are also plotted. \label{ext_metal}}
\end{figure}

In Fig.~\ref{ext_metal}a we study the radial trend between metallicity
and the TIR-to-FUV ratio. Metallicity gradients for 22 SINGS galaxies
were taken from the compilation of Moustakas et al$.$ (2009, in
preparation). These authors compute the oxygen abundances using two
different calibrations: Kobulnicky \& Kewley (2004) and Pilyugin \&
Thuan (2005). The Kobulnicky \& Kewley (2004) values tend to be
smaller than the Pilyugin \& Thuan (2005) by $~0.6$\,dex, on
average. However, to first order the radial abundance gradients
derived for these 22 SINGS galaxies is independent on the adopted
calibration. In this paper we adopt the oxygen abundances derived with
the Kobulnicky \& Kewley (2004) calibration, and refer the reader to
Moustakas et al$.$ (2009, in preparation) for more details on this
subject.

Although our photometric profiles extend up to $\sim 1.5$ times the
optical size of each galaxy, the HII regions where oxygen abundances
were measured do not usually reach beyond the optical radius, and are
also missing in the central regions of some galaxies. In order to
avoid unsafe extrapolations of the metallicity values, we only plot
those data-points lying within the radial range determined by the
innermost and outermost abundance measurements in each galaxy. The
empirical fits of Cortese et al$.$ (2006) for cluster galaxies and of
Boissier et al$.$ (2007) for nearby spirals are shown for reference,
as well as the one of Heckman et al$.$ (1998) for starburst galaxies.

We confirm that the most metal-rich regions also suffer from larger
attenuation. Despite the scatter, we can see that at fixed metallicity
the more inclined galaxies appear to have larger TIR-to-FUV ratios:
from our perspective, UV photons must travel through larger amounts
of dust before reaching us than if the galaxy was to be observed
face-on. We performed a linear fit to those points with $b/a \geq
0.5$, resulting in:
\begin{equation}
12+\log(\mathrm{O/H})=8.70+0.405 \log (L_{\mathrm{TIR}}/L_{\mathrm{FUV}})\label{eq_metal_TIR_FUV}
\end{equation}

In Section~\ref{irxbeta_sfh} we stated that no significant trend with
inclination can be detected in the IRX-$\beta$ diagram, yet the
TIR-to-FUV ratio is clearly larger for nearly edge-on galaxies for a
given metallicity. This apparent riddle is solved in
Fig.~\ref{ext_metal}b, where we plot the metallicity as a function of
the $(\mathrm{FUV}-\mathrm{NUV})$ color for the same elliptical annuli
as in panel (a). Both magnitudes are clearly correlated, and
inclination also plays an important role here, since the observed UV
color is redder at larger inclinations for a given metallicity. The
effect of inclination depicted in panels (a) and (b) is such that more
inclined galaxies appear redder in the UV, but also have larger
TIR-to-FUV ratios. As a consequence, decreasing $b/a$ will not
significantly displace the corresponding data-points out of the main
trend seen in Fig.~\ref{irxbeta}, but would move them along it.

Since the TIR-to-FUV ratio seems to be linearly correlated with
metallicity for face-on galaxies, the distribution of filled circles
in Fig.~\ref{ext_metal}b is analogous to the IRX-$\beta$ plot. The
solid line in that panel is obtained by combining
Eqs.~\ref{eq_metal_TIR_FUV} and \ref{eq_irxbeta}, not from a direct
fit to the data in the plot.

\section{Comparison with physical dust models}\label{models}
\subsection{Description of the models}
In order to derive the physical properties of the dust beyond mere
dust attenuation, we need to assume a given model for the chemical
composition and size distribution of the dust grains, and take into
account their spatial distribution relative to the heating
stars. Modeling the spatial distribution of cold and warm dust,
relative to stars, has been the subject of much attention, especially
in edge on galaxies (see e.g$.$ Xilouris et al$.$ 1999; Popescu et
al$.$ 2000). In this paper we will employ the dust models of DL07,
which constitute an update of those developed by Weingartner \& Draine
(2001) and Li \& Draine (2001). These models were successfully applied
to the integrated photometry of the SINGS galaxies (Draine et al$.$
2007, D07 hereafter). The reader is referred to D07 and DL07 for a
more detailed description than is provided for context in the
following paragraphs.

The DL07 models describe the interstellar dust as a mixture of
carbonaceous grains and amorphous silicate grains, whose size
distributions are chosen to mimic the observed extinction law in the
Milky Way (MW). Although other size distributions can be chosen to
reproduce the extinction curves in the Large and Small Magellanic
Clouds, D07 showed that they did not provide significantly better fits
than the MW ones for the SINGS galaxies.

The physical properties of the smallest carbonaceous grains in the
models are those of PAH particles. The PAH abundance is characterized
by the PAH index, $q_{\mathrm{PAH}}$, defined as the fraction of the
dust mass in the form of PAH grains with less than $10^{3}$ carbon
atoms. Models with different PAH abundances can be generated by
changing the size distribution of the dust grains while keeping
constant the overall average extinction. The fraction of neutral and
ionized PAHs is the one estimated by Li \& Draine (2001) for the
diffuse ISM in the MW. The resulting spectra have fixed band ratios in
the IRAC bands, but those values are representative of normal spiral
galaxies.

One also has to make several assumptions regarding the starlight
heating the dust. In the DL07 models, the specific energy density of
the starlight is set to be a scaled version of the local interstellar
radiation field of the Milky Way estimated by Mathis et al$.$
(1983). The intensity of the incident starlight is then characterized
by a dimensionless scale factor $U$.

Not all the dust within a galaxy is exposed to the same range of
starlight intensities $U$, and the emitted SED will result from the
superposition of the SEDs associated with the diffuse interstellar
medium, star-forming regions, etc. Dale et al$.$ (2001) proposed a
power-law description of the dust heating, and used this model to
describe the global infrared SEDs of the SINGS galaxies in Dale et
al$.$ (2007).

The DL07 models follow a slightly different approach. The majority of
the dust is supposed to be located in the diffuse ISM, being heated by
a radiation field with a constant intensity $U_{\mathrm{min}}$. A
smaller fraction $\gamma$ of the dust is exposed to starlight with
intensities ranging from $U_{\mathrm{min}}$ to $U_{\mathrm{max}}$,
following the power-law expression mentioned above. This term is
intended to represent the dust enclosed in photo-dissociation regions
(PDRs), where the radiation field is much more intense than in the
diffuse component. Therefore, the amount of dust $dM_{\mathrm{dust}}$
exposed to radiation intensities between $U$ and $U+dU$ can be
expressed as:
\begin{equation}
\frac{dM_{\mathrm{dust}}}{dU}=(1-\gamma)M_{\mathrm{dust}}\delta(U-U_{\mathrm{min}})+\gamma M_{\mathrm{dust}}\frac{\alpha-1}{U_{\mathrm{min}}^{1-\alpha}-U_{\mathrm{max}}^{1-\alpha}}U^{-\alpha}
\end{equation}

D07 find that the precise value of $\alpha$ does not have a great
impact on the quality of the fits, given that the relative
contribution of the diffuse and PDR components can be now parametrized
with $\gamma$, and suggest fixing $\alpha=2$, which works well for a
wide range of galaxy properties. They also find that no particular
value of $U_{\mathrm{max}}$ seems to be favored, and hence adopt a
fixed value of $U_{\mathrm{max}}=10^{6}$. In addition, they suggest
using $0.7 \leq U_{\mathrm{min}} \leq 25$ when submillimeter data are
not available, since smaller values correspond to dust temperatures
below $\sim 15$\,K, which cannot be probed with MIPS photometry
alone. This lower cutoff for $U$ prevents the fitting technique from
invoking larger amounts of cold dust heated by weak starlight with
very small values of $U_{\mathrm{min}}$, although this would
underestimate the total dust mass if large amounts of cold dust are
actually present. Note, however, that for those SINGS galaxies with
available submillimeter data, D07 concluded that omitting these data
does not seem to introduce any evident systematic offset in the
derived total dust masses, although the scatter in the resulting dust
masses may increase up to $\sim 50$\%.

For a given PAH abundance, the emitted SED results from the linear
combination of the SEDs associated with the diffuse and PDR
component. Therefore, the shape of the SED is controlled by three
parameters: $q_{\mathrm{PAH}}$, $U_{\mathrm{min}}$ and $\gamma$, and
the dust mass surface density can be derived when normalizing the SED.

Besides the models of DL07, the SINGS galaxies have been used as a
benchmark for other different dust models. For instance, da Cunha et
al$.$ (2008) developed a simple model to interpret the mid- and far-IR
SEDs of galaxies consistently with the emission in the UV, optical and
near-IR. While their prescriptions for the dust emission are simpler
and more empirical than those of DL07, by linking the absortion of
starlight from the UV to the near-IR with dust emission at mid- and
far-IR, they are able to extract the properties of both stars and dust
in a consistent way. The dust properties they derived for the SINGS
galaxies are in good agreement with those presented in D07.

\subsection{Fitting procedure}\label{fitting}
Using our IRAC and MIPS radial profiles, we can therefore estimate
$q_{\mathrm{PAH}}$, $U_{\mathrm{min}}$, $\gamma$ and the dust mass
surface density $\Sigma_{M_{\mathrm{dust}}}$ as a function of the
galactocentric distance, by finding the best-fit model for the SED
within each elliptical annulus. Following DL07, by taking linear
interpolations of the original models we built a grid of models with
different PAH abundances (ranging from $q_{\mathrm{PAH}}=0.4\%$ to
4.6\% in steps of 0.1\%) and values of $\gamma$ (from 0\% to 30\% in
steps of 0.1\%). We looked for the best-fit model by minimizing the
reduced $\chi^{2}$, computed over the IRAC 5.8\,$\micron$ and
8.0\,$\micron$ channels and the three MIPS bands, after correcting
them for stellar emission (see below). We used the quadratic sum of
the photometric uncertainties and the zero-point errors as weights,
plus an additional $\sim 10\%$ error at each band to account for the
limited accuracy of the model, following D07. The 1-$\sigma$
uncertainty for each parameter was derived by projecting the overall
$\chi^{2}$ distribution over the one-dimensional space of that
parameter, and then looking for the values which satisfied
$\chi^{2}=\chi^{2}_{\mathrm{min}}+1$ (Press et al$.$ 1992).

The DL07 models provide the multi-walelength flux densities emitted by
the dust components. Since the IRAC 3.6\,$\micron$ and 4.5\,$\micron$
bands are almost entirely dominated by stellar emission (see e.g$.$
P\'erez-Gonz\'alez et al$.$ 2006), we did not take them explicitly
into account when performing the fits; however, we used the
3.6\,$\micron$ fluxes to estimate the stellar fluxes at the other
bands. We used the stellar populations synthesis models of Bruzual \&
Charlot (2003) to compute the flux ratios $\langle
F_{\nu}\rangle_{\lambda}/\langle F_{\nu}\rangle_{3.6}$. For a fixed
age of 13\,Gyr, and assuming exponential SFHs with timescales $\tau$
ranging from 0.01 to 20\,Gyr, we obtained $\langle
F_{\nu}\rangle_{\lambda}/\langle F_{\nu}\rangle_{3.6}=$0.660, 0.453,
0.269 and 0.032 for $\lambda=$4.5, 5.8, 8.0 and 24\,$\micron$,
respectively, assuming solar metallicity. These values are close to
the ones derived from a 5000\,K blackbody SED, and depend very weakly
on $\tau$, with variations of less than 5\% over the range of
extinction-corrected (FUV$-$3.6\,$\micron$) colors of our data. Larger
deviations can be seen for super-solar metallicities, and also for
very young populations ($\mathrm{age}\lesssim 1$\,Gyr), due to the
contribution of AGB stars. However, the prescription given above
yields good fits for our SEDs.

To check the possible systematic effects introduced by the stellar
emission correction, we fitted the models twice for all galaxies,
using our flux ratios and the ones estimated by Helou et al$.$ (2004)
using stellar population modeling from Starburst99 (Leitherer et
al. 1999). Their scaling factors are somewhat bluer than ours; hence
the derived PAH abundances are systematically larger, since the
estimated stellar emission at the IRAC bands is smaller. Nevertheless,
the median offset in the derived abundances is only $\sim
5$-10\%. There is, however, a much larger difference in those regions
where the PAH contribution to the 8\,$\micron$ flux density is less
than half of the stellar one at that band (e.g$.$ in bulges). The
values of $q_{\mathrm{PAH}}$ might then be unreliable in these cases,
so we have excluded them from our analysis. As for the fraction
$\gamma$ of the dust mass exposed to high-intensity radiation fields,
for $\gamma \lesssim 1$\% the resulting values using Starburst99 are
$\sim 10$\% smaller than the ones obtained when using the factors from
the models of Bruzual \& Charlot (2003), with no difference at larger
values. Finally, the dust mass and luminosity surface densities, as
well as the radiation field scale-factor $U$, do not vary
noticeably. This is not surprising, since these parameters are mostly
determined by the MIPS fluxes, where the stellar contamination is
negligible. For the remainder of the analysis described here we adopt
the factors obtained from the models of Bruzual \& Charlot (2003).

Further systematic effects might result from using a small number of
bands when fitting the models. In Appendix~\ref{app_systematic} we
discuss this issue in detail, and demonstrate that our results are
consistent with the global values obtained by D07 using a more
complete set of IR data.

\subsection{Results from the model}
\subsubsection{Individual profiles and SEDs}
\begin{figure}
\resizebox{0.8\hsize}{!}{\includegraphics[angle=-90]{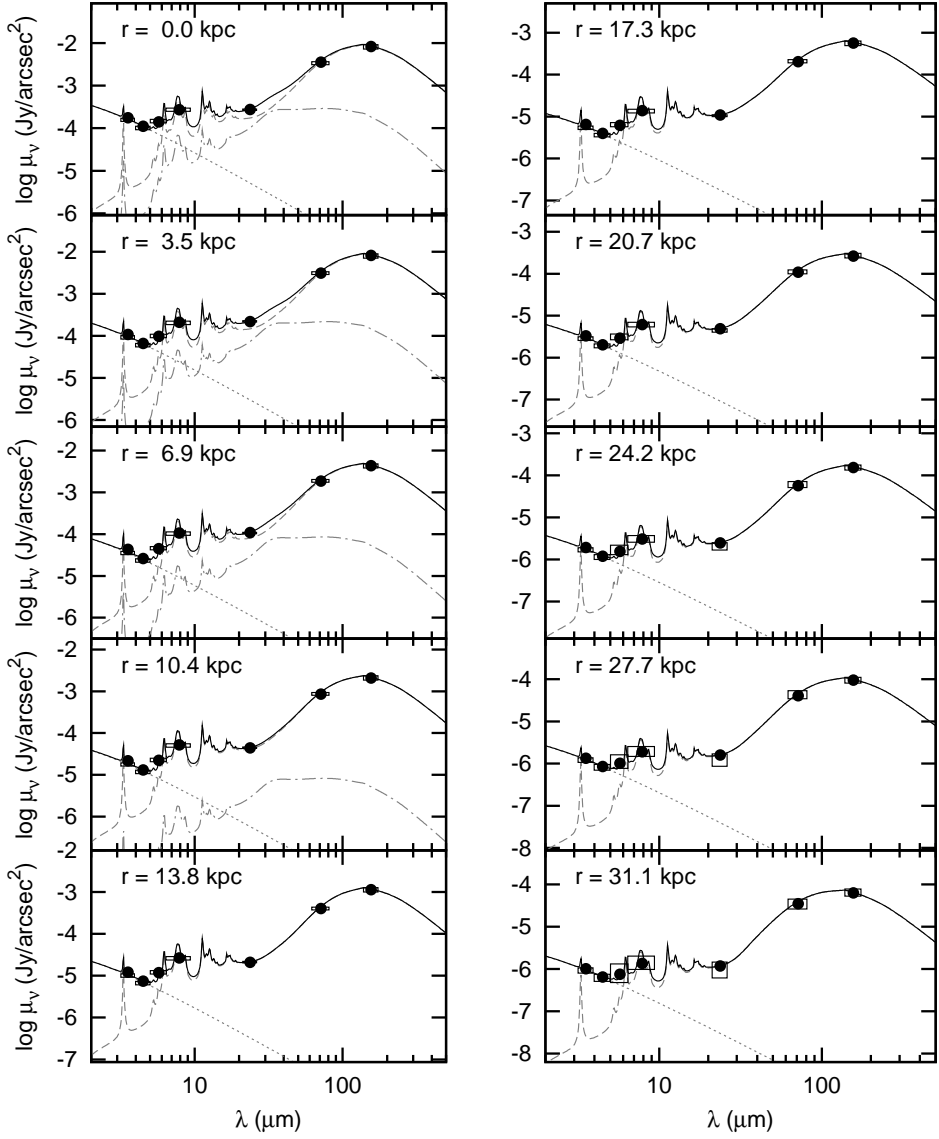}}
\caption{Observed SEDs and model spectra at different galactocentric
distances, without correcting for inclination. A sample plot for
NGC~7331 is shown; additional figures for the remaining galaxies can
be found in the electronic version of the journal. The observed flux
densities are marked with rectangular boxes, their width showing the
corresponding bandpass and their height showing the observational
errors. The best-fitting model spectra (solid line) results from the
combination of a diffuse component (dashed line), the emission from
hot dust (dash-dotted line) and the stellar emission (dotted
line). The filled circles correspond to the model spectra convolved
with each bandpass.\label{seds}}
\end{figure}

In Fig.~\ref{seds} we show the infrared SEDs over a range of
galactocentric distances, along with the best-fitting models, for one
of the sample galaxies: NGC~7331 (see the on-line edition for a
complete figure-set for the whole sample). The corresponding numerical
values of the model parameters are quoted in Table~\ref{model_tab},
which is shown in its entirety in the online edition. Note that the
results for NGC~7331 can be readily compared with those from Thilker
et al$.$ (2007). As explained above, each SED can be decomposed in two
terms: a `diffuse' component (dashed line), heated by a constant
radiation field with $U=U_{\mathrm{min}}$, and a PDR one (dash-dotted
line), with $U_{\mathrm{min}}<U<U_{\mathrm{max}}$. The latter is
particularly important to account for the observed flux density at
24\,$\micron$. The total resulting SED is shown with a solid line, and
includes not only the emission contributed by dust, but also the one
coming from stars (dotted line), estimated from the 3.6\,$\micron$
flux. The observed flux densities (open rectangles) are shown along
with the ones resulting when convolving the model with each band
(filled circles). Although the 3.6 and 4.5\,$\micron$ flux densities
are shown, they were not included when performing the fits, as
explained in Section~\ref{fitting}.

\begin{figure}
\resizebox{1\hsize}{!}{\includegraphics{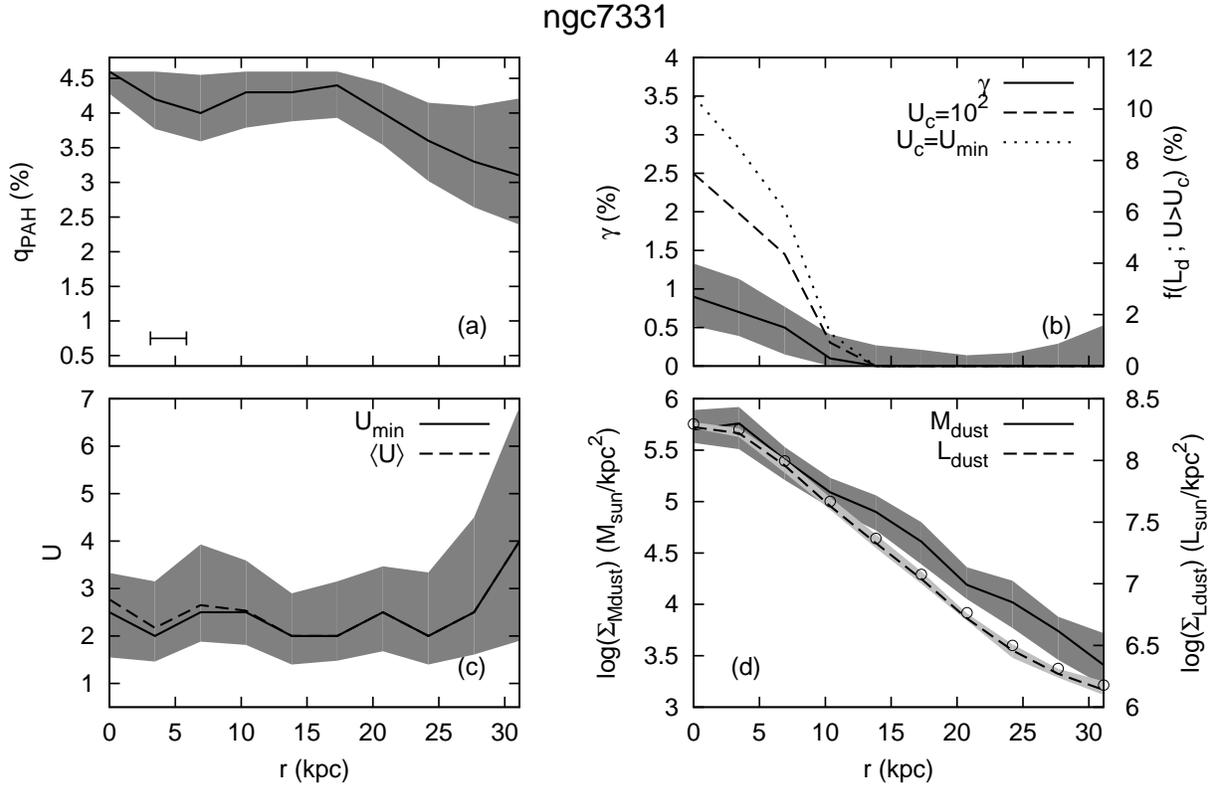}}
\caption{Radial distribution of the different parameters derived from
the dust models. A sample plot is shown for NGC~7331; additional
figures for the remaining galaxies can be found in the electronic
version of the journal. The different lines show the best-fitting
values at each radius, and the gray bands show the estimated
uncertainties. (a): Fraction of the dust mass in the form of
PAHs. Regions in which the stellar emission at 8\,$\micron$ is more
than twice the one from PAHs are marked (when present) with a dotted
line instead of a solid one. The FWHM of the MIPS 160\,$\micron$ band
(38$\arcsec$) is marked with an horizontal segment along the major
axis. (b): Fraction $\gamma$ of the total dust mass heated by very
intense starlight. The right vertical axis shows the fraction of the
dust luminosity contributed by regions exposed to radiation fields
with $U>U_{\mathrm{min}}$ and $U>10^{2}$ (their errors are not shown
for clarity). (c): Scale factor of the minimum ($U_{\mathrm{min}}$)
and average ($\langle U \rangle$) starlight intensity heating the
dust, in units of the local MW radiation field. The errors of $\langle
U \rangle$ are not shown. (d): Dust mass (left) and luminosity (right)
surface densities, both corrected for inclination. The open circles
show the dust luminosity profiles obtained with the photometric
estimator of DL07.\label{radial_params}}
\end{figure}

The radial trends of the different model parameters are shown in
Fig.~\ref{radial_params} for each individual galaxy (only NGC~7331 is
included in the printed version). In all panels, the best-fitting
values are shown with different lines, and the gray bands correspond
to the estimated 1-$\sigma$ uncertainties. Panel (a) shows the radial
variation of the PAH mass fraction, $q_{\mathrm{PAH}}$. Regions where
$q_{\mathrm{PAH}}$ is not reliable are identified with a dotted line
instead of a solid one. In these regions, stellar emission at
8\,$\micron$ is more than twice the one contributed by PAHs, so even
though their fitting errors might be small, they are affected by large
uncertainties resulting from the stellar subtraction. Therefore, these
values will not be considered in subsequent sections. The physical
scale of the MIPS FWHM at 160\,$\micron$ (38\arcsec) at the distance
of each galaxy is marked with a small horizontal line.

In panel (b) we show the fraction of dust mass exposed to high
intensity radiation fields, $\gamma$. The dotted and dashed lines
correspond to the fraction of the dust-luminosity contributed by dust
grains in those regions, heated by starlight with $U>U_{\mathrm{min}}$
and $U>10^{2}$, respectively. Note the different scales for the mass
and luminosity fractions: about 1\% of the dust mass can be
responsible for $\sim 10$\% of the total dust luminosity.

Panel (c) displays the radial trends of the minimum starlight
intensity $U_{\mathrm{min}}$, as well as an average value $\langle U
\rangle$, which depends on both $U_{\mathrm{min}}$ and $\gamma$ (see
Eq.~33 in D07). With this definition, we have $L_{\mathrm{dust}}
\propto \langle U \rangle M_{\mathrm{dust}}$.

Finally, in panel (d) we show the radial profiles of both dust mass
and luminosity surface densities, corrected for inclination using the
semiaxis ratio in Table~\ref{sample_tab}. The open circles show the
dust luminosity computed from the monochromatic fluxes from 8 to
160\,$\micron$, using Eq.~22 from DL07. These values compare well with
the ones obtained by integrating the SED from the model (dashed line).

\subsubsection{Properties of the whole sample}
\begin{figure}
\resizebox{1\hsize}{!}{\includegraphics{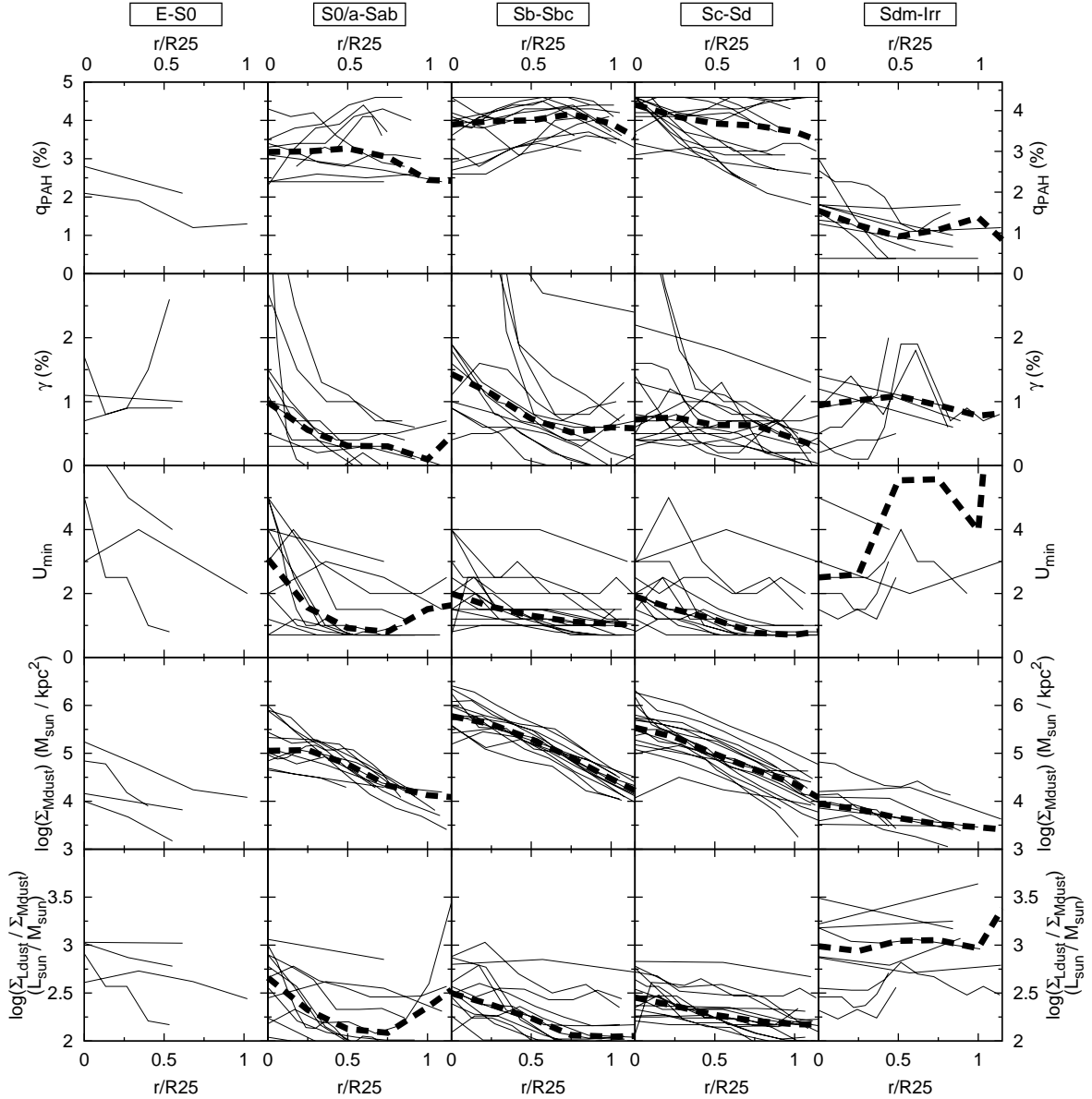}}
\caption{Radial profiles of the different parameters involved in the
dust models. Galaxies have been grouped according to their
morphological type. The different parameters are the PAH abundance
($q_{\mathrm{PAH}}$), the fraction of the dust mass exposed to intense
radiation fields ($\gamma$), the intensity of the `diffuse' heating
starlight, in units of the local MW radiation field
($U_{\mathrm{min}}$), the deprojected dust mass surface density
($\Sigma_{M_{\mathrm{dust}}}$) and the dust light-to-mass ratio
($L_{\mathrm{dust}}/M_{\mathrm{dust}}$). The radial coordinate is
normalized to the optical size of each galaxy. The thick dashed lines
show the median profiles in each panel. No median profiles are shown
for ellipticals and lenticulars, due to the small number of galaxies
in that bin. The median $U_{min}$ profile for Sdm-Im galaxies is above
most of the profiles displayed in that panel due to some galaxies
having $U_{\mathrm{min}}>5$, which are not shown here for
clarity.\label{all_model_profs}}
\end{figure}

In Fig.~\ref{all_model_profs} we show the radial variation of each
parameter in the model for all galaxies in the sample, divided into
bins of morphological type. The radial coordinate is normalized to the
optical radius, R25. The median profiles in each panel are shown as a
thick dashed line.

The top row shows how the PAH abundance changes with the
galactocentric distance. Note that we only plot those values of
$q_{\mathrm{PAH}}$ satisfying the flux criteria defined in
Section~\ref{fitting}; nevertheless, the PAH abundances for the
earliest types should be still considered with caution. The median
profiles of $q_{\mathrm{PAH}}$ in S0/a-Sbc galaxies look somewhat
flat, although some individual profiles exhibit a central depletion of
PAHs, relative to the abundances at larger radii. The overall PAH
abundance in these spirals ranges from 2\% to 4.6\% of the total dust
mass. As we move towards later types, steeper radial gradients become
more frequent, and $q_{\mathrm{PAH}}$ decreases down to $\sim 1.5$\%
or even less for the latest types. As we will see later, this can be
interpreted in terms of a connection between the PAH abundance and
metallicity.

The fraction of the dust mass that is heated by very intense
starlight clearly decreases with radius. This component is mostly
determined by the flux density at 24\,$\micron$, which usually
outlines the spiral pattern in which star-forming regions are
typically arranged. The spatial filling factor of the HII regions is
thus larger at the innermost regions, where they swirl and crowd
together, and where sometimes they coincide with rings, whereas at
larger galactocentric distances they are more dispersed, so their
contribution to the 24\,$\micron$ surface brightness is not so
important in comparison with the diffuse one. Indeed, the radial
decrease in $\gamma$ is steeper in early-type grand-design spirals,
while in late-type flocculent ones the gradient is flatter, since the
spatial distribution of star-forming regions is more uniform. This is
in agreement with the findings of Bendo et al$.$ (2007), who also
concluded that the 24\,$\micron$ emission is more centrally
concentrated in early-type spirals than in late-type ones. In any
case, although the values of $\gamma$ are quite small (typically
around 1\%), such a small fraction of the total dust mass can be
responsible for a non-negligible fraction of the total emitted infrared
power.

The scale-factor $U_{\mathrm{min}}$ describing the intensity of the
diffuse heating starlight also decreases with radius. This quantity is
affected by both young and old stellar populations, and the bulge can
certainly be responsible for part of that heating, raising the central
values of $U_{\mathrm{min}}$, especially in Sa-Sbc
spirals. Additionally, the higher central concentration of
star-forming regions in these galaxies may also increase the diffuse
interstellar radiation field (besides boosting $\gamma$). In the outer
regions of spiral disks, beyond $r\sim 0.5$\,R25, the median value of
$U_{\mathrm{min}}$ is close to 1, meaning that the radiation field has
the same intensity as the local MW one. Noteworthy, in the latest
types both $U_{\mathrm{min}}$ and $\gamma$ present higher median
values, although with more dispersion.

The radial variation of the dust mass surface density also depends on
the Hubble type. The largest amounts of dust are found in Sb-Sd
galaxies, where the surface density decreases exponentially from
$10^6\mathrm{\,M}_{\sun}\mathrm{\,kpc}^{-2}$ in the central regions to
$10^4\mathrm{\,M}_{\sun}\mathrm{\,kpc}^{-2}$ in the outer ones, at the
R25 radius. Conversely, dust is less abundant in Sdm and irregulars,
as well as in ellipticals and lenticulars, with S0/a-Sab spirals lying
in between.

It is worth noting that the profiles of dust mass surface density
appear to be exponential, and the radial scale-length seems to hold
rather constant from Sab to Sd spirals, when expressed in units of the
optical size. To quantify this behavior, we have applied linear fits
to these profiles. We have excluded prominent bulges from the fit, in
the same way as we did in Section~\ref{extinction} with the
attenuation profiles. This is particularly important in many S0/a-Sab
galaxies, in which the central regions usually deviate from the
exponential trend, due to holes or depletions in the amount of dust
(see the full version of Fig.~\ref{radial_params} in the on-line
edition of the journal).

\begin{figure}
\resizebox{1\hsize}{!}{\includegraphics{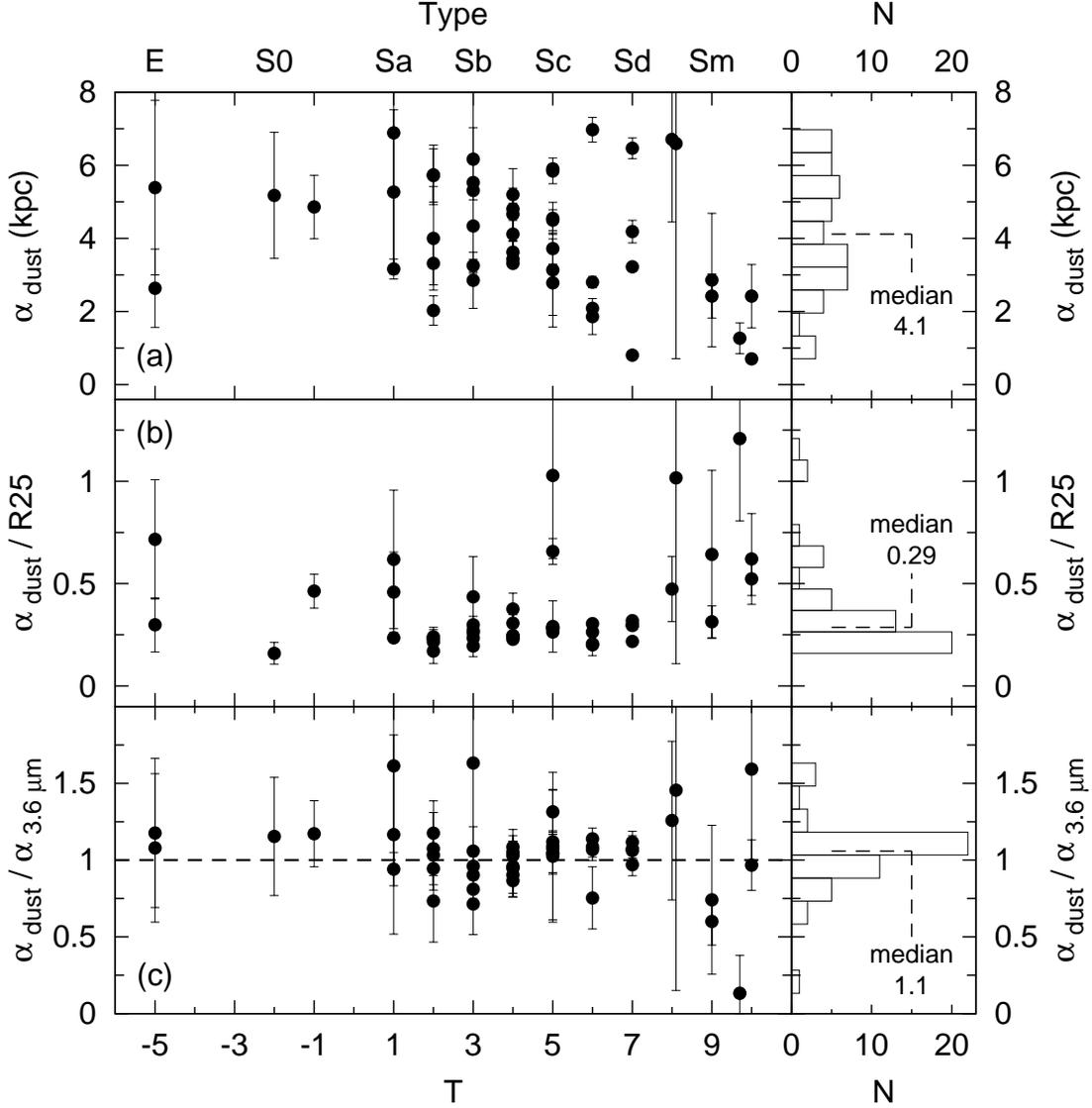}}
\caption{Radial exponential scale-lengths of the dust profiles for
different morphological types. The scale-lengths were derived from
linear fits applied to the profiles after having excluded the bulges
(see text). They are expressed in kpc (a), in units of the optical
radius (b) and in units of the stellar scale-length (c), measured on
the 3.6\,$\micron$ profiles derived from the degraded IRAC images
(bulges excluded). The global distributions for the whole sample, as
well as the corresponding median values, are shown to the
right.\label{alpha_dust}}
\end{figure}

The resulting exponential scale-lengths $\alpha_{\mathrm{dust}}$,
defined so that $\Sigma_{M_{\mathrm{dust}}} \propto
e^{-r/\alpha_{\mathrm{dust}}}$, are shown in
Fig.~\ref{alpha_dust}. When expressed in absolute physical units
(panel a), the dust mass scale-length varies widely across the Hubble
sequence from 1 to 7\,kpc. However, the distribution appears strongly
peaked when normalizing by the optical size (b) and by the
scale-length of the stellar mass profiles (c), the latter obtained
from the profiles measured on the degraded IRAC 3.6\,$\micron$ images,
for consistency. The dust mass radial scale-length is equal to
0.2-0.3\,R25 for Sab-Sd galaxies. The profiles are flatter in later
types (although the dispersion is larger, as these systems might
deviate from the disk-like geometry). In general, the dust
scale-length seems to be $\sim 10$\% larger than the stellar one. By
applying a three-dimensional radiative transfer model to seven edge-on
spirals, Xilouris et al$.$ (1999) concluded that the dust radial
scale-length is $\sim 40$\% larger than the stellar one. The presence
of a cold dust component ($T<15$\,K) with a larger scale-length could
explain this difference, as this component would be detected in
absorption by the radiative transfer modelling, but it would be
unconstrained by our MIPS profiles. Conversely, some of the necessary
geometrical simplifications on which this kind of 3D models rely could
introduce some bias. For instance, the dust profiles are assumed to be
exponential throughout the whole disk, which does not seem to be the
case in many early-type spirals exhibiting central plateaus or holes
in their dust profiles. Indeed, if we do not exclude the bulges when
applying linear fits to our dust profiles, the resulting dust
scale-lengths are $\sim 20$\% larger than the stellar ones for the
whole sample, and $\sim 30$\% larger if we focus on early-type
spirals, whose bulges are more prominent.

The bottom row in Fig.~\ref{all_model_profs} shows the profiles of the
dust light-to-mass ratio, which is essentially proportional to
$\langle U \rangle$. Therefore, these profiles reflect radial changes
in temperature which, in general, seems to decrease with
radius. Interestingly, irregulars can reach larger dust luminosity
surface densities than Sa-Sd spirals, in spite of having considerably
lower dust mass densities, thanks to their dust being hotter (note
their higher values of both $U_{\mathrm{min}}$ and $\gamma$).

Note that we cannot speak of `temperature' in the same way as when
fitting FIR data with a double black-body SED, since the dust models
used here consider the full temperature distribution for all grain
sizes. Although the smallest grains can undergo extreme temperature
changes in small periods of time, large ones dominating the FIR
emission can be considered to have reasonably steady temperatures $T
\approx 17 U^{1/6}$\,K (DL07). As a sanity check, for NGC~7331 we
obtain a roughly flat profile with $U_{\mathrm{min}} \approx 2.5$,
implying $T \approx 20$\,K, the same value found by Thilker et al$.$
(2007) when applying a double black-body fit. By analyzing the IR SEDs
of different regions in M~81, P\'erez-Gonz\'alez et al$.$ (2006) found
that the temperature of the cold component varies from $\approx 22$\,K
in the central regions to $\approx 15$\,K at 10\,kpc from the center,
which is consistent with our $U_{min}$ profile ranging from 5 to 0.7
in the same relative radial range.

\subsubsection{Dust mass surface density and attenuation}
An initial comparison of the dust mass surface density profiles in
Fig.~\ref{all_model_profs} and the attenuation profiles in
Fig.~\ref{all_extprofs} shows that both quantities follow similar
global trends along the Hubble sequence, peaking at Sb-Sbc
spirals. However, the dust mass profiles show a smaller dispersion
than the extinction profiles within each morphological-type bin (see
also the histograms in Figs.~\ref{AFUV_grad} and \ref{alpha_dust}).

\begin{figure}
\resizebox{1\hsize}{!}{\includegraphics{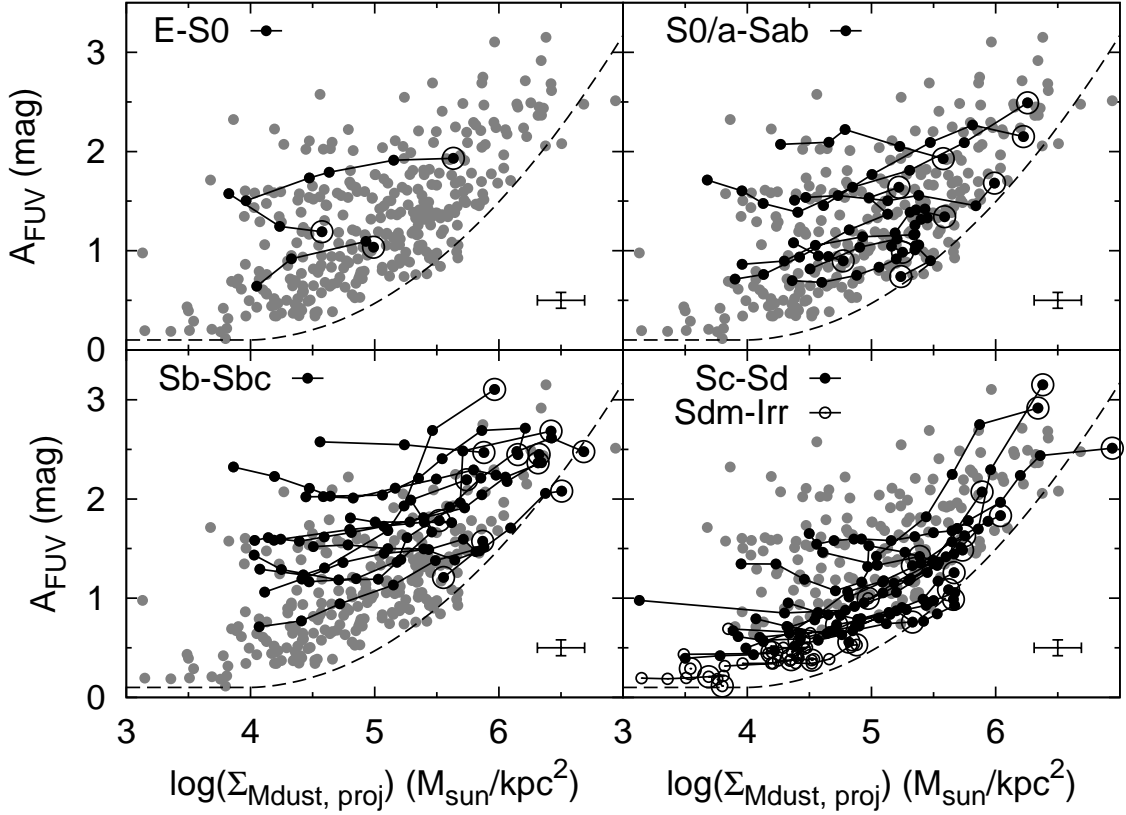}}
\caption{Attenuation in the FUV as a function of the projected dust
mass surface density. Black points in each panel show the profiles of
galaxies with the corresponding Hubble type. The innermost point of
each profile is surrounded by a larger circle. As a reference, the
data for the whole set of annular regions are shown as gray points in
each of the panels. The dashed line is a fit of the lower envelope of
the whole data-set, performed with the boundary-fitting code presented
in Cardiel et al$.$ (2009) (see text for
details).\label{Mdust_ext}}
\end{figure}

To further analyze these differences, in Fig.~\ref{Mdust_ext} we plot
the attenuation in the FUV as a function of the corresponding
projected dust mass surface density (we cannot use the de-projected
one, for the measured values of $A_{\mathrm{FUV}}$ take into account
all the dust along the line of sight). Note that since
$A_{\mathrm{FUV}}$ has been computed with the age-dependent
calibration of Cortese et al$.$ (2008), the varying relative
contribution of young and old stars to the dust heating in galaxies
with different SFHs is being taken into account. It could still be a
problem in ellipticals and lenticulars, though.

This plot reveals that, for a given value of the dust mass density,
there is always a minimum associated attenuation. This value increases
monotonically from almost zero for $\Sigma_{M_{\mathrm{dust,\,proj}}}
\sim 10^3 \mathrm{\,M}_{\sun}\mathrm{\,kpc}^{-2}$ to roughly 3
magnitudes for $\Sigma_{M_{\mathrm{dust,\,proj}}} \gtrsim 10^6
\mathrm{\,M}_{\sun}\mathrm{\,kpc}^{-2}$. This lower envelope for the
distribution of data-points is delineated by Sc galaxies and later,
while for the same amount of dust, earlier spirals present
attenuations 1 or 2\,mags higher. Using the boundary-fitting code
described in Cardiel (2009), we obtained the following analytic
parametrization for the minimum attenuation at a given dust column
density:
\begin{equation}
A_{\mathrm{FUV,\,min}}\approx \left\{ \begin{array}{rl}
5.16-2.57x+0.327x^2 &\mbox{ if $4<x<7$} \\
0.1 &\mbox{ if $3<x<4$}\end{array} \right.
\end{equation}
where $x=\log \Sigma_{M_{\mathrm{dust,\,proj}}}$, in
$\mathrm{\,M}_{\sun}\mathrm{\,kpc}^{-2}$. This fit is shown in
Fig.~\ref{Mdust_ext} with a dashed line.

The scatter in $A_{\mathrm{FUV}}$ found for each given
$\Sigma_{M_{\mathrm{dust,\,proj}}}$ can be attributed to the relative
distribution of stars and dust within each galaxy. It is known that
for geometries more complex than a foreground homogeneous screen of
dust, the distribution of dust grains relative to the heating sources
overcomes the net amount of dust as the main factor determining the
attenuation (Calzetti 2001).

\begin{figure}
\resizebox{1\hsize}{!}{\includegraphics{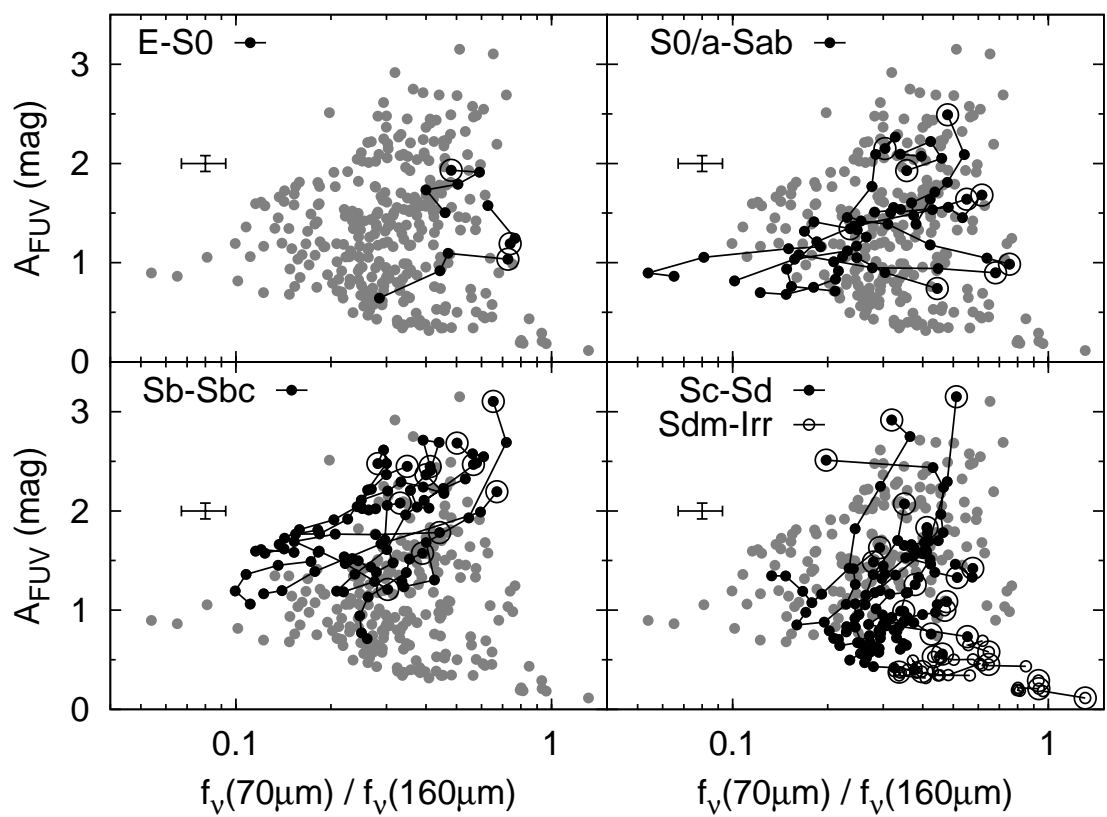}}
\caption{Attenuation in the FUV as a function of the
$f_{\nu}(70\mathrm{\,}\micron)/f_{\nu}(160\mathrm{\,}\micron)$
color. Black points in each panel show the profiles of galaxies with
the corresponding Hubble type. The innermost point of each profile is
surrounded by a larger circle. As a reference, the data for the whole
set of annular regions are shown as gray points in each of the
panels.\label{clumpy}}
\end{figure}

To illustrate this issue, in Fig.~\ref{clumpy} we plot
$A_{\mathrm{FUV}}$ as a function of the far-infrared
color. Data-points are clearly arranged in a wedge-like distribution,
completely analogous to the one found by Dale et al$.$ (2007) using
the integrated photometry of the SINGS galaxies. These authors came to
the conclusion that the location of galaxies in this diagram was
closely related with their morphology in the 24\,$\micron$
band. Galaxies with clumpy morphologies were found to have warm far-IR
colors but low attenuations: while there are significant amounts of
hot dust concentrated around heating sources in star-forming regions,
many UV photons can find their way out of the galaxy through lines of
sight clean of dust, thus decreasing the overall TIR-to-FUV ratio. In
our radial analysis, Sc spirals and later clearly lie in the
lower-right corner of the plot, thus being consistent with having a
clumpy distribution of dust relative to stars. The presence of these
kind of porous star-forming regions has been identified in nearby
galaxies (Roussel et al$.$ 2005), and their contribution to the
TIR-to-FUV ratio may become progressively dominant with increasing
galactocentric distance.

In contrast, Dale et al$.$ (2007) found that the upper-right region of
the plot is populated by galaxies showing centrally concentrated
emission at 24\,$\micron$. In our radially resolved analysis the
central regions of spirals, especially in Sb-Sbc ones, lie in this
area of the diagram. This indicates that UV light has significant
difficulties to avoid dust, thus leading to much larger extinction
values. Finally, regions having cooler far-IR colors tend to exhibit
intermediate attenuations. This area of the plot was identified by
Dale et al$.$ (2007) as corresponding to galaxies with a smooth
24\,$\micron$ morphology. This is also the case for our radial
profiles, since the disks of Sbc spirals and earlier populate this
zone of the plot. While not being as heavily extincted as in the
central regions, UV photons in these disks still find it difficult to
leak through dust, in comparison with clumpy Sc-Sd spirals, in which
the attenuation is lower despite having similar dust mass surface
densities.

\subsubsection{PAH abundance}\label{S_pah}
Over the years, a body of observational evidence has been collected
showing that PAH emission is weak or even suppressed in
low-metallicity systems (see e.g$.$ Boselli et al$.$ 1998; Sturm et
al$.$ 2000; Madden 2000; Madden et al$.$ 2006; Hunt et al$.$ 2005;
Hogg et al$.$ 2005; D07; Engelbracht et al$.$ 2005, 2008). However,
the physical mechanisms driving the observed correlation between PAH
abundance and metallicity remain a subject of investigation.

Several authors have invoked an evolutionary origin for this trend
(see e.g$.$ Galliano et al$.$ 2008; Dwek et al$.$ 2009). PAHs are
thought to form in the envelopes around carbon-rich AGB stars, being
later injected into the ISM through a recycling process that takes
place over timescales of a few Gyr. The remaining dust species are
believed to condense out of ejecta from supernovae in much shorter
timescales. Thus, most of the dust in young, low-metallicity systems
would be made out of SN-condensed species, whereas PAH particles would
still require more time to form. Hence, the different speed of both
processes would naturally lead to a progressive increase in
$q_{\mathrm{PAH}}$ with time, explaining the observed PAH-metallicity
trend. This scenario, however, should be considered cautiously, as
many aspects of chemical evolution and dust formation still remain
unclear. For instance, despite their low metallicities, some dwarf
galaxies may harbor underlying stellar populations a few Gyrs
old. Also, carbon dust might be also produced in the ejecta of
supernovae, and its abundance relative to carbon dust due to AGB stars
should be also taken into account.

Conversely, other possible explanations rely on destructive mechanisms
of PAH particles. Systems with younger stellar populations have harder
UV radiation fields, given their larger fraction of ionizing OB
stars. Furthermore, PAHs are likely to be less shielded by dust from
these hard photons than in more metal-rich systems, since
low-metallicity galaxies usually exhibit lower dust attenuation. There
is observational evidence showing a progressive paucity of aromatic
features as the hardness of the radiation field increases (Engelbracht
et al$.$ 2005, 2008; Madden et al$.$ 2006; Bendo et al$.$ 2006, 2008;
Smith et al$.$ 2007; Gordon et al$.$ 2008).

It is not straightforward to disentangle the relative contribution of
formation and destruction effects to the observed PAH abundances. As
we commented above, predominantly young systems which are expected to
show a paucity of PAHs because of chemical evolution, also exhibit
hard radiation fields that could destroy the already present ones. Besides,
variations in the aromatic emission could be due to changes in the
ionization state of PAHs rather than to a real decrease in terms of
abundance. Wu et al$.$ (2006) found that the equivalent width of the
aromatic features correlates better with a parameter simultaneously
taking into account both the starlight hardness and the metallicity
than with any of those parameters alone.

\begin{figure}
\resizebox{1\hsize}{!}{\includegraphics{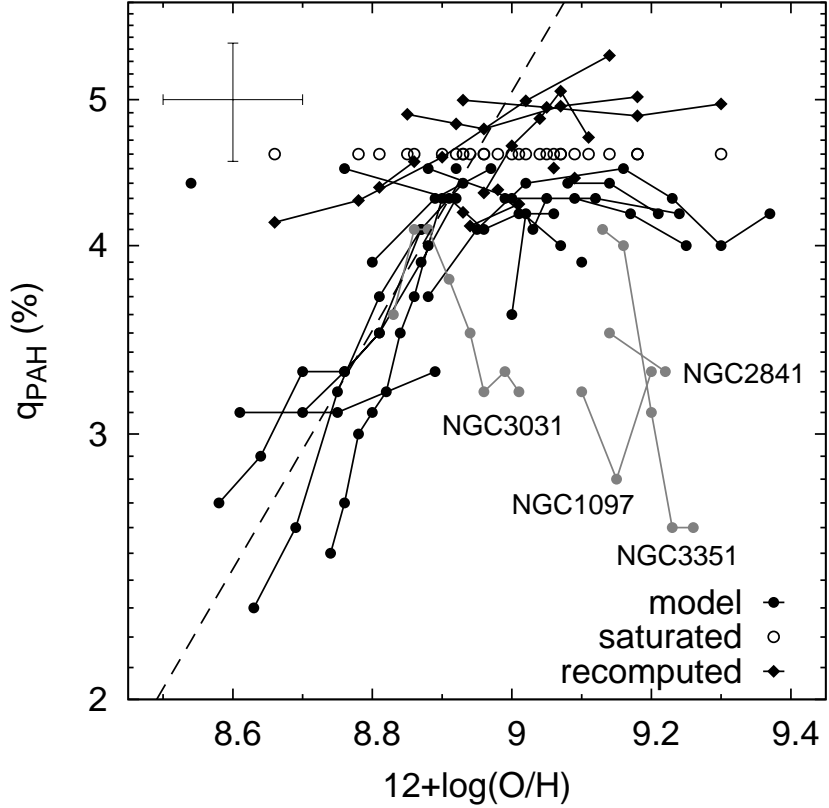}}
\caption{Fraction of the dust mass contributed by PAHs against the
metallicity at each radial bin. Data-points with
$q_{\mathrm{PAH}}=4.6$\% are `saturated', meaning that their PAH
abundances correspond to the current upper limit of the models (see
Section~\ref{S_pah}). For these points we have recomputed
$q_{\mathrm{PAH}}$ from the empirical fit presented in
Appendix~\ref{S_qpah_empirical} (see also
Fig.~\ref{qpah_empirical}). The dashed line is a bisector linear fit
applied to the data-points below $12+\log(O/H)=9$, and corresponds to
$q_{\mathrm{PAH}} \propto \mathrm{(\mathrm{O/H})}^{0.8 \pm 0.3}$. Four
galaxies for which the $q_{\mathrm{PAH}}$ vs. (O/H) trend is reversed
have been highlighted (see text for details). \label{pah_metal}}
\end{figure}

To check whether the PAH-metallicity trend holds true in our radial
analysis, in Fig.~\ref{pah_metal} we plot the PAH abundance at each
galactocentric distance as a function of the corresponding metallicity
at the same radial bin, for the 22 galaxies in common with the sample
of Moustakas et al$.$ (2009, in preparation). As in
Section~\ref{S_ext_metal}, we only trust those metallicities
interpolated within the spatial range covered by HII regions in the
aforementioned study. As explained in Section~\ref{S_ext_metal}, we
adopt the metallicity values based on the Kobulnicky \& Kewley (2004)
calibration. Data points falling outside that range require uncertain
extrapolations of the metallicity gradients, and have been excluded
from this analysis, as well as regions with excessive stellar emission
at 8\,$\micron$ (see Section~\ref{fitting}).

There is a general trend of $q_{\mathrm{PAH}}$ to increase with the
oxygen abundance for $12+\log\mathrm{(O/H)} < 9$. Note that in Sdm and
irregular galaxies, the PAH abundance ranges from 2.5\% to less than
1\%. However, there are not enough individual measurements of HII
regions in these galaxies to derive metallicity gradients, which
precludes making use of those data-points for this plot. The available
abundance data indicate that these galaxies have
$12+\log\mathrm{(O/H)} < 8.6$ (Moustakas et al$.$ 2009, in
preparation), so the monotonic trend between $q_{\mathrm{PAH}}$ and
(O/H) continues down to lower abundances. This general trend is
consistent with that found by D07 in the SINGS galaxies using
galaxy-averaged PAH abundances and metallicities. The dashed line,
which corresponds to a bisector linear fit applied to the data with
$12+\log\mathrm{(O/H)} < 9$, shows that $q_{\mathrm{PAH}} \propto
\mathrm{(O/H)}^{0.8 \pm 0.3}$ in this range of metallicities. The
slope may vary within individual galaxies, and it could be different
for lower oxygen abundances. We call attention to the fact that, up to
now, integrated photometry usually required comparing normal spirals
with dwarf galaxies in order to detect this trend, while radial
profiles make it possible to see it within a more limited range of
galaxy types because of the reduced scatter.

At larger metallicities, however, the slope flattens and even
reverses. In principle, the upper limit of the DL07 models for the PAH
abundance (4.6\%) could be the culprit, but we call attention to the
fact that in most cases this flattening arises before that limit is
reached, and it remains visible when recomputing the PAH abundances
with the empirical recipes presented in
Appendix~\ref{S_qpah_empirical}. Interestingly, some models describing
the stellar evolutionary effects on the chemical composition of dust
predict that such a change in slope might actually occur (Galliano et
al$.$ 2008), and attribute it to the different carbon content of AGB
stars with different masses $-$and hence with different
lifetimes. Stars with masses between 1.2 and 3.0\,M$_{\sun}$ are
thought to end up as carbon-rich post-AGB stars, unlike low-mass
stars, which remain oxygen-rich (see e.g$.$ Garc\'{i}a-Lario 2006 and
references therein). Since the latter have longer lifetimes, they
contribute later to the enrichment of the ISM. Moreover, the C/O ratio
in AGB stars also seems to depend on the stellar metallicity, with
more metal-rich AGBs yielding lower amounts of carbon (see Dwek et al$.$
2009 and references therein). As a result, these models predict that
the formation of carbon dust from AGB stars slows down with time,
while the absolute abundance of SN-condensed dust species keeps
rising.

There are four galaxies in Fig.~\ref{pah_metal} that deserve special
attention, for they have low PAH abundances compared to other galaxies
with similar metallicities; moreover, the trend between those
quantities is actually reversed within each one of these
objects. Interestingly, they are the earliest spirals in the diagram
(Sb and earlier). Therefore, they have presumably undergone rapidly
declining SFHs, with typical timescales of just a few Gyr (Gavazzi et
al$.$ 2002), in contrast to late-type spirals, which exhibit longer
timescales. If SFH is driving the progressive change of slope due to
the different carbon content of AGBs with different masses and/or
metallicities, this effect could be enhanced in systems with rapidly
declining SFHs.

\begin{figure}
\resizebox{!}{0.85\vsize}{\includegraphics{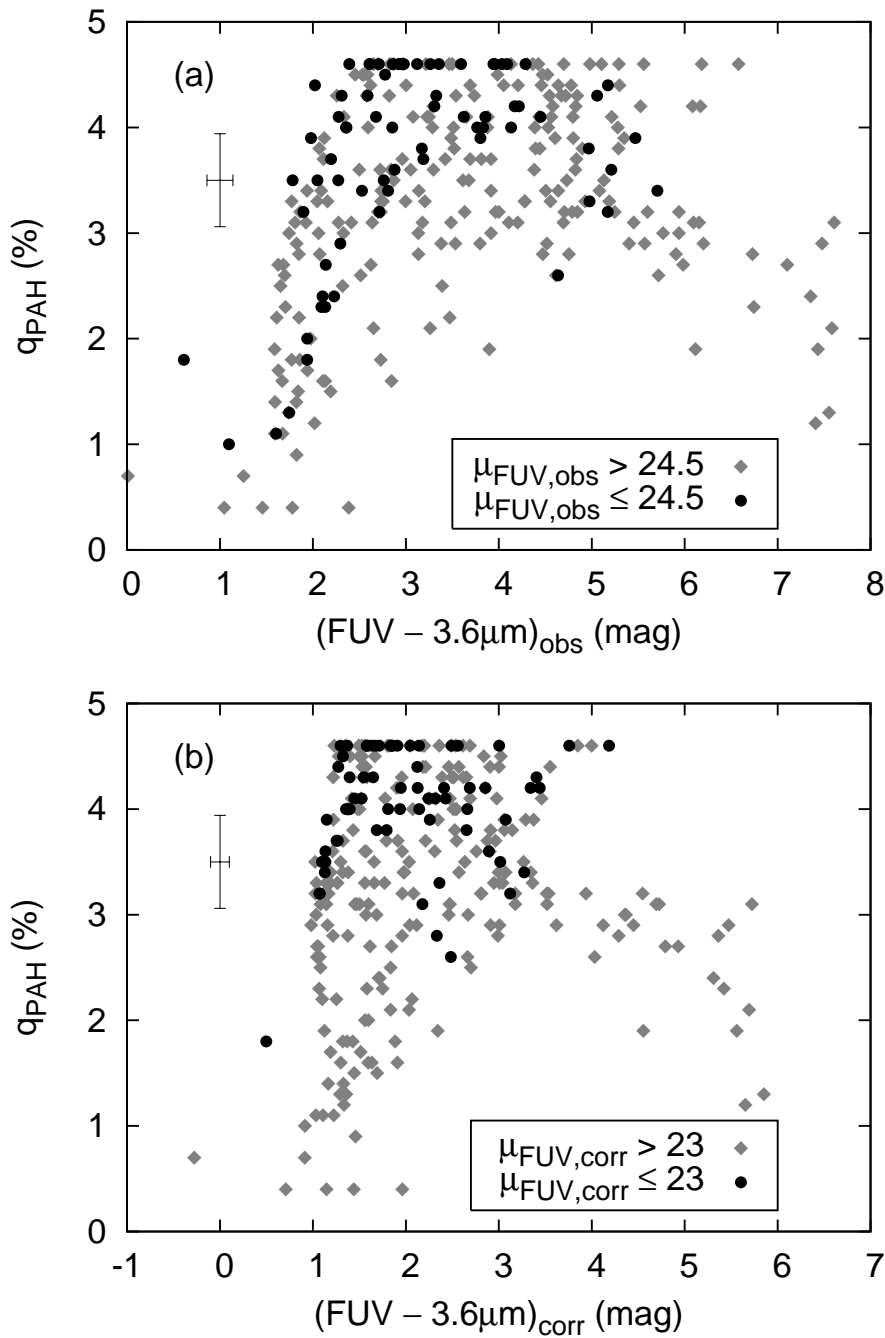}}
\caption{PAH abundance against the observed (a) and
extinction-corrected (b) $(\mathrm{FUV}-3.6\,\micron)$
color. Data-points are sorted out depending on their surface
brightness in the FUV band (in mag\,arcsec$^{-2}$). To allow for a
fair comparison between both panels, an offset of 1.5\,mag has been
applied to the surface brightness value chosen to classify the
data-points, since that is the typical FUV attenuation found in
our sample.\label{pah_fuv_irac}}
\end{figure}

In Fig.~\ref{pah_fuv_irac} we show how the PAH abundance changes with
the observed and extinction-corrected $(\mathrm{FUV}-3.6\,\micron)$
color. The color of the radiation field illuminating the PAHs probably
lies between the observed and extinction-corrected values because it
may be partially reddened by dust. Regions with the reddest colors,
usually the bulges of early-type spirals, exhibit low or intermediate
PAH abundances, as we already saw in Fig.~\ref{all_model_profs}. But
for $(\mathrm{FUV}-3.6\,\micron)_{\mathrm{corr}}$ colors between 0 and
4\,mag (the range in which star-forming disks are typically found, as
shown in Fig.~\ref{irxbeta}), PAHs progressively become less abundant
in the outer and bluer regions, although the trend is noisy and mainly
defined by a lower envelope. If photodestruction of PAH particles by
intense UV fields overcame PAH formation by AGB stars, it would be
conceivable to observe systematic differences in $q_{\mathrm{PAH}}$ by
sorting the data-points according to their FUV surface brightness
($\mu_{\mathrm{FUV}}$) for a fixed $(\mathrm{FUV}-3.6\,\micron)$
color, despite its limitations (see below). UV bright regions should
populate the lower-left area of Fig.~\ref{pah_fuv_irac} if PAHs are
being selectively destroyed in them, but they seem to have comparable
PAH abundances as the remaining regions, since these brightest zones
are usually the central, more evolved ones.

Since we are lacking spatially resolved information from ionization
tracers (such as the [NeIII]/[NeII] ratio), we are relying on the FUV
surface brightness as a rough proxy for the level of ionization. This
choice would be more justified in a pixel-to-pixel analysis, since
$\mu_{\mathrm{FUV}}$ would then allow us to identify regions of
massive star formation, but this analysis is beyond the scope of this
paper. However, the usefulness of $\mu_{\mathrm{FUV}}$ in radial
profiles as a ionization tracer is much more limited. It tends to
select the central and brightest zones of the disks, where the
evolutionary origin reflected by metallicity seems to be more
important. In the outer regions, however, the azimuthally averaged FUV
surface brightness is fainter, even if localized massive star
formation is present. In other words, the declining FUV profiles do
not imply that the UV field within individual star-forming regions
becomes softer as we move away from the center of the galaxy. Indeed,
it is worth noting that in M~33 the hardness of the UV radiation field
inside single HII regions is seen to increase with galactocentric
distance (Rubin et al$.$ 2008). Since our PAH abundances are not
limited to individual HII regions, but take into account all the
aromatic emission within elliptical annuli, the corresponding
ionization tracer should reflect the average radiation field heating
all that dust, which is likely to be softer than the one present in
the core of star-forming regions. With the available data,
$\mu_{\mathrm{FUV}}$ is possibly our best choice, but in any case the
results concerning the relative importance of PAH destruction are not
conclusive and should be treated with caution.

The whole analysis presented in this section supports the idea that,
at least in normal spiral galaxies and in spatial scales larger than
those of single HII regions, the observed trends between the PAH mass
fraction and metallicity or $(\mathrm{FUV}-3.6\,\micron)$ color may
have an evolutionary origin. We emphasize, however, that while the
varied timescales in which different elements are injected in the ISM
may be partly responsible of the observed dependence of the PAH mass
fraction on O/H, other processes governing the balance of dust
formation and destruction may be correlated with metallicity as well.

In their analysis of individual HII regions in M~101, Gordon et al$.$
(2008) find that the equivalent width of the aromatic features is
better correlated with the level of ionization than with metallicity,
and this seems to hold for the starburst galaxies studied by
Engelbracht et al$.$ (2008). Rather than representing a contradiction
with other studies (including ours), these results reflect diverse
behaviors associated with the different regimes of physical conditions
being probed. While photodestruction of PAH particles by hard UV
photons may overcome evolutionary effects in the core of HII regions
and in starburst galaxies, it does not seem to play such an important
role at larger spatial scales in our galaxies. Indeed, Gordon et al$.$
(2008) note that this trend with the hardness of the radiation field
only appears above a certain level of ionization. Below this level the
trend seems to vanish, implying that PAHs are not being destroyed
faster than the remaining dust species. This is likely the case in our
azimuthally averaged profiles, where we are measuring aromatic
features emerging not only from the very inside of HII regions, but
also from the diffuse ISM, where PAHs are not necessarily exposed to
such levels of ionization. From the SED decomposition shown in
Fig.~\ref{seds} we can see that, in general, the dust emission at 5.8
and 8.0\,$\micron$ is dominated by the diffuse component. Therefore,
at these spatial scales chemical evolution seems to act as the main
driver of the observed trends, rather than selective PAH
destruction. Recent analysis of the dust emission in NGC~0300 (Helou
et al$.$ 2004), NGC~4631 (Bendo et al$.$ 2006) and a subset of 15
SINGS spirals (Bendo et al$.$ 2008) seem to further support this
general picture. As in Gordon et al$.$ (2008), these authors found
that the PAH emission at 8\,$\micron$ and the hot dust radiation at
24\,$\micron$ are correlated in scales of a few kpc, but this
correlation breaks down within individual star-forming regions, where
the PAH emission is less centrally peaked than the one contributed by
very hot dust.

\subsubsection{Photometric 8\,$\micron$ equivalent width}

Another important issue to address is how the PAH abundance $-$as a
fraction of the total dust mass$-$ relates to the equivalent width of
the observed spectral aromatic features. Although 2D IRS spectral maps
are available for most of the SINGS galaxies, they do not usually
cover the full spatial extent of these objects. In this regard,
Engelbracht et al$.$ (2008) propose a photometric method to estimate
the equivalent width of the 7.7\,$\micron$ feature using the
stellar-subtracted flux at 8.0\,$\micron$ band, and a logarithmic
interpolation of the stellar-subtracted fluxes at 4.5 and
24\,$\micron$ to estimate the underlying continuum. Note that, as
pointed out by these authors, the 8\,$\micron$ equivalent width (EW,
either the spectroscopic or its photometric proxy) is a measurement of
the abundance of PAHs relative to small grains or very hot larger
ones, which contribute to the underlying continuum at those
wavelengths. Translating the 8\,$\micron$ EW into an abundance
relative to the {\it total} dust mass can be challenging without a
prior knowledge of the temperature distribution of the dust grains, as
we will see below.

It is not easy to obtain this photometric estimator for the
8\,$\micron$ EW directly from our data. Although recovering the
dust-only emission at 8.0 and 24\,$\micron$ by means of the stellar
factors presented in Section~\ref{fitting} does not pose any problem,
computing the non-stellar flux at 4.5\,$\micron$ is not
straightforward, since the emission at that band is almost entirely
dominated by stars (the sample studied by Engelbracht et al$.$ (2008)
consists of starburst galaxies usually exhibiting less stellar
contamination at that band). Indeed, in our sample the stellar
emission at 4.5\,$\micron$, extrapolated from the flux at
3.6\,$\micron$, is sometimes slightly larger than the actual observed
values (partly due to photometric uncertainties, and to the fact that
these stellar factors are average values). Alternatively, here we use
the non-stellar fluxes corresponding to the best-fitting model for
each galactocentric distance, instead of recovering them from the
observed data.

\begin{figure}
\resizebox{1\hsize}{!}{\includegraphics{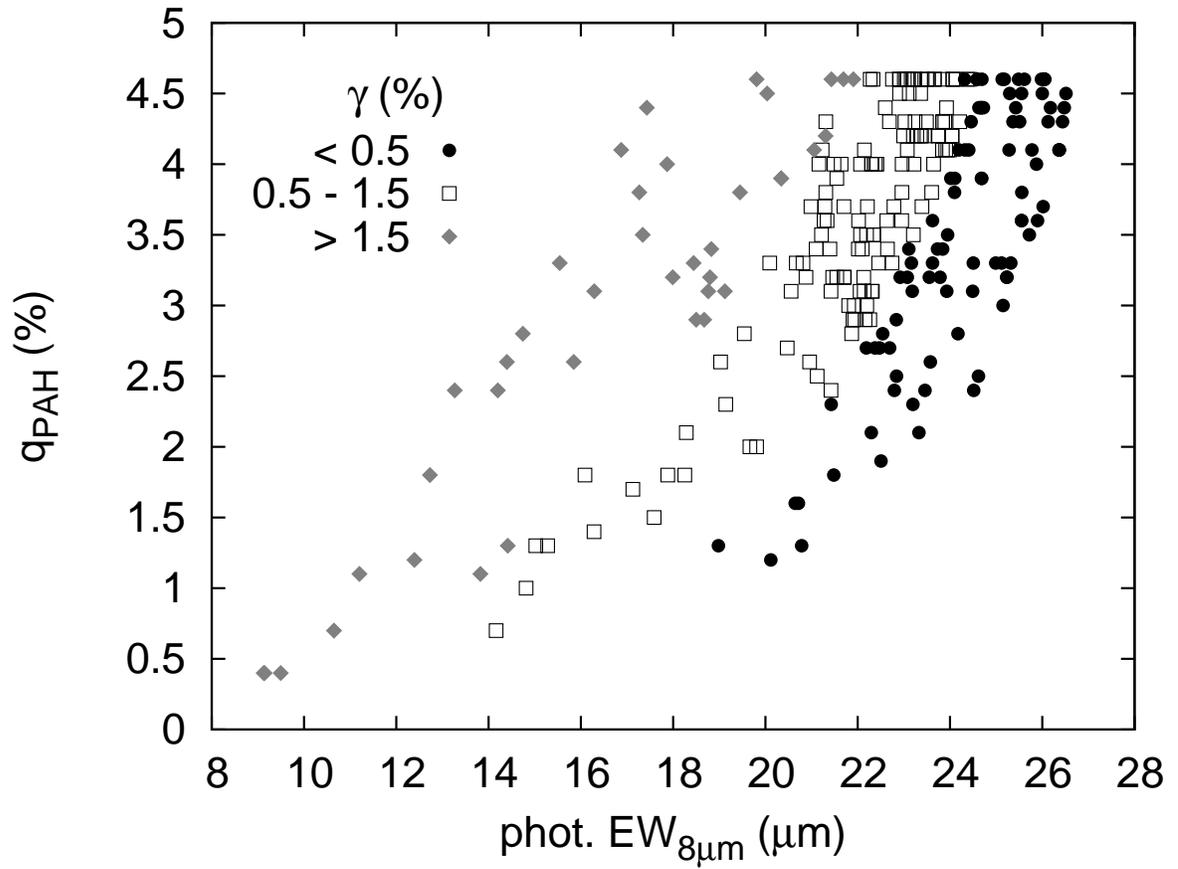}}
\caption{PAH abundance against the photometric estimator of the
8\,$\micron$ equivalent width, computed following the prescriptions
of Engelbracht et al$.$ (2008). Different symbols correspond to
different fractions of the total dust mass heated by very intense
starlight.\label{EW8um}}
\end{figure}

In Fig.~\ref{EW8um} we plot the PAH abundance derived from the models
against the photometric equivalent width estimated following the
prescriptions given by Engelbracht et al$.$ (2008). Both quantities
are positively correlated, but there is a clear dependence on the
fraction of dust exposed to intense starlight: at fixed
$q_{\mathrm{PAH}}$, the photometric proxy for the equivalent width
decreases as the fraction $\gamma$ of hot dust increases. This is most
likely due to the effect that $\gamma$ has on the flux at
24\,$\micron$: as seen in previous sections, exposing a small amount
of dust to very intense radiation fields can significantly increase
the observed emission at that band, and thus increase the interpolated
continuum at 8\,$\micron$. Therefore, the resulting equivalent width
will result in smaller values. In conclusion, the 8\,$\micron$ EW may
be used to compare the PAH mass fraction in samples of galaxies not
presenting large variations in their FIR SEDs, but should be used with
caution if this is not the case.

\subsubsection{Dust-to-gas ratio}\label{S_dust2gas}
The dust-to-gas ratio is an important quantity when studying the
chemical enrichment of the ISM, since it tells us about the amount of
metals that get locked up in the dust through the stellar
yields. Indeed, a correlation has been widely shown between the
dust-to-gas ratio and the gas-phase oxygen abundance (e.g$.$ Lisenfeld
\& Ferrara 1998; Edmunds 2001; James et al$.$ 2002; Hirashita et al$.$
2002; D07). In their analysis of a small sample of 8 nearby galaxies
(including the MW), Issa et al$.$ (1990) concluded that the
dust-to-gas ratio (as traced by the $A_{V}/N_{H}$ ratio) decreases
with radius, following a trend with metallicity. A similar result was
found by Boissier et al$.$ (2004), who confirmed that a similar trend
arises when the dust mass is directly estimated from the FIR emission,
rather than from the attenuation.

In order to obtain total gas profiles, we combined HI data from the
THINGS survey with CO profiles compiled from the literature (see
Table~\ref{sample_tab}), which were transformed into H$_{2}$ profiles
by means of the metallicity-dependent CO-to-H$_{2}$ conversion factor
of Boselli et al$.$ (2002). While this approach certainly constitutes
an improvement over using a constant conversion factor, the derived
H$_{2}$ profiles should be still treated cautiously. The CO-to-H$_{2}$
conversion factor may depend on parameters like the cloud density and
the excitation temperature of the CO, both of which could poentially
vary systematically with galactocentric radius. Also, note that
whereas the HI profiles were obtained in a consistent way with respect
to the UV and IR data by measuring on the HI intensity maps degraded
to the 160\,$\micron$ resolution, the CO profiles were directly
interpolated at different radii from published profiles. Hence, the CO
profiles might not necessarily match exactly the same spatial
regions. Note also that since no metallicity gradients are available
for NGC~4826 (Sab) and NGC~3627 (Sb), we used the CO-to-H$_{2}$
conversion factor corresponding to the global $H$-band luminosity of
each object (Boselli et al$.$ 2002).

\begin{figure}
\resizebox{0.85\hsize}{!}{\includegraphics{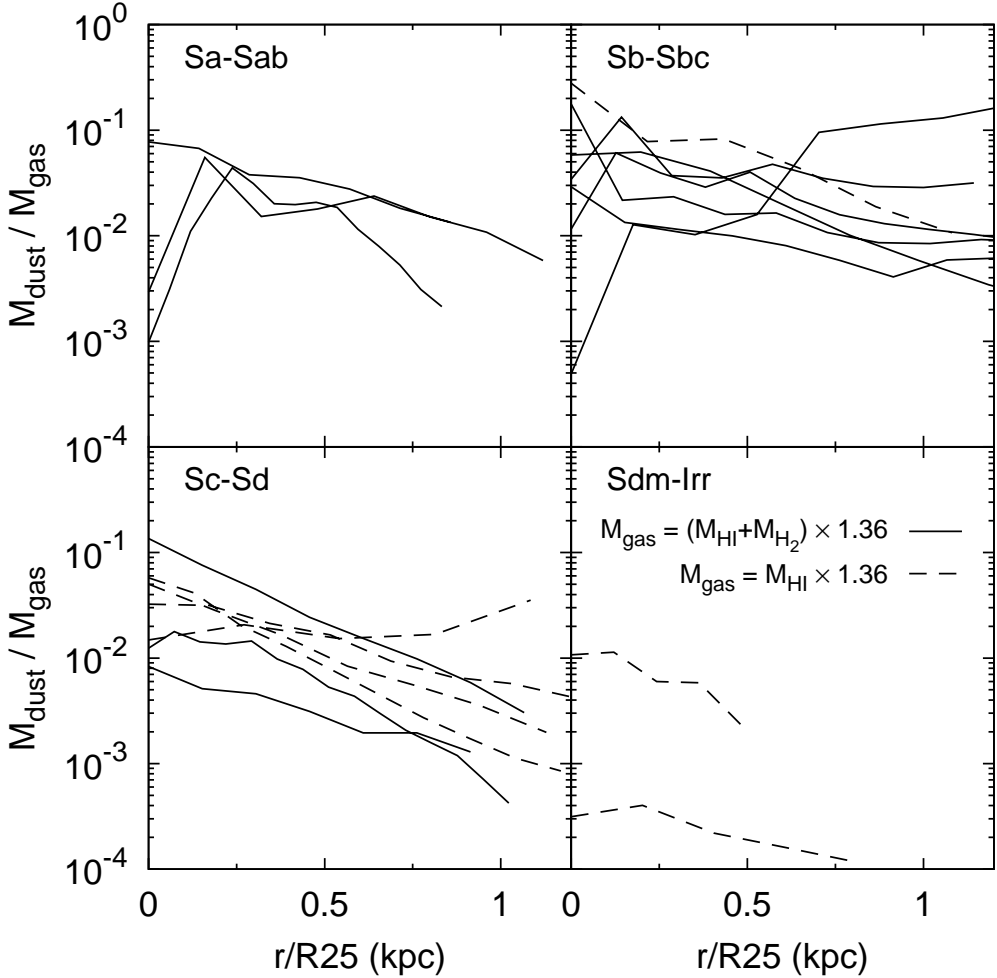}}
\caption{Radial variation of the dust-to-gas ratio for the
SINGS-THINGS galaxies, divided into different bins of morphological
types. The radial axis is normalized to the optical size of each
galaxy, as given by the R25 radius. Galaxies with available HI and CO
profiles are shown with solid lines, whereas those lacking CO data
appear as dashed lines (their profiles are then upper limits, although
most of the molecular gas is expected to be located in the innermost
regions). In all but two galaxies a metallicity-dependent
CO-to-H$_{2}$ conversion factor changing with radius was used (see
text). In all plots throughout this paper explicitly involving (O/H),
we only trust metallicity values within a restricted spatial range
defined by the HII regions used to determine the metallicity
gradients. In this figure, however, the accuracy of the oxygen
abundance is less critical, since it constitutes a second-order effect
in comparison with the assumption of a constant CO-to-H$_{2}$
factor. Therefore, in this particular plot we have not applied these
restrictions to the metallicity values when computing the H$_{2}$ mass
surface densities.\label{dust2gas_prof}}
\end{figure}

In Fig.~\ref{dust2gas_prof} we show the radial dust-to-gas profiles
for the SINGS-THINGS galaxies, sorted by morphological type. Those
galaxies for which only HI data are available are plotted with a
dashed-line, and their dust-to-gas ratios are therefore just upper
limits, at least in the innermost regions where molecular gas is more
likely to be found.

In most objects the dust-to-gas ratio decreases with radius. In the
central regions of Sb-Sd spirals it ranges between $10^{-1}$ and
$10^{-2}\mathrm{\,M}_{\sun}\mathrm{\,kpc}^{-2}$, and it decreases
faster in Sc-Sd ones, reaching values close to
$10^{-3}\mathrm{\,M}_{\sun}\mathrm{\,kpc}^{-2}$ or lower at the
optical radius. NGC~3627 constitutes a notable exception, for its
dust-to-gas ratio increases with radius. It is an interacting Sb
spiral in the Leo triplet that gets redder in the outer regions,
possibly because of SF quenching as a result of gas removal or
exhaustion. NGC~2976 is a Sc galaxy whose dust to gas ratio remains
roughly constant across the entire optical size. This quite unusual
object exhibits two bright knots of intense star formation located at
almost symmetrical positions with respect to the center of the galaxy.

\begin{figure}
\resizebox{1\hsize}{!}{\includegraphics{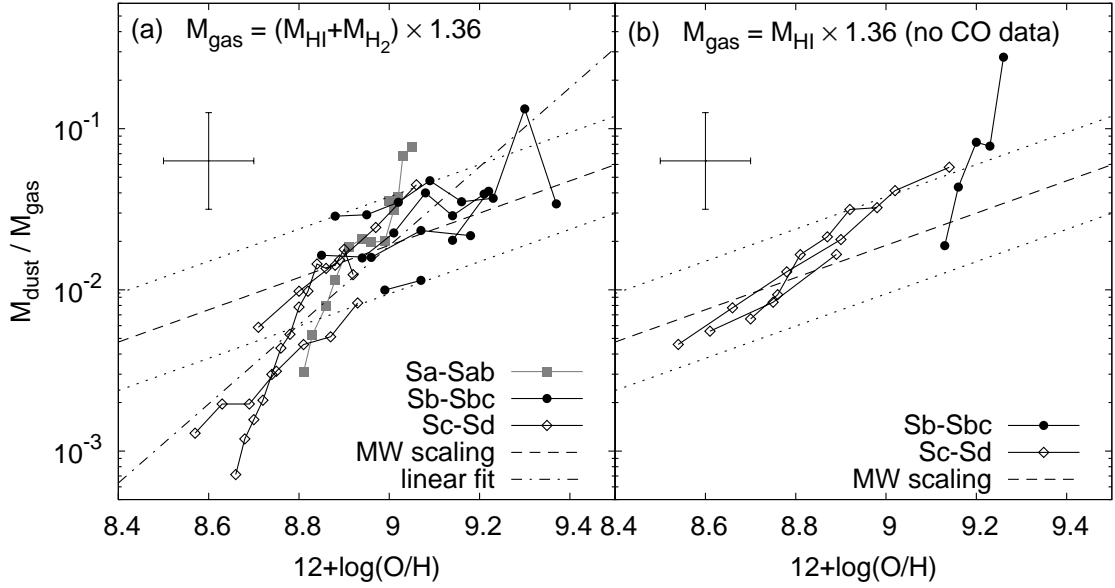}}
\caption{Dust-to-gas ratio for the SINGS-THINGS galaxies against the
metallicity at different galactocentric distances. Data-points
belonging to the same galaxy have been connected. Different symbol
shapes correspond to different morphological types. The dashed line
marks the simple scaling law described by Eq.~\ref{MW_scaling},
whereas the dotted lines show variations of a factor of 2 around
it. The dot-dashed line results from a linear fit to the data. (a):
Galaxies for which both HI and CO data are available. (b): Objects
lacking CO data. The values at the innermost (i.e$.$ more metallic)
regions are upper limits.\label{dust2gas_metal}}
\end{figure}

In Fig.~\ref{dust2gas_metal}a we plot the dust-to-gas ratio against
the gas-phase oxygen abundance, for all the SINGS galaxies with
available spatially resolved metallicities, HI and CO data. Points
belonging to the same object have been connected for clarity. If we
were to assume that the abundances of all heavy elements were
proportional to the oxygen abundance and that all heavy elements
condensed to form dust in the same way as in the MW, then the
dust-to-gas ratio should scale proportionally to the oxygen abundance
(D07):
\begin{equation}
\frac{M_{\mathrm{dust}}}{M_{\mathrm{gas}}} \approx \frac{0.01}{1.36}\frac{(\mathrm{O/H})}{(\mathrm{O/H})_{\mathrm{MW}}}\label{MW_scaling}
\end{equation}
where 0.01 is the dust-to-hydrogen ratio of the MW and the factor 1.36
accounts for helium and heavier elements. The dashed line in
Fig.~\ref{dust2gas_metal}a marks this relation and shows that, at
least to first order and for $12+\log(\mathrm{O/H})>8.9$, the radial
correlation between the dust-to-gas ratio and the metallicity follows
this simple scaling law within a factor of $\sim 2$. In
Fig.~\ref{dust2gas_metal}b we plot the galaxies for which only HI data
are available. Their values of $M_{\mathrm{dust}}/M_{\mathrm{gas}}$
might be overestimated if significant amounts of molecular gas are
present (which is likely the case in the central regions).

We call attention to the fact that at low oxygen abundances the
observed trend in panel (a) seems to be steeper than the one predicted
by Eq.~\ref{MW_scaling}. A linear fit to the whole data-set yields:
\begin{equation}
\log(M_{\mathrm{dust}}/M_{\mathrm{gas}})=5.63+2.45 \times \log(\mathrm{O/H})
\end{equation}

In their analysis of the integrated photometry of the SINGS galaxies,
D07 found that galaxies with and without available submillimeter data
followed slightly different trends in the
$M_{\mathrm{dust}}/M_{\mathrm{gas}}$ vs. metallicity plots, with the
slope being a bit steeper when the dust masses had been obtained
without submillimeter data.

It is not clear, therefore, if the apparent departure from the MW
scaling law seen in Fig.~\ref{dust2gas_metal} at the lowest
metallicities is real, or due to our cool dust masses being
underestimated (the metallicity gradients might be also
uncertain). Should this trend be real, it could link the behavior of
the outermost, lower metal-abundant regions of spirals with the
properties of dwarf galaxies, which exhibit lower dust-to-gas ratios
than those predicted by extrapolating Eq.~\ref{MW_scaling}. Detailed
models describing dust evolution in dwarf galaxies (see Edmunds 2001;
Hirashita et al$.$ 2002) usually invoke several mechanisms to either
delay dust injection in the ISM or to eliminate it, via outflows
caused by collective SN-driven winds. However, the extremely low
dust-to-gas ratios of many dwarf galaxies could be partly due to their
extended HI envelopes, where no dust emission is apparently seen (see
e.g$.$ Walter et al$.$ 2007). Indeed, when
$M_{\mathrm{dust}}/M_{\mathrm{gas}}$ is computed inside regions
detected both in FIR and HI, the values derived are more consistent
with Eq.~\ref{MW_scaling} (D07).

The extent to which these explanations proposed for dwarf galaxies are
also valid for the outer regions of spirals remains unclear. The
radial decrease of the star-formation efficiency found in nearby
spirals (Thilker et al$.$ 2007; Leroy et al$.$ 2008) could also partly
explain the low dust-to-gas ratios seen in
Fig.~\ref{dust2gas_metal}. The presence of large reservoirs of gas
which has not been yet transformed into stars (and hence into dust)
would drop the dust-to-gas ratio in the outer regions of these
galaxies.

\section{Summary and conclusions}\label{summary}
We have measured multi-wavelength surface brightness profiles for the
galaxies included in the SINGS sample, using UV, IR and HI data from
GALEX, {\it Spitzer} (from SINGS) and VLA (from the THINGS survey),
complemented with published CO data and metallicity gradients. The
analysis of this panchromatic, spatially resolved data-set reveals
important results regarding the dust properties of nearby galaxies:
\begin{enumerate}
\item We have derived radial extinction profiles by means of a
SFH-dependent calibration of the TIR-to-FUV ratio. In most galaxies
the attenuation decreases with the galactocentric distance. It is
largest in Sb-Sbc galaxies, ranging from $A_{\mathrm{FUV}} \sim
2.5$\,mag in the central regions to $\sim 1.5$\,mag at the optical
radius R25. As we move towards earlier or later Hubble types, the
attenuation decreases and its radial variation becomes less
pronounced. S0/a-Sab spirals typically have $A_{\mathrm{FUV}} \sim
1.5$\,mag, and in Sdm and irregular galaxies the extinction drops
below 0.5\,mag. However, the dispersion is rather large ($\sim
1-2$\,mag) within each bin of morphological types. This is the result
of the different spatial distribution of stars and dust (see below).

\item The TIR-to-FUV excess is correlated with the UV spectral slope
$\beta$, following an IRX-$\beta$ relation shifted with respect to the
one found in starburst galaxies. The main driver of this departure
seems to be the SFH, with more quiescent regions having redder UV
slopes for a given TIR-to-FUV ratio. Indeed, the intrinsic
$(\mathrm{FUV}-3.6\micron)$ color $-$a metric tracer of the
current-to-past SFR$-$ correlates well with the distance from the
relation for starbursts. Our radial analysis reveals that this shift
is particularly evident in bulges, as well as in the outer regions of
anemic spirals with clear signs of star-formation quenching. However,
second-order factors might also play a role, since the dependence on
the SFH gets completely blurred unless a wide range of SFHs is
explored. We also provide a calibration to estimate the UV attenuation
from the UV slope when FIR data are unavailable, which is
statistically valid for `normal' star-forming galaxies.

\item By combining our data with oxygen abundance gradients, we have
analyzed the influence of metallicity on attenuation. Both quantities
are clearly correlated, with metal-richer regions exhibiting larger
attenuations. Inclination plays an important role here; at fixed
metallicity, the extinction is larger in galaxies that are closer to
edge-on. The same happens with the UV color: more inclined galaxies
appear redder than nearly face-on ones. Both effects are roughly
balanced when combining both quantities in the IRX-$\beta$ plot, so
that data-points are just shifted along the IRX-$\beta$ relationship.

\item We have fit the FIR profiles with the dust models of Draine \&
Li (2007), deriving the radial distribution of the PAH abundance, the
dust mass and luminosity surface densities and the properties of the
heating starlight. The fraction of the dust mass in the form of PAHs
($q_{\mathrm{PAH}}$) varies widely along the Hubble sequence, reaching
a maximum of $\sim 3-4.5$\% in Sb-Sbc spirals, and almost disappearing
($\lesssim 1$\%) in the outer regions of Sdm and irregular
galaxies. Indeed, the PAH abundance typically decreases with radius in
Sc spirals and later; in earlier types, however, the profiles tend to
flatten and even reverse. The PAH abundance increases with metallicity
for $8.5<12+\log(\mathrm{O/H})<9$, and the trend progressively
flattens and eventually reverses at larger oxygen abundance. This is
in qualitative agreement with the predictions of models coupling
stellar evolution with dust formation. Although this dependence on
metallicity has already been found when comparing normal spirals with
dwarf galaxies, using radial profiles leads to a considerable
reduction of the dispersion and carrying out the study in a narrower
range in metallicity, without the need to include low-metallicity
dwarfs.

\item The average intensity of the heating starlight decreases
monotonically with the galactocentric distance, and in the outer
regions it usually reaches similar values as the local MW radiation
field, corresponding to cold dust temperatures of $\sim 20$\,K. In
order to explain the observed 24\,$\micron$ flux densities, it is
required that about 1\% of the total dust mass must be exposed to very
intense radiation fields. Such a small amount of dust can account for
$\sim 10$\% of the total dust IR luminosity.

\item The dust mass surface density is largest in Sb-Sbc spirals,
varying from $10^{6}$ to
$10^{4}\mathrm{\,M}_{\sun}\mathrm{\,kpc}^{-2}$ from the center to the
R25 radius. Dust is less abundant in earlier spirals, typically having
surface densities below
$10^{5}\mathrm{\,M}_{\sun}\mathrm{\,kpc}^{-2}$, and also in Sdm and
later, where it drops below
$10^{4}\mathrm{\,M}_{\sun}\mathrm{\,kpc}^{-2}$. The dust profiles are
exponential, and have a radial scale-length that holds remarkably
constant (0.2-0.3\,R25) from Sb to Sd galaxies. The dust profiles in
S0/a and Sab spirals usually present a central depletion, but have
similar radial scale-lengths than Sb-Sd ones in the outer exponential
regions. Compared to the radial distribution of old stars, as traced
by the 3.6\,$\micron$ profiles, the dust radial scale-length is, on
average, $\sim 10$\% larger than the stellar one.

\item The relative spatial distribution of stars and dust plays a
crucial role in determining the attenuation. There is a wide range of
attenuations (1-2\,mag) that can be found for a given observed (i.e$.$
projected) dust mass surface density. The minimum values of
attenuation are always found in Sc galaxies and later, while for the
same amount of dust earlier spirals (especially Sb-Sbc ones) exhibit
larger extinctions. In late-type spirals, dust and stars are arranged
in a clumpy geometry: while some heating sources can be heavily
attenuated, many UV photons can leak through lines of sight clean of
dust, thus decreasing the overall attenuation. This does not happen so
easily in the disks of Sbc spirals and earlier, where the same dust
surface density yields larger attenuations.

\item By merging the dust mass profiles with HI data from the THINGS
survey and CO data from the literature, we have studied the radial
variation of the dust-to-gas ratio, which is seen to decrease by an
order of magnitude from the center to the edge of the optical disk of
each galaxy. Typical values range from $0.1$\,-\,$0.01$ in the central
regions to $0.01$\,-\,$0.001$ in the outer ones, with early-type
spirals having larger amounts of dust than late-type ones with a
similar gas content. The dust-to-gas ratio is clearly correlated with
the gas-phase oxygen abundance, being larger in the central,
metal-rich zones of the disks than in the outer ones. To first order,
for $12+\log(\mathrm{O/H}>8.9)$ the relation can be described with a
simple scaling law: $M_{\mathrm{dust}}/M_{\mathrm{gas}} \propto
\mathrm{(O/H)}$, which would imply that most condensable elements tend
to form dust in a roughly similar way in spiral galaxies of different
morphological types. However, at lower metallicities the dust-to-gas
ratios are systematically below this simple relation. This could imply
that large amounts of gas that has not yet undergone star-formation
activity reside in the outer regions of spiral disks, although the
presence of dust cooler than 15\,K could increase our dust-to-gas
ratios.

\end{enumerate}

\acknowledgments

JCMM acknowledges the receipt of a Formaci\'on del Profesorado
Universitario fellowship from the Spanish Ministerio de Educaci\'on y
Ciencia. JCMM, AGdP, JZ, PGP and JG are partially financed by the
Spanish Programa Nacional de Astronom\'{\i}a y Astrof\'{\i}sica under
grant AYA2006-02358. AGdP is also financed by the MAGPOP EU Marie
Curie Research Training Network. We thank the anonymous referee for
valuable comments that have improved the paper. We have made use of
the NASA/IPAC Extragalactic Database (NED), which is operated by the
Jet Propulsion Laboratory, California Institute of Technology
(Caltech) under contract with NASA. GALEX (Galaxy Evolution Explorer)
is a NASA Small Explorer, launched in 2003 April. We gratefully
acknowledge NASA's support for construction, operation, and science
analysis for the GALEX mission, developed in cooperation with the
Centre National d'\'Etudes Spatiales of France and the Korean Ministry
of Science and Technology. This work is part of SINGS, the {\it
Spitzer} Infrared Nearby Galaxies Survey. The {\it Spitzer} Space
Telescope is operated by the Jet Propulsion Laboratory, Caltech, under
NASA contract 1403. This work also makes use of data from THINGS, The
HI Nearby Galaxies Survey. We thank A$.$ Leroy and F$.$ Walter for
kindly providing the HI radial profiles. We also thank L$.$ Cortese
for providing his fit for Fig.~\ref{ext_metal}.

{\it Facilities:} \facility{GALEX}, \facility{{\it Spitzer}}, \facility{VLA}

\appendix
\section{Empirical estimations of the model parameters}\label{app_empirical}
Fitting the infrared SEDs of star-forming galaxies with detailed
models for the dust emission constitutes the proper way to analyze the
physical properties of the interstellar dust. However, it is also
desirable to have empirical calibrations from which one can roughly
estimate these properties directly from observed quantities. Here we
provide some basic equations for that purpose. In
Table~\ref{recipes_tab} we provide the residual rms of each one of
these recipes.

\subsection{Estimating $\gamma$}
Draine \& Li (2007) already provide a useful fit to estimate
$f(L_{d};U>10^{2})$, that is, the fraction of the total dust
luminosity contributed by high-intensity regions. They define the
following quantity:
\begin{equation}
P_{24}-0.14P_{8.0} \equiv \frac{\langle \nu F_{\nu}^{\mathrm{ns}} \rangle_{24}-0.14\langle \nu F_{\nu}^{\mathrm{ns}} \rangle_{8.0}}{\langle \nu F_{\nu} \rangle_{70}+\langle \nu F_{\nu} \rangle_{160}}
\end{equation}
where the nonstellar flux densities $F_{\nu}^{\mathrm{ns}}$ can be
estimated using the stellar emission corrections given in
Section~\ref{fitting}. The numerator is the 24\,$\micron$ flux coming
from large grains heated by very intense radiation fields. The factor
$0.14\langle \nu F_{\nu}^{\mathrm{ns}} \rangle_{8.0}$ accounts for the
emission at 24\,$\micron$ contributed by stochastically heated PAHs,
since this emission is not necessarily linked to high-intensity
regions. DL07 showed that $f(L_{d};U>10^{2})$ can be estimated using
the following fitting function:
\begin{equation}
f(L_{d};U>10^{2})(\%)=105(P_{24}-0.14P_{8.0}-0.035)^{0.75}
\end{equation}

\begin{figure}
\resizebox{1\hsize}{!}{\includegraphics{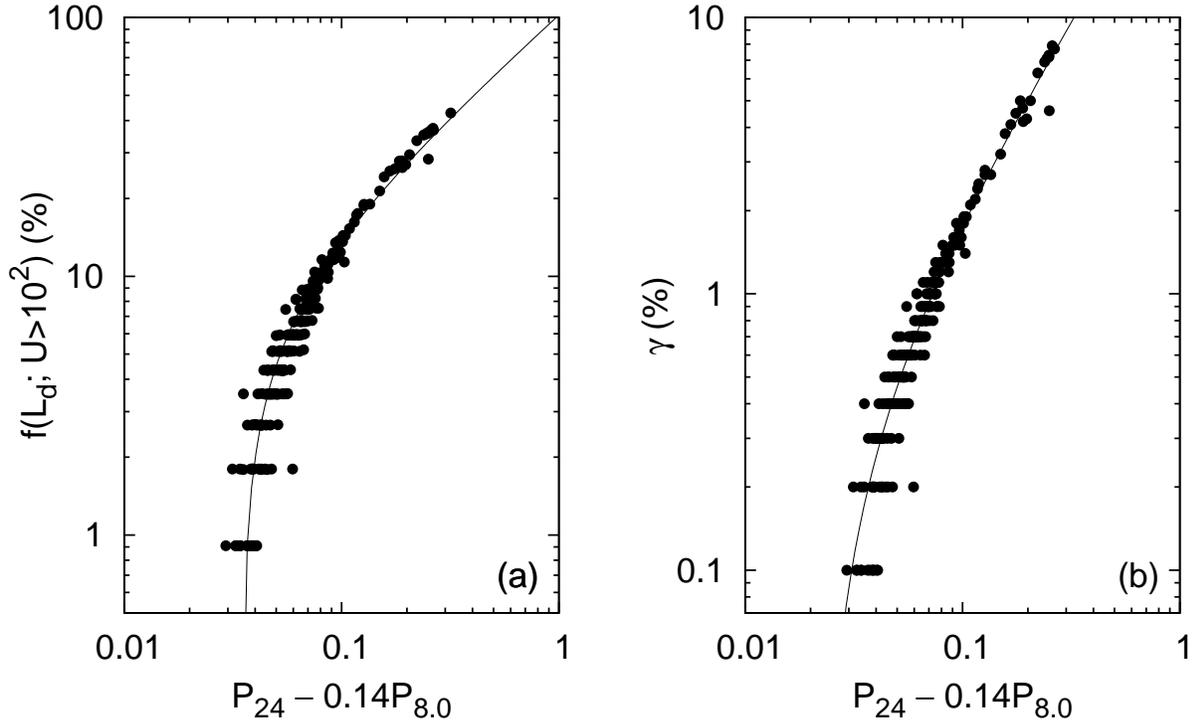}}
\caption{(a): Fraction of the total dust luminosity contributed by
grains in regions where $U>10^{2}$ as a function of the observable
$P_{24}-0.14P_{8.0}$ (see text). The solid line is the theoretical
fitting function provided by Draine \& Li (2007). (b): Relation
between the fraction of the dust mass enclosed in high-intensity
regions and $P_{24}-0.14P_{8.0}$. The solid line is the best
fit-function of our radial profiles.\label{gamma_empirical}}
\end{figure}

While DL07 showed that this function is certainly a good proxy for the
fraction of the dust luminosity coming from high-intensity regions,
Fig.~\ref{gamma_empirical}a shows that this is also the case for our
radial profiles. The dispersion in the lower region is due to
differences in $U_{\mathrm{min}}$ and $q_{\mathrm{PAH}}$.

The corresponding mass fraction enclosed in these regions is also
tightly correlated with $P_{24}-0.14P_{8.0}$, as can be seen in panel
(b). Following DL07, we performed a similar fit to obtain an estimator
for $\gamma$:
\begin{equation}
\gamma\ (\%)=46.2(P_{24}-0.14P_{8.0}-0.023)^{1.28}
\end{equation}

\subsection{Estimating $q_{\mathrm{PAH}}$}\label{S_qpah_empirical}
\begin{figure}
\resizebox{1\hsize}{!}{\includegraphics{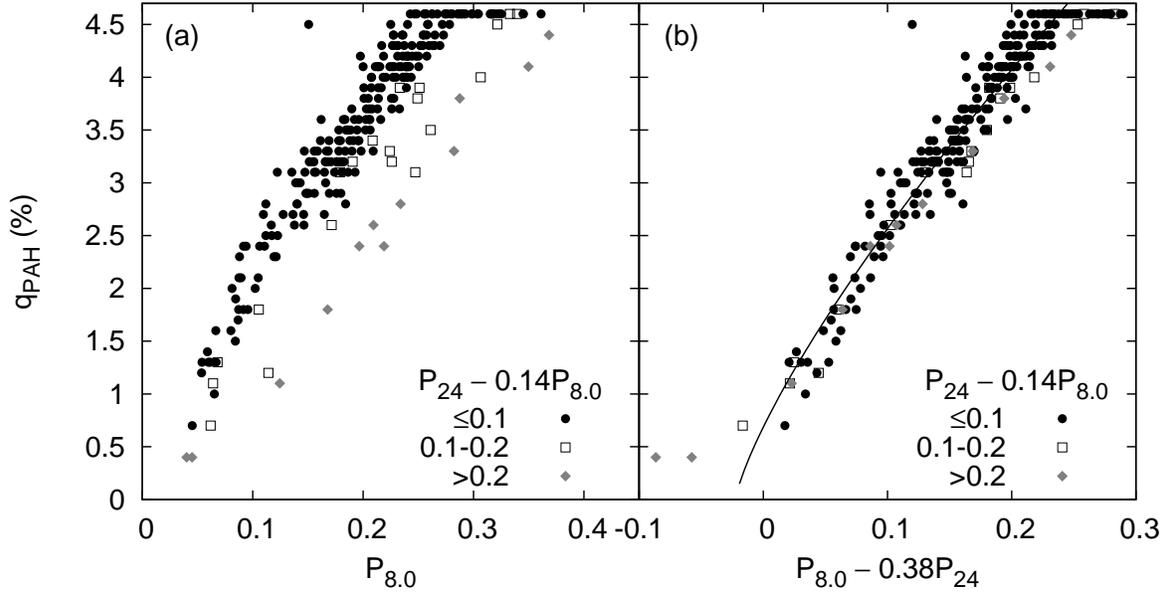}}
\caption{(a): Fraction of the dust mass contributed by PAHs as a
function of $P_{8.0}$ (see text). The data-points have been divided
into three different bins, according to their value of
$P_{24}-0.14P_{8.0}$, which is a proxy for the fraction of dust mass
heated by intense starlight. (b): PAH abundance as a function of a
linear combination of $P_{8.0}$ and $P_{24}$, resulting from a
Principal Component Analysis of the data. The best-fit line is also
shown. Data-points whose values of $q_{\mathrm{PAH}}$ saturate at the
lower and upper limits of the dust model (0.4\% and 4.6\%,
respectively) were not taken into account when performing the
fit. \label{qpah_empirical}}
\end{figure}

The emission at 8.0\,$\micron$ depends on both the PAH abundance and
the amount of dust exposed to high-intensity starlight. Therefore, one
would expect that $q_{\mathrm{PAH}}$ could be obtained from $P_{8.0}$
and $P_{24}-0.14P_{8.0}$, the latter accounting for differences in
$\gamma$, as seen above. This is confirmed in
Fig.~\ref{qpah_empirical}a, where we can see that the PAH abundance
correlates well with $P_{8.0}$, with a second order dependence on
$P_{24}-0.14P_{8.0}$. As expected, at fixed $q_{\mathrm{PAH}}$ the
emission at 8.0\,$\micron$ $-$relative to the emission at longer
wavelengths$-$ increases with $\gamma$.

A close inspection of data-points distribution reveals that a certain
linear combination of $P_{8.0}$ and $P_{24}$ would allow estimating
$q_{\mathrm{PAH}}$ with a single fit. We computed this linear
combination by means of a Principal Component Analysis. The results
are shown in panel (b), where we can see that the PAH abundance can be
readily estimated using the following fitting function:
\begin{equation}
q_{\mathrm{PAH}}\ (\%)=13.03(P_{8.0}-0.38P_{24}+0.022)^{0.77}
\end{equation}

Since the largest value of the PAH abundance derived from the models
is $q_{\mathrm{PAH}}=4.6$\%, there are many data-points saturating
that region of the plot $-$their actual PAH abundances being possibly
larger$-$ so we excluded those points in this analysis. For the same
reason, points with $q_{\mathrm{PAH}}=0.4$\% were also excluded, since
this is the lower limit of the dust models. Nevertheless, our fit
should still allow us to recover PAH abundances a few percent larger
than the current upper limit, since there is no reason a priori for
this trend to break just above that value.

\subsection{Estimating $\langle U \rangle$}\label{S_Umean_empirical}
\begin{figure}
\resizebox{1\hsize}{!}{\includegraphics{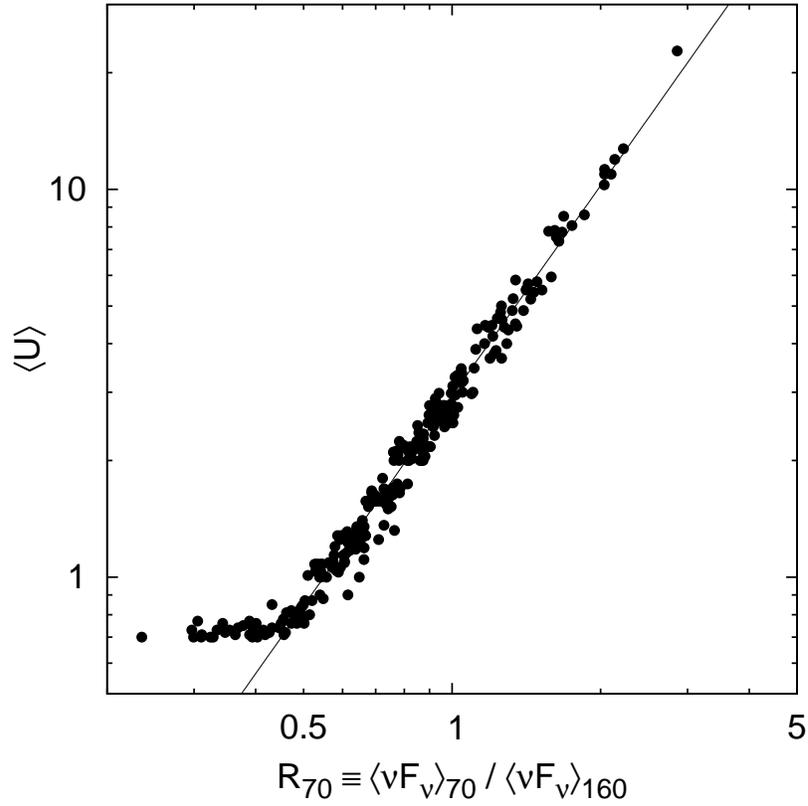}}
\caption{Scale factor of the dust-weighted mean starlight intensity
$-$in units of the local MW radiation field$-$ against the ratio of
the luminosities at 70 and 160\,$\micron$. The best fitting line is
also shown. Data-points with $\langle U \rangle \approx 0.7$ have been
excluded from the fit (see text for details).\label{Umean_empirical}}
\end{figure}

The scale factor of the dust-weighted mean starlight intensity,
$\langle U \rangle$, depends strongly on the shape of the FIR SED. In
Fig.~\ref{Umean_empirical} we show that this parameter is linearly
correlated $-$in logarithmic scale$-$ with $R_{70} \equiv \langle \nu
F_{\nu} \rangle_{70} / \langle \nu F_{\nu} \rangle_{160}$, which is
sensitive to the temperature of the largest grains dominating the FIR
emission. A linear fit yields the following empirical function:

\begin{equation}
\log \langle U \rangle = 0.468+1.801 \log R_{70}\label{eq_Umean}
\end{equation}

The starlight intensity of the diffuse component, $U_{\mathrm{min}}$,
is not so tightly correlated with $R_{70}$, since the effect of
$\gamma$ must be taken into account. Once $\langle U \rangle$ and
$\gamma$ have been estimated using the formulae above,
$U_{\mathrm{min}}$ can be computed from Eq.~33 in DL07.

Note that, as explained in Section~\ref{models}, the smallest value of
$U_{\mathrm{min}}$ considered in the model-fitting procedure is 0.7,
as DL07 suggest to proceed when lacking submillimeter data. As a
consequence, the resulting fits also have $\langle U \rangle \geq
0.7$. For this reason, some data-points are clustered in the lower
region of Fig.~\ref{Umean_empirical}, and have not been taken into
account when performing the fit.

\subsection{Estimating the dust mass}\label{S_Mdust_empirical}
We can estimate the dust mass from the mean starlight intensity
$\langle U \rangle$ and the dust luminosity, since
\begin{equation}
L_{\mathrm{dust}}=P_{0} \langle U \rangle M_{\mathrm{dust}}\label{eq_Ldust}
\end{equation}
where $P_{0}$ is the power per unit of mass radiated by the dust, when
exposed to a radiation field equal to the local MW one (i.e$.$ with
$U=1$). According to the DL07 models, differences in the PAH abundance
can introduce variations of $\sim 2\%$ in this
parameter. Nevertheless, a constant average value of $P_{0} \approx
137 L_{\sun}/M_{\sun}$ can be safely adopted.

The dust luminosity can be estimated directly from the photometric
data using the weighted sum proposed by DL07. By substituting that sum
and the empirical fit for $\langle U \rangle$ (Eq.~\ref{eq_Umean})
into Eq.~\ref{eq_Ldust}, we can derive the dust mass from the IRAC and
MIPS flux densities and the distance to the source:
\begin{eqnarray}
\frac{M_{\mathrm{dust}}}{M_{\sun}}&=&\frac{4 \pi}{1.616 \times 10^{-13}} \left(\frac{D}{\mathrm{Mpc}}\right)^{2} \left(\frac{\langle \nu F_{\nu} \rangle_{70}}{\langle \nu F_{\nu} \rangle_{160}}\right)^{-1.801} \nonumber\\
&& \times \frac{0.95 \langle \nu F_{\nu}^{ns} \rangle_{8.0} + 1.150 \langle \nu F_{\nu}^{ns} \rangle_{24} + \langle \nu F_{\nu} \rangle_{70} + \langle \nu F_{\nu} \rangle_{160}}{\mathrm{erg\ s^{-1}\ cm^{-2}}}\label{eq_Mdust}
\end{eqnarray}

\begin{figure}
\resizebox{1\hsize}{!}{\includegraphics{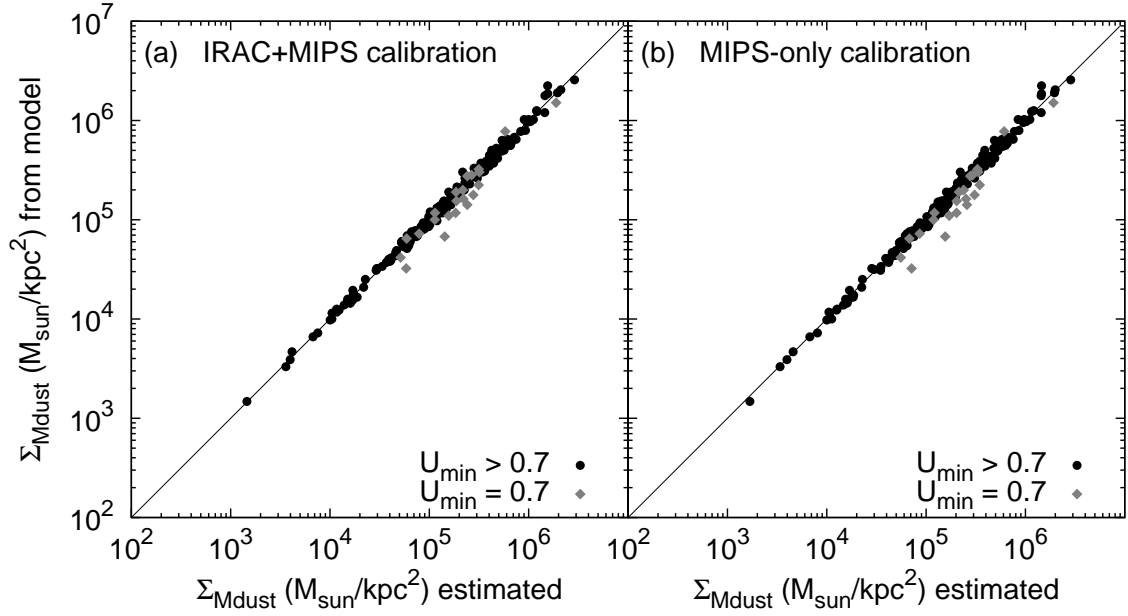}}
\caption{Comparison of the dust mass surface densities obtained from
the model-fitting technique and from the estimator given in
Appendix~\ref{S_Mdust_empirical}, either with the IRAC+MIPS
calibration (Eq.~\ref{eq_Mdust}, panel a) or with the MIPS-only one
(Eq.~\ref{eq_Mdust_no8um}, panel b). Note that since we are dealing
with surface densities, the results do not depend on the distance to
the sources. Regions for which the fitting procedure `saturates' at
the lowest value of the diffuse starlight intensity $U_{\mathrm{min}}$
have been marked.\label{Mdust_empirical}}
\end{figure}

As in previous sections, the non-stellar flux densities $F^{ns}_{\nu}$
can be obtained from the observed ones by extrapolating the stellar
continuum from the observed 3.6\,$\micron$ flux density (see
Section~\ref{fitting}). To check the validity of this estimator, we
have applied it to our radial profiles. In Fig.~\ref{Mdust_empirical}a
we compare the dust mass surface densities obtained from the
model-fitting procedure with the ones estimated through
Eq.~\ref{eq_Mdust}. The agreement is excellent, with a scatter of only
9\%. The most deviant points are those with the smallest value of the
diffuse starlight intensity considered throughout this work (see
Appendix~\ref{S_Umean_empirical}). Note that there is a small offset
between the dust masses derived from Eq.~\ref{eq_Mdust} and the actual
ones obtained from the model fitting, due to the fact that the
$L_{\mathrm{dust}}$ estimator slightly overestimates the actual
$L_{\mathrm{dust}}$ from the model for the particular range of values
of $\langle U \rangle$ and $q_{\mathrm{PAH}}$ in our
galaxies. Although Eq.~\ref{eq_Mdust} is intended to be more general,
if one wishes to account for this fact, the dust masses derived from
Eq.~\ref{eq_Mdust} should be multiplied by 0.95 (see
Table~\ref{recipes_tab}).

In the absence of IRAC data, we can still compute the dust mass simply
by substituting the $L_{\mathrm{TIR}}$ estimator of DL07 (i.e$.$ the
numerator of the last term in Eq.~\ref{eq_Mdust}) with the MIPS-only
calibration of Dale \& Helou (2002). We can also skip the stellar
emission correction for the 24\,$\micron$ band, since it is almost
negligible. Hence, we have:
\begin{eqnarray}
\frac{M_{\mathrm{dust}}}{M_{\sun}}&=&\frac{4 \pi}{1.616 \times 10^{-13}} \left(\frac{D}{\mathrm{Mpc}}\right)^{2} \left(\frac{\langle \nu F_{\nu} \rangle_{70}}{\langle \nu F_{\nu} \rangle_{160}}\right)^{-1.801} \nonumber \\
&& \times \frac{1.559 \langle \nu F_{\nu} \rangle_{24} + 0.7686 \langle \nu F_{\nu} \rangle_{70} + 1.347 \langle \nu F_{\nu} \rangle_{160}}{\mathrm{erg\ s^{-1}\ cm^{-2}}}\label{eq_Mdust_no8um}
\end{eqnarray}

In Fig.~\ref{Mdust_empirical}b we see that this expression also
constitutes a good metric tracer for the dust mass. Note, however,
that the dust masses derived in this paper have been obtained in the
absence of spatially resolved submillimeter data, and therefore the
amount of dust colder than 15\,K might not be strongly
constrained. When analyzing the integrated properties of the SINGS
galaxies, D07 showed that omitting submillimeter data can lead to an
uncertainty of $\sim 50\%$ in the dust mass, and this should be taken
into account when using the previous equations to directly compute the
dust masses.

\section{Possible systematic effects in the dust models}\label{app_systematic}
When using the DL07 dust models to fit our radial infrared SEDs, we
have relied only on IRAC and MIPS data. Hence, it is worth analyzing
the possible systematics effects that may arise from using a limited
number of bands to constrain the models. D07 made use of a larger set
of multiwavelength data to study the global properties of the SINGS
galaxies. Besides IRAC and MIPS observations, they included in their
fit data from 2MASS, IRAS, SCUBA (for 17 galaxies) and the IRS ``blue
peak-up'' detector array onboard {\it Spitzer} (for 9 galaxies). While
useful for global analysis, the use of these additional data sets in a
spatially resolved analysis poses some problems regarding resolution,
depth and heterogeneous data quality.

To assess the validity of our methodology, we have compared our values
of the model parameters with the ones derived by D07 for each galaxy
as a whole. In principle, given a radial profile for each parameter in
a certain galaxy, it is not straightforward to reduce `a posteriori'
the spatially resolved information into a single data-point
representative of the whole galaxy. A more reliable and robust
procedure consists of applying the empirical formulae presented in
Appendix~\ref{app_empirical} to the IRAC and MIPS integrated
photometry used in D07. These recipes are affected by the same
systematic effects $-$if any$-$ as the values directly yielded by our
model fitting procedure, since they were explicitly calibrated using
those data. Should there be any strong biases in our values, they
should arise when comparing our estimates with the values computed by
D07.

\begin{figure}
\resizebox{1\hsize}{!}{\includegraphics{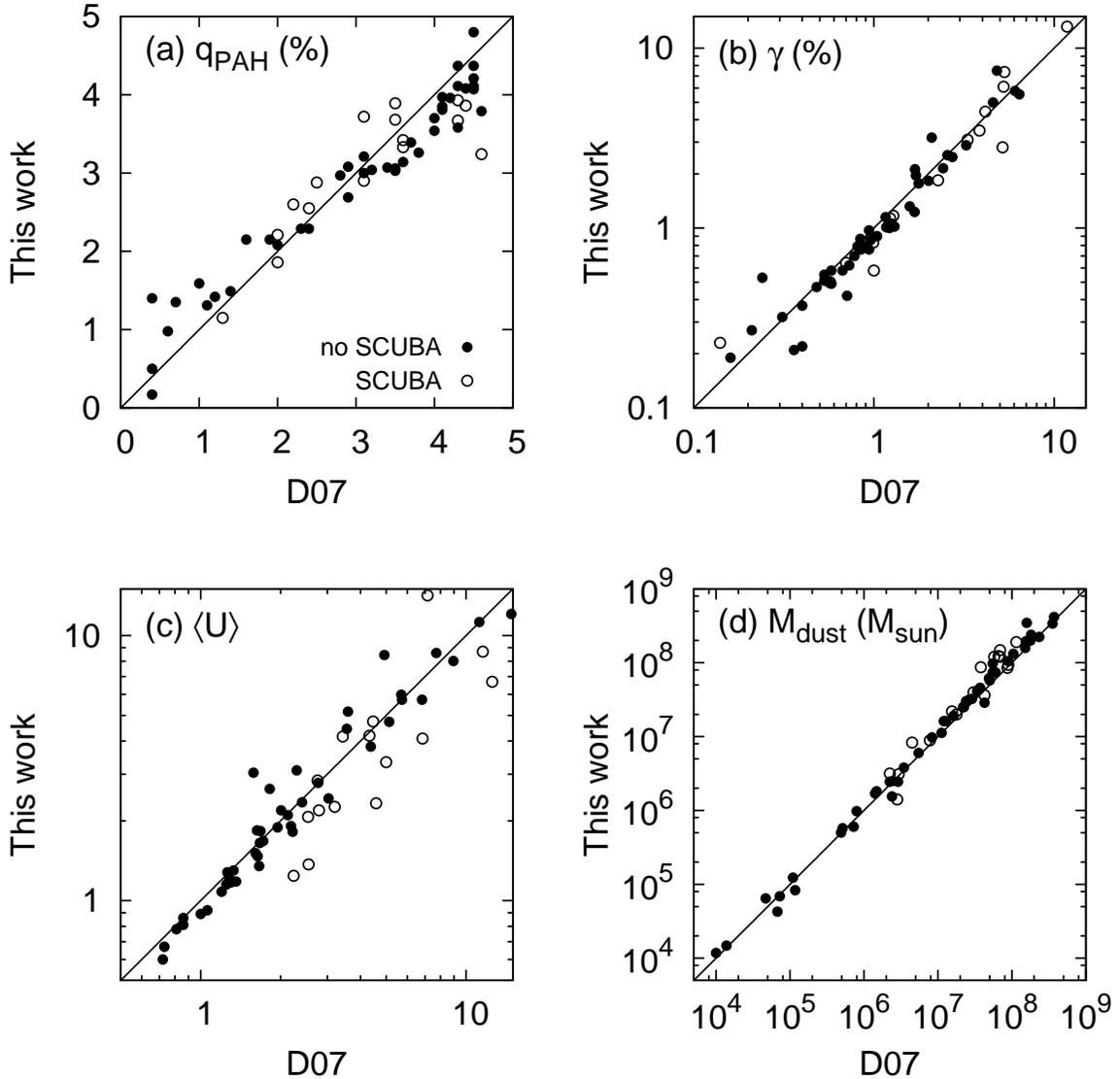}}
\caption{Comparison of the values of the dust model parameters derived
by D07 with the ones estimated using our recipes. The former were
obtained by using global photometry from 2MASS, IRAS and {\it Spitzer}
to constrain the dust models. Our estimates were computed by applying
the empirical formulae described in Appendix~\ref{app_empirical} to
the integrated IRAC and MIPS data alone. Open circles correspond to
galaxies for which submillimeter data from SCUBA were used by D07. The
mean offsets and scatter are quoted in Table~\ref{recipes_tab}. (a):
Abundance of PAHs. NGC~1291, NGC~1316, NGC~4125 and NGC~4594 have been
excluded, since their IRAC 8.0\,$\micron$ flux densities are highly
contaminated by stellar emission, leading to very uncertain values of
$q_{\mathrm{PAH}}$. (b): Fraction of the dust mass heated by intense
starlight. (c): Dust-weighted mean starlight intensity, in units of
the local MW radiation field. (d): Total dust mass.\label{comp_D07}}
\end{figure}

Table~\ref{recipes_tab} quotes the offsets and scatter between our
values and those published in D07. Note that part of the scatter is
due to the intrinsic rms of each recipe. In Fig.~\ref{comp_D07} we
compare the D07 values with the ones estimated using our
recipes. Panel (a) shows that our PAH abundances might be
overestimated at very low abundances, and underestimated at the
largest ones. The overall relative scatter is $\sim 30$\%.

In panel (b) we compare the values of the fraction of dust mass heated
by very intense starlight. The correlation is excellent, although our
values are systematically 6\% lower, with a rms of $\sim 25$\%.

The average intensity of the radiation field heating the dust that we
get from our SEDs is just 3\% lower than the one from D07, with a
scatter of 22\%, as can be seen in panel (c). The most deviant points
contributing to the dispersion are those with available SCUBA data,
for which we get lower dust temperatures.

As a result, our dust masses appear to be just slightly overestimated
in comparison with those obtained by D07 [see panel (d)]. For galaxies
lacking SCUBA data, our masses are only 6\% larger, with a rms of
$\sim 22$\%. In galaxies with available submillimeter data, the offset
increases up to 25\%, with a scatter of 40\%.

In brief, the correlations shown in Fig.~\ref{comp_D07} are good
enough to assert that the results presented in this paper do not
suffer from any strong bias due to our restricted wavelength range
that could compromise the validity of our conclusions.

\clearpage
\begin{deluxetable}{lrrrrrcrrccc}
\tabletypesize{\scriptsize}
\tablecolumns{12}
\rotate
\tablecaption{Sample\label{sample_tab}}
\tablewidth{0pt}
\tablehead{
\colhead{Object name} & \colhead{RA$_{2000}$} & \colhead{DEC$_{2000}$} & \colhead{$2a$} & \colhead{$2b$} &\colhead{P.A.} & \colhead{E(B$-$V)} & \colhead{$T$} & \colhead{dist} & \colhead{GALEX} & \colhead{THINGS} & \colhead{CO} \\
\colhead{} & \colhead{(h:m:s)} & \colhead{(d:m:s)} & \colhead{(arcmin)} & \colhead{(arcmin)} & \colhead{(deg)} & \colhead{(mag)} & \colhead{type} & \colhead{(Mpc)} & \colhead{data} & \colhead{data} & \colhead{data}\\
\colhead{(1)} & \colhead{(2)} & \colhead{(3)} & \colhead{(4)} & \colhead{(5)} & \colhead{(6)} & \colhead{(7)} & \colhead{(8)} & \colhead{(9)} & \colhead{(10)} & \colhead{(11)} & \colhead{(12)}}
\startdata
   NGC~0024 & 00 09 56.5 & $-$24 57 47.3 &   5.8 &   1.3 & 46 & 0.020 & 5 & 8.2 & yes & no & \nodata\\
   NGC~0337 & 00 59 50.1 & $-$07 34 40.7 &   2.9 &   1.8 & 310 & 0.112 & 7 &  25 & yes & no & \nodata\\
   NGC~0628 & 01 36 41.8 & 15 47 00.5 &  10.5 &   9.5 & 25 & 0.070 & 5 &  11 & yes & yes & 1\\
   NGC~0855 & 02 14 03.6 & 27 52 37.8 &   2.6 &   1.0 & 60 & 0.072 & $-$5 & 9.7 & yes & no & \nodata\\
   NGC~0925 & 02 27 16.9 & 33 34 45.0 &  10.5 &   5.9 & 282 & 0.076 & 7 & 9.3 & yes & yes & 2\\
   NGC~1097 & 02 46 19.1 & $-$30 16 29.7 &   9.3 &   6.3 & 310 & 0.027 & 3 &  15 & yes & no & 3\\
   NGC~1291 & 03 17 18.6 & $-$41 06 29.1 &   9.8 &   8.1 & 345 & 0.013 & 0 & 9.7 & yes & no & \nodata\\
   NGC~1316 & 03 22 41.7 & $-$37 12 29.6 &  12.0 &   8.5 & 50 & 0.021 & $-$2 &  19 & yes & no & \nodata\\
   NGC~1512 & 04 03 54.3 & $-$43 20 55.9 &   8.9 &   5.6 & 90 & 0.011 & 1 &  10 & yes & no & \nodata\\
   NGC~1566 & 04 20 00.4 & $-$54 56 16.1 &   8.3 &   6.6 & 60 & 0.009 & 4 &  17 & yes & no & 4\\
   NGC~1705 & 04 54 13.5 & $-$53 21 39.8 &   1.9 &   1.4 & 50 & 0.008 & 11 & 5.1 & yes & no & \nodata\\
   NGC~2403 & 07 36 51.4 & 65 36 09.2 &  21.9 &  12.3 & 307 & 0.040 & 6 & 3.2 & yes & yes & 3\\
      Holmberg~II & 08 19 05.0 & 70 43 12.1 &   7.9 &   6.3 & 15 & 0.032 & 10 & 3.4 & yes & yes & \nodata\\
    DDO~053 & 08 34 07.2 & 66 10 54.0 &   1.5 &   1.3 & 300 & 0.037 & 10 & 3.6 & yes & yes & \nodata\\
   NGC~2841 & 09 22 02.6 & 50 58 35.5 &   8.1 &   3.5 & 327 & 0.016 & 3 &  14 & yes & yes & 5\\
   NGC~2915 & 09 26 11.5 & $-$76 37 34.8 &   1.9 &   1.0 & 309 & 0.275 & 90 & 3.8 & yes & no & \nodata\\
       Holmberg~I & 09 40 32.3 & 71 10 56.0 &   3.6 &   3.0 & 360 & 0.048 & 10 & 3.8 & yes & yes & \nodata\\
   NGC~2976 & 09 47 15.5 & 67 54 59.0 &   5.9 &   2.7 & 323 & 0.069 & 5 & 3.6 & yes & yes & \nodata\\
   NGC~3049 & 09 54 49.7 & 09 16 17.9 &   2.2 &   1.4 & 25 & 0.038 & 2 &  22 & NUV & no & \nodata\\
   NGC~3031 & 09 55 33.2 & 69 03 55.1 &  26.9 &  14.1 & 337 & 0.080 & 2 & 3.6 & yes & yes & 2\\
   NGC~3190 & 10 18 05.6 & 21 49 55.0 &   4.4 &   1.5 & 305 & 0.025 & 1 &  17 & yes & no & \nodata\\
   NGC~3184 & 10 18 17.0 & 41 25 28.0 &   7.4 &   6.9 & 135 & 0.017 & 6 & 8.6 & yes & no & \nodata\\
   NGC~3198 & 10 19 54.9 & 45 32 59.0 &   8.5 &   3.3 & 35 & 0.012 & 5 &  17 & yes & yes & \nodata\\
    IC~2574 & 10 28 23.5 & 68 24 43.7 &  13.2 &   5.4 & 50 & 0.036 & 9 & 4.0 & yes & yes & \nodata\\
   NGC~3351 & 10 43 57.7 & 11 42 13.0 &   7.4 &   5.0 & 13 & 0.028 & 3 &  12 & yes & yes & \nodata\\
   NGC~3521 & 11 05 48.6 & $-$00 02 09.1 &  11.0 &   5.1 & 343 & 0.058 & 4 & 9.0 & yes & yes & 1\\
   NGC~3621 & 11 18 16.5 & $-$32 48 50.6 &  12.3 &   7.1 & 339 & 0.080 & 7 & 8.3 & yes & yes & \nodata\\
   NGC~3627 & 11 20 15.0 & 12 59 29.6 &   9.1 &   4.2 & 353 & 0.032 & 3 & 9.1 & yes & yes & 1\\
   NGC~3938 & 11 52 49.4 & 44 07 14.6 &   5.4 &   4.9 & 15 & 0.021 & 5 &  12 & NUV & no & \nodata\\
   NGC~4125 & 12 08 06.0 & 65 10 26.9 &   5.8 &   3.2 & 275 & 0.019 & $-$5 &  21 & NUV & no & \nodata\\
   NGC~4236 & 12 16 42.1 & 69 27 45.3 &  21.9 &   7.2 & 342 & 0.015 & 8 & 4.5 & yes & no & \nodata\\
   NGC~4254 & 12 18 49.6 & 14 24 59.4 &   5.4 &   4.7 & 35 & 0.039 & 5 &  17 & NUV & no & \nodata\\
   NGC~4321 & 12 22 54.9 & 15 49 20.6 &   7.4 &   6.3 & 30 & 0.026 & 4 &  18 & NUV & no & \nodata\\
   NGC~4450 & 12 28 29.6 & 17 05 05.8 &   5.2 &   3.9 & 355 & 0.028 & 2 &  17 & NUV & no & \nodata\\
   NGC~4536 & 12 34 27.1 & 02 11 16.4 &   7.6 &   3.2 & 310 & 0.018 & 4 &  15 & yes & no & \nodata\\
   NGC~4559 & 12 35 57.7 & 27 57 35.1 &  10.7 &   4.4 & 330 & 0.018 & 6 &  17 & yes & no & \nodata\\
   NGC~4569 & 12 36 49.8 & 13 09 46.3 &   9.5 &   4.4 & 23 & 0.046 & 2 &  17 & yes & no & 1,6\\
   NGC~4579 & 12 37 43.6 & 11 49 05.1 &   5.9 &   4.7 & 275 & 0.041 & 3 &  17 & yes & no & \nodata\\
   NGC~4594 & 12 39 59.4 & $-$11 37 23.0 &   8.7 &   3.5 & 90 & 0.051 & 1 & 9.1 & yes & no & \nodata\\
   NGC~4625 & 12 41 52.7 & 41 16 25.4 &   2.2 &   1.9 & 330 & 0.018 & 9 & 9.5 & yes & no & \nodata\\
   NGC~4631 & 12 42 08.0 & 32 32 29.4 &  15.5 &   2.7 & 86 & 0.017 & 7 & 9.0 & yes & no & \nodata\\
   NGC~4725 & 12 50 26.6 & 25 30 02.7 &  10.7 &   7.6 & 35 & 0.012 & 2 &  17 & yes & no & \nodata\\
   NGC~4736 & 12 50 53.1 & 41 07 13.6 &  11.2 &   9.1 & 285 & 0.018 & 2 & 5.2 & yes & yes & 1\\
   NGC~4826 & 12 56 43.8 & 21 40 51.9 &  10.0 &   5.4 & 295 & 0.041 & 2 &  17 & yes & yes & 1\\
   NGC~5033 & 13 13 27.5 & 36 35 38.0 &  10.7 &   5.0 & 170 & 0.011 & 5 &  13 & yes & no & \nodata\\
   NGC~5055 & 13 15 49.3 & 42 01 45.4 &  12.6 &   7.2 & 285 & 0.018 & 4 & 8.2 & yes & yes & 1\\
   NGC~5194\ \ddag & 13 29 52.7 & 47 11 42.6 &  11.2 &   9.0 & 360 & 0.035 & 4 & 8.4 & yes & yes & 7\\
     TOL~89 & 14 01 21.6 & $-$33 03 49.6 &   2.8 &   1.7 & 352 & 0.066 & 8.1 &  16 & yes & no & \nodata\\
   NGC~5408 & 14 03 20.9 & $-$41 22 40.0 &   1.6 &   0.8 & 12 & 0.069 & 9.7 & 4.5 & no & no & \nodata\\
   NGC~5474 & 14 05 01.6 & 53 39 44.0 &   4.8 &   4.3 & 360 & 0.011 & 6 & 6.8 & yes & no & \nodata\\
   NGC~5713 & 14 40 11.5 & $-$00 17 21.2 &   2.8 &   2.5 & 10 & 0.039 & 4 &  27 & yes & no & \nodata\\
   NGC~5866 & 15 06 29.6 & 55 45 47.9 &   4.7 &   1.9 & 308 & 0.013 & $-$1 &  15 & yes & no & \nodata\\
    IC~4710 & 18 28 38.0 & $-$66 58 56.0 &   3.6 &   2.8 &  5 & 0.089 & 9 & 8.5 & yes & no & \nodata\\
   NGC~6822 & 19 44 56.6 & $-$14 47 21.4 &  15.5 &  13.5 &  7 & 0.236 & 10 & 0.60 & yes & no & \nodata\\
   NGC~6946 & 20 34 52.3 & 60 09 14.2 &  11.5 &   9.8 & 75 & 0.342 & 6 & 5.5 & yes & no & \nodata\\
   NGC~7331 & 22 37 04.1 & 34 24 56.3 &  10.5 &   3.7 & 351 & 0.091 & 3 &  15 & yes & yes & 5\\
   NGC~7793 & 23 57 49.8 & $-$32 35 27.7 &   9.3 &   6.3 & 278 & 0.019 & 7 & 2.0 & yes & yes & \nodata\\
\enddata
\tablecomments{Sample. (1): Galaxy name. (2): RA(J2000) of the galaxy center. (3): DEC(J2000) of the galaxy center. (4),(5): Apparent major and minor isophotal diameters at $\mu_{B}$=25 mag arcsec$^{-2}$ from the RC3 catalog. (6): Position angle from RC3. \ddag The PA and axis ratio for NGC~5194 differ from those in the RC3, which are affected by the presence of NGC~5195. (7): Galactic color excess from Schlegel et al$.$ (1998). (8): Morphological type $T$ as given in the RC3 catalog. (9): Distance to the galaxy, rounded to the nearest Mpc when larger than 10 Mpc, taken from Gil de Paz et al$.$ (2007) and Kennicutt et al$.$ (2003). (10): Available GALEX images. (11): Available HI maps from THINGS. (12): References for the CO data. 1: Regan et al$.$ (2001). 2: Sage (1993). 3: Young et al$.$ (1995). 4: Bajaja et al$.$ (1995). 5: Young \& Scoville (1982). 6: Kenney \& Young (1988). 7: Paglione et al$.$ (2001).}
\end{deluxetable}

\begin{deluxetable}{rrrrrrrrrrr}
\tabletypesize{\scriptsize}
\tablecolumns{11}
\setlength{\tabcolsep}{0.045in}
\tablecaption{Degraded GALEX, IRAC and MIPS profiles.\label{phot_profiles_tab}}
\tablewidth{0pt}
\tablehead{
\colhead{$r$} & \colhead{$r$} & \colhead{$\mu_{\mathrm{FUV}}$} & \colhead{$\mu_{\mathrm{NUV}}$} & \colhead{$\mu_{3.6\micron}$} & \colhead{$\mu_{4.5\micron}$} & \colhead{$\mu_{5.8\micron}$} & \colhead{$\mu_{3.6\micron}$} & \colhead{$\mu_{24\micron}$} & \colhead{$\mu_{70\micron}$} & \colhead{$\mu_{160\micron}$}\\
\colhead{(arcsec)} & \colhead{(kpc)} & \colhead{(mag/\sq)} & \colhead{(mag/\sq)} & \colhead{(mag/\sq)} & \colhead{(mag/\sq)} & \colhead{(mag/\sq)} & \colhead{(mag/\sq)} & \colhead{(mag/\sq)} & \colhead{(mag/\sq)} & \colhead{(mag/\sq)} \\
\colhead{(1)} & \colhead{(2)} & \colhead{(3)} & \colhead{(4)} & \colhead{(5)} & \colhead{(6)} & \colhead{(7)} & \colhead{(8)} & \colhead{(9)} & \colhead{(10)} & \colhead{(11)}}
\startdata
\cutinhead{NGC~7331}
    0 &   0.0 &24.99$\pm$0.02  &23.74$\pm$0.01  &18.41$\pm$0.01  &18.88$\pm$0.01  &18.48$\pm$0.01  &17.81$\pm$0.01  &17.82$\pm$0.01  &15.08$\pm$0.01  &14.11$\pm$0.01  \\
   48 &   3.5 &25.08$\pm$0.01  &24.01$\pm$0.00  &18.99$\pm$0.01  &19.45$\pm$0.01  &18.90$\pm$0.01  &18.12$\pm$0.01  &18.04$\pm$0.01  &15.16$\pm$0.01  &14.15$\pm$0.01  \\
   96 &   6.9 &25.31$\pm$0.01  &24.53$\pm$0.01  &20.03$\pm$0.01  &20.47$\pm$0.01  &19.74$\pm$0.01  &18.86$\pm$0.01  &18.83$\pm$0.01  &15.71$\pm$0.01  &14.82$\pm$0.01  \\
  144 &  10.4 &25.62$\pm$0.01  &24.97$\pm$0.01  &20.78$\pm$0.01  &21.22$\pm$0.01  &20.52$\pm$0.02  &19.65$\pm$0.01  &19.78$\pm$0.01  &16.56$\pm$0.01  &15.60$\pm$0.01  \\
  192 &  13.8 &26.07$\pm$0.01  &25.47$\pm$0.01  &21.39$\pm$0.01  &21.84$\pm$0.02  &21.20$\pm$0.04  &20.36$\pm$0.02  &20.59$\pm$0.02  &17.39$\pm$0.03  &16.28$\pm$0.01  \\
  240 &  17.3 &26.52$\pm$0.01  &25.99$\pm$0.02  &22.06$\pm$0.02  &22.50$\pm$0.03  &21.88$\pm$0.07  &21.07$\pm$0.04  &21.34$\pm$0.04  &18.10$\pm$0.05  &17.03$\pm$0.02  \\
  288 &  20.7 &27.30$\pm$0.01  &26.74$\pm$0.04  &22.77$\pm$0.04  &23.20$\pm$0.06  &22.66$\pm$0.14  &21.93$\pm$0.09  &22.28$\pm$0.10  &18.79$\pm$0.09  &17.82$\pm$0.04  \\
  336 &  24.2 &28.14$\pm$0.03  &27.51$\pm$0.08  &23.32$\pm$0.07  &23.75$\pm$0.10  &23.33$\pm$0.26  &22.68$\pm$0.17  &23.11$\pm$0.22  &19.44$\pm$0.16  &18.43$\pm$0.07  \\
  384 &  27.7 &28.88$\pm$0.05  &28.15$\pm$0.15  &23.69$\pm$0.09  &24.12$\pm$0.14  &23.79$\pm$0.39  &23.18$\pm$0.28  &23.69$\pm$0.37  &19.81$\pm$0.22  &18.95$\pm$0.22  \\
  432 &  31.1 &29.43$\pm$0.07  &28.58$\pm$0.22  &23.98$\pm$0.12  &24.42$\pm$0.18  &24.15$\pm$0.54  &23.53$\pm$0.38  &23.96$\pm$0.47  &20.06$\pm$0.28  &19.38$\pm$0.17  \\
\enddata
\tablecomments{Surface brightness profiles measured on the GALEX, IRAC and MIPS images, after matching their PSFs to that of the 160\,$micron$ band. Only the profiles for NGC~7331 are shown as an example in the printed version of the jornal. Check the on-line edition for the full table. (1): Radius along the semi-major axis in arcseconds. (2): Radius along the semimajor axis in kpc. (3)-(11): Surface brightness at different wavelengths. All values are in AB mag arcsec$^{-2}$. The uncertainties include photometric and background errors, but not zero-point uncertainties, which are added in quadrature when needed.}
\end{deluxetable}

\clearpage
\begin{deluxetable}{rrrrrrrrr}
\tabletypesize{\scriptsize}
\tablecolumns{9}
\tablecaption{Extinction profiles\label{extprof_tab}}
\tablewidth{0pt}
\tablehead{
\colhead{} & \colhead{} & \colhead{} & \colhead{} & \multicolumn{2}{c}{B05} & & \multicolumn{2}{c}{C08}\\
\cline{5-6}  \cline{8-9}
\colhead{$r$} & \colhead{$r$} & \colhead{$\log(L_{\mathrm{TIR}}/L_{\mathrm{FUV}})$} & \colhead{$\log(L_{\mathrm{TIR}}/L_{\mathrm{NUV}})$} & \colhead{$A_{\mathrm{FUV}}$} &\colhead{$A_{\mathrm{NUV}}$} & & \colhead{$A_{\mathrm{FUV}}$} & \colhead{$A_{\mathrm{NUV}}$}\\
\colhead{(arcsec)} & \colhead{(kpc)} & \colhead{} & \colhead{} & \colhead{(mag)} & \colhead{(mag)} & & \colhead{(mag)} & \colhead{(mag)}\\
\colhead{(1)} & \colhead{(2)} & \colhead{(3)} & \colhead{(4)} & \colhead{(5)} & \colhead{(6)} & & \colhead{(7)} & \colhead{(8)}}
\startdata
\cutinhead{NGC~7331}
   0 &  0.0 &  1.76 $\pm$  0.07 &  1.44 $\pm$  0.07 &	 3.52$^{+ 0.11}_{- 0.11}$ &  2.35$^{+ 0.10}_{- 0.10}$ & &	 2.45$^{+ 0.10}_{- 0.10}$ &  1.95$^{+ 0.10}_{- 0.10}$ \\
  48 &  3.5 &  1.76 $\pm$  0.07 &  1.50 $\pm$  0.07 &	 3.50$^{+ 0.11}_{- 0.11}$ &  2.48$^{+ 0.10}_{- 0.10}$ & &	 2.71$^{+ 0.11}_{- 0.11}$ &  2.21$^{+ 0.11}_{- 0.11}$ \\
  96 &  6.9 &  1.59 $\pm$  0.07 &  1.45 $\pm$  0.07 &	 3.16$^{+ 0.10}_{- 0.10}$ &  2.38$^{+ 0.10}_{- 0.10}$ & &	 2.69$^{+ 0.11}_{- 0.11}$ &  2.26$^{+ 0.12}_{- 0.12}$ \\
 144 & 10.4 &  1.39 $\pm$  0.07 &  1.30 $\pm$  0.07 &	 2.75$^{+ 0.10}_{- 0.10}$ &  2.09$^{+ 0.09}_{- 0.09}$ & &	 2.41$^{+ 0.10}_{- 0.10}$ &  2.01$^{+ 0.10}_{- 0.10}$ \\
 192 & 13.8 &  1.27 $\pm$  0.07 &  1.21 $\pm$  0.07 &	 2.51$^{+ 0.09}_{- 0.09}$ &  1.91$^{+ 0.09}_{- 0.09}$ & &	 2.21$^{+ 0.10}_{- 0.10}$ &  1.83$^{+ 0.10}_{- 0.10}$ \\
 240 & 17.3 &  1.16 $\pm$  0.07 &  1.12 $\pm$  0.07 &	 2.30$^{+ 0.09}_{- 0.09}$ &  1.75$^{+ 0.09}_{- 0.09}$ & &	 2.04$^{+ 0.09}_{- 0.09}$ &  1.69$^{+ 0.09}_{- 0.09}$ \\
 288 & 20.7 &  1.16 $\pm$  0.07 &  1.11 $\pm$  0.07 &	 2.30$^{+ 0.09}_{- 0.09}$ &  1.74$^{+ 0.09}_{- 0.09}$ & &	 2.03$^{+ 0.10}_{- 0.10}$ &  1.66$^{+ 0.10}_{- 0.10}$ \\
 336 & 24.2 &  1.23 $\pm$  0.07 &  1.15 $\pm$  0.08 &	 2.44$^{+ 0.10}_{- 0.10}$ &  1.82$^{+ 0.11}_{- 0.11}$ & &	 2.11$^{+ 0.10}_{- 0.10}$ &  1.70$^{+ 0.11}_{- 0.11}$ \\
 384 & 27.7 &  1.34 $\pm$  0.08 &  1.22 $\pm$  0.10 &	 2.65$^{+ 0.12}_{- 0.12}$ &  1.94$^{+ 0.14}_{- 0.14}$ & &	 2.23$^{+ 0.12}_{- 0.12}$ &  1.79$^{+ 0.14}_{- 0.14}$ \\
 432 & 31.1 &  1.43 $\pm$  0.09 &  1.26 $\pm$  0.12 &	 2.82$^{+ 0.14}_{- 0.14}$ &  2.01$^{+ 0.18}_{- 0.18}$ & &	 2.32$^{+ 0.14}_{- 0.14}$ &  1.83$^{+ 0.18}_{- 0.18}$ \\
\enddata
\tablecomments{Radial extinction profiles. (1): Radius along semimajor axis in arcsec. (2): Radius along semimajor axis in kpc. (3): TIR-to-FUV ratio. (4): TIR-to-NUV ratio. (5), (6): Extinction at FUV and NUV computed with the SFH-independent fits of Buat et al$.$ (2005). (7), (8): Extinction at FUV and NUV computed with the SFH-dependent prescriptions of Cortese et al$.$ (2008).}
\end{deluxetable}

\clearpage
\begin{deluxetable}{rrlrrrrrrl}
\tabletypesize{\scriptsize}
\tablecolumns{10}
\tablecaption{Model parameters profiles\label{model_tab}}
\tablewidth{0pt}
\tablehead{
\colhead{$r$} & \colhead{$r$} & \colhead{$q_{\mathrm{PAH}}$} & \colhead{$\gamma$} & \colhead{$f(U>10^{2})$} & \colhead{$U_{\mathrm{min}}$} & \colhead{$\langle U \rangle$} & \colhead{$\log \Sigma_{L_{\mathrm{dust}}}$} & \colhead{$\log \Sigma_{M_{\mathrm{dust}}}$} & \colhead{$\log (M_{\mathrm{dust}}/M_{\mathrm{gas}})$}\\
\colhead{(arcsec)} & \colhead{(kpc)} & \colhead{(\%)} & \colhead{(\%)} & \colhead{(\%)} & \colhead{} & \colhead{} & \colhead{($L_{\sun}\mathrm{\,kpc^{-2}}$)} & \colhead{$(M_{\sun}\mathrm{\,kpc^{-2}})$} & \colhead{}\\
\colhead{(1)} & \colhead{(2)} & \colhead{(3)} & \colhead{(4)} & \colhead{(5)} & \colhead{(6)} & \colhead{(7)} & \colhead{(8)} & \colhead{(9)} & \colhead{(10)}}
\startdata
\cutinhead{NGC~7331}
   0 &  0.0 & 4.6$^{+0.0}_{-0.3}$ &   0.9$^{+ 0.4}_{- 0.4}$ &  7.5$^{+ 4.3}_{- 6.3}$ &  2.5$^{+ 0.8}_{- 0.9}$ &  2.8$^{+ 0.7}_{- 1.1}$ &  8.27$^{+ 0.04}_{- 0.01}$ &  5.70$^{+ 0.19}_{- 0.13}$ & $-$1.55$^{+ 0.19}_{- 0.13}$\\
  48 &  3.5 & 4.2$^{+0.4}_{-0.4}$ &   0.7$^{+ 0.4}_{- 0.3}$ &  5.9$^{+ 5.2}_{- 5.9}$ &  2.0$^{+ 1.1}_{- 0.5}$ &  2.2$^{+ 1.2}_{- 0.6}$ &  8.22$^{+ 0.04}_{- 0.03}$ &  5.76$^{+ 0.16}_{- 0.25}$ & $-$1.88$^{+ 0.16}_{- 0.25}$\\
  96 &  6.9 & 4.0$^{+0.6}_{-0.4}$ &   0.5$^{+ 0.3}_{- 0.3}$ &  4.3$^{+ 6.2}_{- 4.3}$ &  2.5$^{+ 1.4}_{- 0.6}$ &  2.6$^{+ 1.3}_{- 0.8}$ &  7.96$^{+ 0.05}_{- 0.06}$ &  5.41$^{+ 0.12}_{- 0.20}$ & $-$1.94$^{+ 0.12}_{- 0.20}$\\
 144 & 10.4 & 4.3$^{+0.3}_{-0.5}$ &   0.1$^{+ 0.3}_{- 0.1}$ &  0.9$^{+ 2.9}_{- 0.9}$ &  2.5$^{+ 1.1}_{- 0.7}$ &  2.5$^{+ 1.1}_{- 0.7}$ &  7.63$^{+ 0.07}_{- 0.03}$ &  5.09$^{+ 0.14}_{- 0.15}$ & $-$2.00$^{+ 0.15}_{- 0.16}$\\
 192 & 13.8 & 4.3$^{+0.3}_{-0.4}$ &   0.0$^{+ 0.3}_{- 0.0}$ &  0.0$^{+ 2.4}_{- 0.0}$ &  2.0$^{+ 0.9}_{- 0.6}$ &  2.0$^{+ 1.2}_{- 0.5}$ &  7.33$^{+ 0.07}_{- 0.04}$ &  4.90$^{+ 0.16}_{- 0.18}$ & $-$2.09$^{+ 0.18}_{- 0.20}$\\
 240 & 17.3 & 4.4$^{+0.2}_{-0.5}$ &   0.0$^{+ 0.2}_{- 0.0}$ &  0.0$^{+ 2.0}_{- 0.0}$ &  2.0$^{+ 1.1}_{- 0.5}$ &  2.0$^{+ 1.3}_{- 0.4}$ &  7.04$^{+ 0.06}_{- 0.04}$ &  4.61$^{+ 0.19}_{- 0.22}$ & $-$2.23$^{+ 0.25}_{- 0.27}$\\
 288 & 20.7 & 4.0$^{+0.4}_{-0.5}$ &   0.0$^{+ 0.1}_{- 0.0}$ &  0.0$^{+ 1.6}_{- 0.0}$ &  2.5$^{+ 1.0}_{- 0.8}$ &  2.5$^{+ 1.1}_{- 0.8}$ &  6.72$^{+ 0.04}_{- 0.02}$ &  4.19$^{+ 0.17}_{- 0.14}$ & $-$2.39$^{+ 0.29}_{- 0.27}$\\
 336 & 24.2 & 3.6$^{+0.6}_{-0.6}$ &   0.0$^{+ 0.2}_{- 0.0}$ &  0.0$^{+ 2.4}_{- 0.0}$ &  2.0$^{+ 1.3}_{- 0.6}$ &  2.0$^{+ 1.5}_{- 0.6}$ &  6.46$^{+ 0.05}_{- 0.06}$ &  4.02$^{+ 0.21}_{- 0.25}$ & $-$2.23$^{+ 0.33}_{- 0.35}$\\
 384 & 27.7 & 3.3$^{+0.8}_{-0.7}$ &   0.0$^{+ 0.3}_{- 0.0}$ &  0.0$^{+ 4.4}_{- 0.0}$ &  2.5$^{+ 2.0}_{- 0.9}$ &  2.5$^{+ 2.0}_{- 0.8}$ &  6.27$^{+ 0.04}_{- 0.03}$ &  3.74$^{+ 0.14}_{- 0.28}$ & $-$2.21$^{+ 0.30}_{- 0.38}$\\
 432 & 31.1 & 3.1$^{+1.1}_{-0.7}$ &   0.0$^{+ 0.5}_{- 0.0}$ &  0.0$^{+ 8.6}_{- 0.0}$ &  4.0$^{+ 2.8}_{- 2.1}$ &  4.0$^{+ 1.3}_{- 2.1}$ &  6.14$^{+ 0.07}_{- 0.04}$ &  3.41$^{+ 0.31}_{- 0.20}$ & $-$2.26$^{+ 0.45}_{- 0.38}$\\
\enddata
\tablecomments{Radial profiles of the model parameters. (1): Radius along semimajor axis in arcsec. (2): Radius along semimajor axis in kpc. (3): Fraction of the dust mass in the form of PAHs. This value is not reliable in regions marked with a dagger (\dag), where the dust contribution to the observed 8\,$\micron$ flux is lees than half of the stellar emission at that band. (4): Fraction of the dust mass heated by very intense starlight. (5): Fraction of the dust luminosity contributed by regions heated by starlight with $U>10^{2}$. (6): Minimum value for the starlight scale factor, in units of the local MW radiation field. (7): Dust-weighted average scale factor for the starlight intensity. (8): Dust luminosity surface density, corrected for inclination. (9): Dust mass surface density, corrected for inclination. (10): Dust-to-gas ratio. The dust mass is computed as $M_{\mathrm{gas}}=1.36 \times (M_{\mathrm{HI}}+M_{\mathrm{H_{2}}})$, or $M_{\mathrm{gas}}=1.36 \times M_{\mathrm{HI}}$ when CO data are not available. In the latter case, the values are marked with a double dagger (\ddag), and should be treated as upper limits (at least in the inner regions).}
\end{deluxetable}

\clearpage
\begin{deluxetable}{crrrrrrrrr}
\tabletypesize{\scriptsize}
\tablecolumns{10}
\tablecaption{Accuracy of the recipes\label{recipes_tab}}
\tablewidth{0pt}
\tablehead{
\colhead{} & \colhead{} & \colhead{} & \colhead{} & \colhead{} & \multicolumn{5}{c}{Comparison with D07}\\
\cline{6-10}
\colhead{Parameter$^1$} & \colhead{} & \multicolumn{2}{c}{Empirical fit$^2$} & \colhead{} & \multicolumn{2}{c}{no SCUBA$^3$} & \colhead{} &\multicolumn{2}{c}{SCUBA$^4$}\\
\cline{3-4}
\cline{6-7}
\cline{9-10}
\colhead{} & \colhead{} & \colhead{offset} & \colhead{rms} & \colhead{} & \colhead{offset} & \colhead{rms} & \colhead{} & \colhead{offset} & \colhead{rms}
}
\startdata
$q_{\mathrm{PAH}}$  & & $0\%$  & $\pm 11\%$  & & $+2.7\%$ & $\pm 30\%$ & & $-1.1\%$ & $\pm 14\%$\\
$\gamma$            & & $0\%$  & $\pm 25\%$  & & $-5.7\%$ & $\pm 23\%$ & & $-6.4\%$ & $\pm 27\%$\\
$\langle U \rangle$ & & $0\%$  & $\pm 9.2\%$ & & $-2.9\%$ & $\pm 22\%$ & & $-22\%$  & $\pm 35\%$\\
$M_{\mathrm{dust}}$ & & $+5\%$ & $\pm 9.6\%$ & & $+11\%$  & $\pm 22\%$ & & $+30\%$  & $\pm 39\%$\\
$0.95 \times M_{\mathrm{dust}}$ & & $0\%$ & $\pm 9.6\%$ & & $+5.8\%$ & $\pm 22\%$ & & $+25\%$  & $\pm 39\%$\\
\enddata

\tablecomments{Accuracy of the empirical recipes for each
model parameter (Appendix~\ref{app_empirical}) and comparison with the
results of Draine et al$.$ (2007) (Appendix~\ref{app_systematic}).
(1): Each one of the dust model parameters. (2): Offset and rms of the
empirical fits. By definition, the offset is zero for all parameters,
since the corresponding estimators are derived from fits applied to
our data. The only exception is the dust mass, for which the empirical
recipe is directly obtained by combining the estimators for $\langle U
\rangle$ and $L_{\mathrm{dust}}$. This leads to dust masses 5\% larger
than the ones yielded by the model (see
Appendix~\ref{S_Mdust_empirical}). If desired, this can be accounted
for by multiplying by 0.95 (last row). (3): Comparison between the
values published by D07, obtained by fitting the global SEDs including
data from 2MASS, IRAC, MIPS and IRAS, and the ones we estimate by
applying our recipes to their IRAC and MIPS data alone. The first
number indicates the relative offset (our value/D07 value) and the
second one the rms. (4): Same as (3), but for the galaxies for which
submillimeter data from SCUBA were added to the data-set used in D07.}
\end{deluxetable}

\clearpage

\clearpage

\clearpage

\clearpage

\clearpage

\clearpage

\clearpage

\clearpage

\clearpage

\clearpage

\clearpage

\clearpage

\clearpage

\clearpage

\clearpage

\clearpage

\clearpage

\clearpage

\clearpage

\clearpage

\clearpage

\end{document}